\documentclass[useAMS,usenatbib]{mn2e}
\usepackage{times}
\usepackage{natbib}
\usepackage{graphics}
\usepackage{graphicx}
\usepackage{color}
\usepackage{verbatim}
\usepackage{amsmath}
\usepackage{amssymb}
\usepackage{mathtools}
\usepackage{subfigure}
\usepackage{float}
\usepackage{tabularx}
\usepackage{supertabular}
\usepackage{rotating}
\usepackage{longtable}
\usepackage{dpfloat, booktabs}
\usepackage{color}
\usepackage{soul}
\usepackage{multicol}
\usepackage{graphicx}
\usepackage{multirow}               
\usepackage{amsfonts}

\usepackage{pifont}

\usepackage{wrapfig}
\usepackage{subfig}   

\definecolor{Mygrey}{gray}{0.75}

\title[CO TFR of $z=0.05$ -- $0.3$ galaxies] {CO Tully-Fisher relation
  of star-forming galaxies at $z=0.05$ -- $0.3$}
\author[S.\ Topal et al.]  {Sel\c{c}uk Topal,$^{1,2}$\thanks{E-mail:
    selcuktopal@yyu.edu.tr}
  Martin Bureau,$^{2}$ Alfred L. Tiley,$^{2,3}$ Timothy A.\ Davis,$^{4}$ and Kazufumi Torii$^{5}$\\
  $^{1}$Department of Physics, Van Yuzuncu Yil University, Van, 65080,
  Turkey\\
   $^{2}$Sub-department of Astrophysics, University of Oxford, Denys
  Wilkinson Building, Keble Road, Oxford OX1~3RH, U.K.\\
  $^{3}$Centre for Extragalactic Astronomy, Department of Physics,
  Durham University, South Road, Durham, DH1 3LE, U.K.\\
  $^{4}$School of Physics \&\ Astronomy, Cardiff University, Queens
  Buildings, The Parade, Cardiff, CF24 3AA, UK,\\
  $^{5}$Nobeyama Radio Observatory, 462-2 Nobeyama Minamimaki-mura,
  Minamisaku-gun, Nagano 384-1305, Japan\\}

\begin{document}
\date{Accepted . Received ; in original form }
\pagerange{\pageref{firstpage}-\pageref{lastpage}} \pubyear{2014}
\maketitle
\label{firstpage}
%
%
\begin{abstract}
  The Tully-Fisher relation (TFR) is an empirical relation between
  galaxy luminosity and rotation velocity. We present here the first
  TFR of galaxies beyond the local Universe that uses carbon monoxide
  (CO) as the kinematic tracer. Our final sample includes
  $25$ isolated, non-interacting star-forming
  galaxies with double-horned or boxy CO integrated
    line profiles located at redshifts $z\le0.3$, drawn from a larger
  ensemble of $67$ detected objects. The best reverse
  $K_{\text{s}}$-band, stellar mass and baryonic mass
    CO TFRs are respectively
    $M_{K_{\text{s}}}=(-8.4\pm2.9)\left[\log{\left(\frac{W_{50}/\rm{km~s^{-1}}}{\sin{i}}\right)}-2.5\right]\,+\,(-23.5\pm0.5)$,
    $\log{\left(M_{\star} /
        M_\odot\right)}=(5.2\pm3.0)\left[\log{\left(\frac{W_{50}/\rm{km~s^{-1}}}{\sin{i}}\right)}-2.5\right]\,+\,(10.1\pm0.5)$
    and
    $\log{\left(M_{\rm b} /
        M_\odot\right)}=(4.9\pm2.8)\left[\log{\left(\frac{W_{50}/\rm{km~s^{-1}}}{\sin{i}}\right)}-2.5\right]\,+\,(10.2\pm0.5)$,
    where $M_{K_{\text{s}}}$ is the total absolute $K_{\text{s}}$-band
    magnitude of the objects, $M_{\star}$ and $M_{\rm b}$ their total
    stellar and baryonic masses, and $W_{50}$ the width of their line
    profile at $50\%$ of the maximum. Dividing the sample into
    different redshift bins and comparing to the TFRs of a sample of
    local ($z=0$) star-forming galaxies from the literature, we find
    no significant evolution in the slopes and zero-points of the TFRs
    since $z\approx0.3$, this in either luminosity or mass. In
  agreement with a growing number of CO TFR studies of nearby
  galaxies, we more generally find that CO is a suitable and
  attractive alternative to neutral hydrogen (H\,{\small I}). Our work
  thus provides an important benchmark for future higher redshift CO
  TFR studies.
  
\end{abstract}
%
%
\section{Introduction}
\label{sec:intro}
The Tully-Fisher relation (TFR; \citealt{tu77}) is a well-established
empirical correlation between the total stellar luminosity of a galaxy
(tracing its total stellar mass) and its rotation velocity (tracing
its total mass). It has been widely studied in the local Universe at
both optical and near-infrared wavelengths, exhibiting a relatively
small intrinsic scatter. Although the existence of a correlation
between the stellar luminosity and the width of the neutral hydrogen
(H\,{\small I}) line (roughly twice the maximum rotation velocity) of
late-type galaxies (spirals and irregulars) was suggested before
\citep{bal74}, \citet{tu77} showed that the relation could
also be used for distance measurements. It also holds
across a wide range of galaxy environments
\citep[e.g.][]{m93,w98,tp00}. The TFR relation is therefore a useful
tool to indirectly probe the
connection between the total mass-to-light ratio
$M/L$ and the total galaxy mass, and when studied as
a function of redshift to test theories of galaxy formation
\citep[e.g.][]{ste99}.

When a suitable kinematic tracer is available, it has been shown that
the TFR also holds for early-type galaxies (ETGs, lenticulars and
ellipticals; e.g.\ \citealt{nei99,mag01,ger01,de07,w10,d11,hei15}).
Crucially, if lenticulars are ``dead'' spirals (i.e.\ spirals for
which star formation has ceased), then their masses should remain
roughly constant over time while their luminosity decreases. This
would lead to an increase of their $M/L$ and thus a shift of the TFR
zero-point compared to that of spirals. Although some past works were
unable to find such an offset \citep[e.g.][]{dres83,ne99,hi01,hi03},
the results of other studies over the last decade or so do indicate
one \citep[e.g.][]{mat02,ben06,w10,d11,hei15}. In particular,
\citet{ben06} found that the TFR of lenticular galaxies lies about
$1.2$~mag below the spiral TFR with a scatter of $1.0$~mag in the
$K_{\textrm{s}}$-band, the largest offset found to date.

The H\,{\small I} emission line has been used heavily as the kinematic
tracer for TFR studies \citep[e.g.][]{tu77,tp00,piz07}. However,
carbon-monoxide (CO) has also been shown to be an excellent kinematic
tracer for TFR studies, as long as the CO emission extends beyond the
peak of the galaxy rotation curve
\citep{dic92,ss94,tu97,la98,tu01,ho07,d11,ti16}.

It is worth reflecting on the advantages of using CO
  line widths for TFR studies, compared with the widely used
  H\,{\small I} and optical emission lines.  First and foremost, we
  can detect CO to much greater distances than H\,{\small I}; CO is
  routinely detected in normal star-forming galaxies at intermediate
  redshifts ($z\approx1$-–$3$; e.g.\ \citealt{tac10, gen15}) and in
  starbursting galaxies up to $z\approx7$ \citep[e.g.][]{rie09,
    wang11}. Second, the beam size of CO observations is typically
  much smaller than that for H\,{\small I}, both for single-dish and
  interferometric observations, easily allowing to spatially resolve
  galaxies (and individual members within clusters of galaxies) even
  at high redshifts. Third, CO is less extended radially and more
  tightly correlated with the stellar component, and thus less
  affected by interactions between galaxies \citep{la98}. Finally,
  H$\alpha$ probes warmer gas than CO and is primarily emitted from
  star-forming regions. CO is therefore a convenient and robust tracer
  to probe the TFR as a function of redshift, encompassing many galaxy
  morphologies and providing an independent test of other established
  measures.

CO line profiles show a wide variety of shapes for
  several reasons: inner velocity field and/or CO distribution
  differences, beam response along the disc, pointing errors (for
  single-dish observations), opacity effects, etc
  \citep{la97}. Nevertheless, even in ETGs (where the radial extent of
  the molecular gas can be very limited; \citealt{dav13}), \citet{d11}
  have shown that galaxies with a double-horned or boxy integrated CO
  profile do yield accurate measurements of the maximum circular
  velocity (see also \citealt{ti16}), whereas galaxies with a
  single-peaked profile often do not.

We note that although $z>0$ TFR studies based on optical observations
exist \citep[e.g.][]{cons05,flo06,kas07,pu08}, to our knowledge there
is as yet no CO TFR work beyond the local Universe. Although studying
the TFR of high-$z$ disc galaxies comes at a price, e.g.\ the
increased difficulty to determine exact galaxy morphologies and
axial ratios (and therefore inclinations), much can be learnt about
their formation and evolution if successful. For example, does the
$M/L$ ratios of distant galaxies differ from those of their local
counterparts? If so, how are the stellar populations evolving, and
what is the relative growth rate of luminous and dark matter? Our
goals in this paper are thus twofold. First, to probe whether there is
any evolution of the CO TFR as a function of redshift up to $z=0.3$,
by comparing the TFR of galaxies within our sample and from the
literature \citep{tp00,ti16}. Second, to provide a local benchmark for
future higher redshift CO TFR studies. Our work is the first attempt
to construct a TFR for galaxies beyond the local Universe using CO
emission as the kinematic tracer.

This paper is structured as follows. Section~\ref{sec:data} describes
the data used, while Section~\ref{sec:samples} discusses the sample
selection. Section~\ref{sec:tfr} presents the velocity measurements
and TFR fits. The results are discussed in
Section~\ref{sec:discussion} and we conclude briefly in
Section~\ref{sec:conclusions}.
%
%
\section{Data and Parent Sample}
\label{sec:data}
\subsection{EGNoG CO sample}
\label{subsec:egnog}
The Evolution of molecular Gas in Normal Galaxies (EGNoG) survey is a
CO(1-0) survey of $31$ galaxies at $z\approx0.05$--$0.5$ by
\citet{ba13}. All galaxies were selected from the Sloan Digitized Sky
Survey Data Release~7 (SDSS DR7; \citealt{york00,str02,kev09}) and the
Cosmic Evolution Survey (COSMOS; \citealt{sco07}) to be as
representative as possible of the main sequence of star-forming
galaxies (a correlation between star formation rate, SFR, and stellar
mass, $M_{\star}$) at the redshifts concerned.

First, only galaxies with a spectroscopic redshift (essential for
follow-up CO observations) as well as
$4\leq M_{\star}\leq30\times10^{10}$~$M_{\odot}$ and
$4\leq {\textrm{SFR}}\leq100$~$M_{\odot}$~yr$^{-1}$ (to restrict the
sample to main sequence objects) were selected. Galaxies harbouring an
active galactic nucleus (AGN) were then rejected, as diagnosed from
standard emission line ratios measured in the SDSS spectra (see
\citealt{kauf03} and Section~\ref{subsec:agn}). Interacting galaxies
were also excluded via a visual inspection of the SDSS images,
although we revisit this issue in Section~\ref{sec:samples}.

The galaxies to be observed in CO were selected randomly from all the
galaxies meeting the above selection criteria. CO(1-0) observations of
all $31$ EGNoG galaxies were obtained using the Combined Array for
Research in Millimeter-wave Astronomy (CARMA) and were spatially
integrated to generate total spectra. The core of our sample is
composed of the $24$ EGNoG galaxies that were reliably detected
according to \citet{ba13}, all at $z\approx0.05$--$0.3$ and all from
SDSS (i.e.\ none of the COSMOS galaxies at $z\approx0.5$ was reliably
detected in CO). A few of these galaxies are luminous infrared
galaxies (LIRGs, with infrared luminosities
$10^{11}<L_{\textrm{IR}}<10^{12}$~$L_{\odot}$), but none is an
ultra-luminous infrared galaxy (ULIRG, with
$L_{\textrm{IR}}>10^{12}$~$L_{\odot}$). See \citet{ba13} for more
details of the sample selection, observations and data reduction.
\subsection{Additional CO data}
\label{subsec:literature}
Additional CO data for galaxies within the EGNoG redshift range were
taken from the literature. \citet{mir90} detected CO in $19$ LIRGs and
$9$ ULIRGs at $z=0.01$--$0.1$, complementing the work of \citet{san91}
who published CO line profiles for an additional $52$ LIRG and $8$
ULIRG detections at $z=0.01$--$0.1$. \citet{tu00} detected $10$ LIRGs
and $3$ ULIRGS at $z=0.05$--$0.2$. In addition, \citet{mat12} detected
CO in $7$ galaxies at $z=0.08$--$0.2$ and \citet{luca17} published the
CO line profiles of $5$ galaxies at $z\approx0.2$. See the related
papers for more details of the observations and data
reduction. Overall, we thus obtained the CO profiles of an additional
$113$ galaxies from the literature, $43$ of which are in the redshift
range $z=0.05$--$0.3$ we aim to study, while the rest are located much
closer at $z<0.05$.

Combining the EGNoG CO detections ($24$ galaxies) with the additional
literature detections ($43$ galaxies), we obtain a
  parent sample of $67$ galaxies with integrated CO profiles at
$z=0.05$--$0.3$.

Homogenous samples of galaxies are necessary for TFR
  studies. To build such samples, environmental effects (e.g.\
  interactions and mergers) and intrinsic properties (e.g.\
  inclination and luminosities) have to be considered first.  See
  Section~\ref{sec:samples} for detailed explanations of the selection
  criteria applied to our parent sample, to construct the more
  homogenous samples of galaxies necessary for TFR studies.
\subsection{Near-infrared photometry}
\label{subsec:photo}
Stellar luminosities are also required to construct
TFRs. Near-infrared photometry is superior to that at shorter
wavelengths as it is less affected by dust extinction. This is
particularly crucial for highly inclined galaxies and dusty
high-redshift objects. For example, if left uncorrected dust
extinction can cause an error $\gtrsim1$~mag at optical wavelengths
(e.g.\ $B$-band or $\approx440$~nm), while the uncertainties at
$K$-band ($\approx2.2~\micron$) are much less ($\approx0.1$~mag;
\citealt{noo07}). Longer wavelengths (e.g.\ mid-infrared) are affected
by dust emission and are thus also inappropriate. Partially as a
results of these effects, but also because the $M/L$ of stellar
populations varies the least at $K$-band \citep[e.g.][]{mar05}, the
scatter of the TFR is correspondingly minimised in this band
\citep{ver01}. The $K$-band is therefore the optimal choice of
passband to measure the galaxy luminosities.

The total apparent $K_{\textrm{s}}$-band magnitudes of all our sample
galaxies were obtained from the Two Micron All Sky Survey (2MASS;
\citealt{j00,sk06}) and are listed in Table~\ref{tab:general}. For
most galaxies, we adopted the {\it k\_m\_ext} parameter from the 2MASS
Extended Source Catalog (XSC; \citealt{j00}), i.e.\ the integrated
$K_{\textrm{s}}$-band magnitude from an extrapolated fit. For galaxies
that are not extended in 2MASS and thus not in the XSC, we adopted the
{\it k\_m} parameter, i.e.\ the default integrated $K_{\text{s}}$-band
magnitude from the 2MASS Point Source Catalog (PSC; \citealt{sk06}),
measured in a $4\farcs0$ radius aperture. The width of the 2MASS
optical system point spread function (PSF) meant that $2$--$15\%$ of
the total fluxes fell outside this aperture, but after applying
curve-of-growth corrections the standard aperture measurements
accurately reflect the fluxes within ``infinite'' apertures capturing
all of the sources' emission \citep{sk06}.

For the four galaxies in our sample that do not have
  2MASS data available, we adopted the {\it kAperMag6} parameter,
  i.e.\ the $K_{\text{s}}$-band $5\farcs7$ aperture integrated
  magnitude from the United Kingdom Infrared Telescope (UKIRT)
  Infrared Deep Sky Survey (UKIDSS; \citealt{ukirt1,ukirt2}). All
magnitudes quoted in this paper are Vega magnitudes (see
Table~\ref{tab:general}).
%
%
\begin{table*}
  \caption{General galaxy parameters for the initial and final sub-samples.}
  \begin{tabular}{llrrcccl}
    \hline
    Galaxy&SDSS name&$z$&$m_{K_{\textrm{s}}}$ (mag)&$\log(M_\star/M_\odot)$&$\log(M_{\rm b}/M_\odot)$&BPT class&Notes\\
    (1)&(2)&(3)&(4)&(5)&(6)&(7)&(8)\\
    \hline
    \multicolumn{7}{c}{Final sub-sample galaxies}\\
    G1&SDSSJ091957.00+013851.6&$0.176$&$14.84\,\pm\,0.15$&$10.9\,\pm\,0.1$&$11.1\,\pm\,0.1$&$1$&PSC, \citet{luca17}\\	
    G2&SDSSJ140522.72+052814.6&$0.195$&$14.84\,\pm\,0.02$&$11.0\,\pm\,0.1$&$11.1\,\pm\,0.1$&$2$&UKIRT, \citet{luca17}\\
    G3&SDSSJ111628.07+291936.1&$0.046$&$11.83\,\pm\,0.08$&$11.0\,\pm\,0.1$&$11.0\,\pm\,0.1$&$1$&\citet{tu00}\\
    G4&SDSSJ231332.46+133845.3&$0.081$&$13.29\,\pm\,0.15$&$11.0\,\pm\,0.1$&$11.1\,\pm\,0.1$&$1$&\\
    G5&SDSSJ141906.70+474514.8&$0.072$&$12.43\, \pm\,0.09$&$10.9\,\pm\,0.1$&$10.9\,\pm\,0.1$&$1$&\citet{tu00}\\
    G6&SDSSJ233455.23+141731.0&$0.062$&$12.11\,\pm\,0.06$&$11.0\,\pm\,0.1$&$11.1\,\pm\,0.1$&$1$&\\
    G7&SDSSJ221938.11+134213.9&$0.084$&$12.59\,\pm\,0.12$&$11.2\,\pm\,0.1$&$11.3\,\pm\,0.1$&$1$&\\
    G8&SDSSJ223528.63+135812.6&$0.183$&$13.51\,\pm\,0.18$&$11.4\,\pm\,0.1$&$11.5\,\pm\,0.1$&$2$&\\ 
    G9&SDSSJ100518.63+052544.2&$0.166$&$14.65\,\pm\,0.11$&$10.8\,\pm\,0.1$&$10.8\,\pm\,0.1$&$2$&PSC\\
    G10&SDSSJ105527.18+064015.0&$0.173$&$14.52\,\pm\,0.10$&$11.0\,\pm\,0.1$&$11.1\,\pm\,0.1$&$2$&PSC\\
    G11&SDSSJ124252.54+130944.2&$0.175$&$14.84\,\pm\,0.13$&$10.8\,\pm\,0.2$&$10.9\,\pm\,0.2$&$1$&PSC\\
    G12&SDSSJ091426.24+102409.6&$0.176$&$13.19\,\pm\,0.19$&$11.5\,\pm\,0.1$&$11.5\,\pm\,0.1$&$2$&\\
    G13&SDSSJ114649.18+243647.7&$0.177$&$14.74\,\pm\,0.09$&$11.1\,\pm\,0.1$&$11.1\,\pm\,0.1$&$2$&PSC\\
    G14&SDSSJ092831.94+252313.9&$0.283$&$15.06\,\pm\,0.14$&$11.2\,\pm\,0.2$&$11.3\,\pm\,0.1$&$1$&PSC\\
    G15&SDSSJ133849.18+403331.7&$0.285$&$14.04\,\pm\,0.19$&$11.3\,\pm\,0.2$&$11.4\,\pm\,0.1$&$1$&\\
    G16&SDSSJ142735.69+033434.2&$0.246$&$14.58\,\pm\,0.02$&$11.3\,\pm\,0.1$&$11.3\,\pm\,0.1$&$2$&UKIRT, \citet{luca17}\\ 
    G17&SDSSJ144518.88+025012.3&$0.190$&$14.86\,\pm\,0.15$&$11.2\,\pm\,0.1$&$11.2\,\pm\,0.1$&$1$&PSC, \citet{luca17}\\ 
    G18&SDSSJ151337.28+041921.1&$0.175$&$15.43\,\pm\,0.03$&$10.8\,\pm\,0.1$&$10.8\,\pm\,0.1$&$2$&UKIRT, \citet{luca17}\\ 
    G19&SDSSJ095904.41+024957.8&$0.119$&$15.35\,\pm\,0.20$&$10.4\,\pm\,0.1$&$10.7\,\pm\,0.1$&$1$&PSC, \citet{mat12}\\ 
    G20&SDSSJ100107.15+022519.5&$0.121$&$15.20\,\pm\,0.19$&$10.0\,\pm\,0.2$&$10.2\,\pm\,0.2$&$0$&PSC, \citet{mat12}\\ 
    G21&2MASXJ17320995+2007424&$0.050$&$12.20\,\pm\,0.07$&$10.4\,\pm\,0.2$&$10.6\,\pm\,0.1$&$0$&\citet{tu00}\\ 
    G22&SDSSJ145114.64+164143.6&$0.050$&$11.74\,\pm\,0.06$&$10.7\,\pm\,0.1$&$10.8\,\pm\,0.0$&$3$&\citet{tu00}\\ 
    G23&2MASXJ16381190-6826080&$0.050$&$10.93\,\pm\,0.04$&$10.6\,\pm\,0.2$&$11.0\,\pm\,0.1$&$0$&\citet{mir90}\\  
    G24&2MASXJ10200023+0813342&$0.050$&$13.08\,\pm\,0.16$&$10.6\,\pm\,0.1$&$10.9\,\pm\,0.0$&$5$&\citet{san91}\\ 
    G25&SDSSJ135751.77+140527.3&$0.099$&$12.91\,\pm\,0.11$&$10.8\,\pm\,0.1$&$10.8\,\pm\,0.1$&$1$&\\
    \multicolumn{7}{c}{Remaining initial sub-sample galaxies}\\
    G26&2MASX J01385289-1027113&$0.050$&$12.78\,\pm\,0.12$&$9.9\,\pm\,0.2$&$10.4\,\pm\,0.1$&$0$&\citet{mir90}\\ 
    G27&SDSSJ100318.58+025504.8&$0.105$&$15.51\,\pm\,0.21$&$10.1\,\pm\,0.1$&$10.2\,\pm\,0.1$&$1$&PSC, \citet{mat12}\\ 
    G28&SDSSJ095933.75+014905.8&$0.133$&$14.69\,\pm\,0.11$&$10.5\,\pm\,0.1$&$10.7\,\pm\,0.1$&$1$&PSC, \citet{mat12}\\ 
    G29&SDSSJ100051.21+014027.1&$0.166$&$15.04\,\pm\,0.14$&$10.5\,\pm\,0.1$&$10.6\,\pm\,0.1$&$1$&PSC, \citet{mat12}\\ 
    G30&SDSSJ100045.29+013847.4&$0.220$&$14.63\,\pm\,0.02$&$11.2\,\pm\,0.1$&$11.2\,\pm\,0.1$&$3$&UKIRT, \citet{mat12}\\ 
    G31&SDSSJ234311.26+000524.3&$0.097$&$13.52\,\pm\,0.17$&$10.7\,\pm\,0.1$&$10.8\,\pm\,0.1$&$1$&\\
    G32&SDSSJ211527.81-081234.4&$0.091$&$12.94\,\pm\,0.12$&$10.6\,\pm\,0.1$&$10.6\,\pm\,0.1$&$1$&\\
    G33&2MASXJ02211866+0656431&$0.098$&$12.63\,\pm\,0.11$&$10.4\,\pm\,0.2$&$10.9\,\pm\,0.1$&$0$&\citet{tu00}\\ 
    G34&SDSSJ105733.59+195154.2&$0.077$&$13.10\,\pm\,0.12$&$10.7\,\pm\,0.1$&$10.8\,\pm\,0.1$&$1$&\\
    G35&SDSSJ141601.21+183434.1&$0.055$&$12.27\,\pm\,0.09$&$11.1\,\pm\,0.1$&$11.1\,\pm\,0.1$&$2$&\\
    G36&SDSSJ100559.89+110919.6&$0.076$&$12.92\,\pm\,0.13$&$11.0\,\pm\,0.1$&$11.0\,\pm\,0.1$&$1$&\\
    G37&SDSSJ002353.97+155947.8&$0.192$&$13.12\,\pm\,0.16$&$11.3\,\pm\,0.1$&$11.4\,\pm\,0.1$&$1$&\\
    G38&SDSSJ134322.28+181114.1&$0.178$&$14.11\,\pm\,0.14$&$11.3\,\pm\,0.1$&$11.4\,\pm\,0.1$&$2$&\\
    G39&SDSSJ130529.30+222019.8&$0.190$&$14.42\,\pm\,0.08$&$11.0\,\pm\,0.1$&$11.0\,\pm\,0.1$&$1$&PSC\\
    G40&SDSSJ090636.69+162807.1&$0.301$&$15.30\,\pm\,0.15$&$11.2\,\pm\,0.2$&$11.3\,\pm\,0.2$&$1$&PSC\\
    \hline
  \end{tabular}		
  \label{tab:general}
  \parbox[t]{0.95\textwidth}{\textit{Notes:} Column~3: redshift,
    taken from \cite{ba13} for EGNoG galaxies and from the
    original paper otherwise (see Sections~\ref{subsec:egnog} and
    \ref{subsec:literature}). Column~4: total Vega apparent magnitude, 
    taken from the 2MASS PSC survey \citep{j00} of UKIRT
    \citep{ukirt1,ukirt2} for the galaxies so 
    noted in Column~8 and from the 2MASS XSC survey \citep{sk06}
    otherwise (see Section~\ref{subsec:photo}). Column~5: stellar
    mass, taken from MPA-JHU DR8 (see
    Section~\ref{subsec:mass}). Column~6: baryonic mass,
    as determined in Section~\ref{subsec:barmass}. 
    Column~7: BPT class, following the
    original \citet{bald81} classification (see
    Section~\ref{subsec:agn}). Column~8: source of the data
    for sample galaxies not belonging to EGNoG, as well as galaxies
    not found in the 2MASS XSC survey (PSC and UKIRT).}
\end{table*}
\subsection{Inclinations}
\label{subsec:imeasure}
The inclination of each galaxy is necessary to deproject its measured
velocity width. This was calculated using each galaxy's axial ratio
from its SDSS $r$-band image (specifically the {\it expAB\_r} and {\it
  expABErr\_r} parameters from the SDSS Data Release~12 catalogue;
\citealt{al15}) and the standard expression \citep{holm58}
\begin{equation}
  i_{b/a} = \cos^{-1}\,\left(\sqrt{\frac{q^2-q_{0}^2}{1-q_{0}^2}}\,\,\right)\,\,\,,
\end{equation}
where $q$ is the ratio of the semi-minor ($b$) to the semi-major ($a$)
axis of the galaxy, $q_{0}$ is the intrinsic axial ratio when the
galaxy is seen edge-on ($q_{0}\equiv c/a$), and $q_{0}=0.2$ is assumed
here (appropriate for late-type systems; \citealt{tu77,tp88}).
\subsection{Stellar masses}
\label{subsec:mass}
The most common technique for measuring stellar masses
  is to fit observed spectral energy distributions (SEDs) to templates
  generated from stellar population synthesis models. However, each
  method has its own degeneracies. \citet{mob15} tested the
  consistency of stellar masses measured using different methods
  (including \citealt{br03}, from which the stellar masses in this
  study are derived) and found good agreement between the input and
  estimated stellar masses when using the median of the stellar masses
  of individual galaxies derived from different methods.

In our study, the stellar mass of each galaxy was
taken from the Max Planck Institute for Astrophysics-Johns Hopkins
University Data
Release~8\footnote{https://www.sdss3.org/dr10/spectro/galaxy\_mpajhu.php}
(MPA-JHU DR8). Each mass was derived by fitting the galaxy SDSS
$ugriz$ photometry to a grid of models from the \citet{br03} stellar
population synthesis code, encompassing a wide range of star formation
histories. The mass and its uncertainty are defined as the median of
the probability distribution and half the difference between the
$16$th and $84$th percentiles of the distribution ($1\sigma$ error),
respectively.

For the five galaxies that do not have their stellar
  mass calculated by MPA-JHU (indicated by BPT class $0$ in
  Table~\ref{tab:general}), we obtained stellar masses from {\it
    kcorrect} (see Section~\ref{subsec:kcor}), that also uses
  \citeauthor{br03}'s (\citeyear{br03}) stellar population synthesis
  code. Population synthesis codes can change stellar masses by around
  $0.2$~dex \citep[e.g.][]{mob15,roe15}. We therefore assumed the same
  $0.2$~dex uncertainty for those five galaxies. The stellar masses
of all the galaxies in our sample (both EGNoG and others) are listed
in Table~\ref{tab:general}.

As our galaxies are located across the redshift range
  $z=0.05$--$0.3$, redshift effects and photometric uncertainties
  (that both tend to increase with redshift) must also be
  considered. Only $4$ galaxies in our sample are located at
  $z\approx0.3$, while the rest are located at
  $z\le0.2$. \citet{mob15} found no redshift-dependent bias at
  $z=0$--$4$ for stellar masses measured with the same input
  parameters but using different methods/codes.  However, the
  signal-to-noise ratio (S/N) of photometric data can also introduce
  further scatter in the stellar masses. Although this effect becomes
  dominant for faint galaxies with low photometric S/N ratios,
  \citet{mob15} found that when the input parameters are left free,
  there is an offset in the stellar masses at high S/N ratios for most
  of the methods. This indicates that the errors in the stellar masses
  are not necessarily caused by photometric uncertainties
  \citep{mob15}.
\subsection{Baryonic masses}
\label{subsec:barmass}
The baryonic mass of a galaxy consists of all visible
  components, i.e.\ both gas and stars. The molecular gas masses,
  $M_{\rm H_{2}}$, of all galaxies in our sample were taken from the
  related papers \citep{mir90,san91,tu00,mat12,ba13,luca17}, whereas
  the stellar masses were derived as described in the previous
  sub-section. Finally, to estimate the atomic gas masses,
  $M_{\rm H\small{I}}$, we used the molecular-to-atomic gas mass
  relation of \citet[i.e.\ $M_{\rm H_{2}}$ / $M_{\rm H\small{I}}$, see
  their Table~4]{sain11}. The baryonic masses, $M_{\rm b}$, of all
  galaxies are listed in Table~\ref{tab:general}.
\subsection{Absolute magnitudes and K corrections}
\label{subsec:kcor}
Because our galaxies span the redshift range $z=0.01$--$0.3$, the
portion of their spectra intercepted by the $K_{\text{s}}$ filter
varies from object to object, and we must correct the apparent
magnitudes measured to rest-frame ($z=0$) measurements. This so-called
K-correction is fully described in \citet{hogg02}, and it was applied
to our data using the publicly available code {\it
  kcorrect}\footnote{http://kcorrect.org/} version~4
\citep{bla07}. Using the spectroscopic redshifts provided (see
Table~\ref{tab:general}), {\it kcorrect} finds the intrinsic spectrum
that best represents the observed galaxy SED (here SDSS $ugriz$ and
2MASS or UKIRT $JHK$ total apparent magnitudes) by fitting templates
from the \citet{br03} stellar population synthesis code. The templates
have been optimised to minimise the residuals between the observed and
modelled galaxy fluxes.

The {\it kcorrect} routine determines absolute magnitudes for each
galaxy by calculating the distance modulus, accounting for the angular
diameter distance and cosmological surface brightness dimming. We
adopt here the cosmological parameters from the Planck results
\citep{planck15}. A Galactic extinction correction is also applied
using the extinction maps of \citet{sch98}. The 2MASS Vega magnitudes
were transformed to AB magnitudes to use {\it kcorrect}, but were then
transformed back to Vega magnitudes for use in this paper following
the application of K-correction. Fully corrected total absolute
$K_{\text{s}}$-band Vega magnitudes for all our sample galaxies are
listed in Table~\ref{tab:TFR}.
%
%
%
\begin{table*}
  \caption{TFR galaxy parameters for the initial and final sub-samples.}
  \begin{tabular}{lcccccl}
    \hline
    Galaxy&$M_{K_{\textrm{s}}}$&$W_{50}$&$b/a$&$i_{b/a}$&$W_{\rm 50}/\sin i$&Notes\\
          &(mag)&(km~s$^{-1}$)&&($\degr$)&(km~s$^{-1}$)&\\
    (1)&(2)&(3)&(4)&(5)&(6)&(7)\\
    \hline
    \multicolumn{7}{c}{Final sub-sample galaxies}\\
    G1&$-24.87\,\pm\,0.03$&$189.7\,\pm\,\phantom{00}4.0$&$0.66\,\pm\,0.01$&$49.68\,\pm\,0.02$&$248.7\,\pm\,\phantom{00}6.6$&\\
    G2&$-24.66\,\pm\,0.04$&$512.3\,\pm\,\phantom{0}12.9$&$0.46\,\pm\,0.01$&$64.70\,\pm\,0.01$&$566.7\,\pm\,\phantom{0}14.7$&\\
    G3&$-24.72\,\pm\,0.08$&$176.1\,\pm\,\phantom{0}20.1$&$0.68\,\pm\,0.01$&$48.71\,\pm\,0.01$&$234.4\,\pm\,\phantom{0}26.8$&\\
    G4&$-24.49\,\pm\,0.15$&$217.3\,\pm\,\phantom{0}43.1$&$0.63\,\pm\,0.01$&$52.72\,\pm\,0.01$&$273.1\,\pm\,\phantom{0}54.3$&\\
    G5&$-25.06\,\pm\,0.09$&$230.2\,\pm\,\phantom{0}10.6$&$0.63\,\pm\,0.01$&$52.10\,\pm\,0.01$&$291.5\,\pm\,\phantom{0}13.6$&\\
    G6&$-25.11\,\pm\,0.07$&$417.4\,\pm\,\phantom{0}61.7$&$0.59\,\pm\,0.01$&$55.27\,\pm\,0.01$&$507.8\,\pm\,\phantom{0}75.1$&\\
    G7&$-25.24\,\pm\,0.12$&$483.2\,\pm\,\phantom{0}36.5$&$0.62\,\pm\,0.01$&$53.57\,\pm\,0.01$&$600.5\,\pm\,\phantom{0}45.5$&\\
    G8&$-26.06\,\pm\,0.18$&$586.4\,\pm\,\phantom{0}18.5$&$0.44\,\pm\,0.01$&$66.40\,\pm\,0.02$&$639.9\,\pm\,\phantom{0}20.7$&\\ 
    G9&$-24.59\,\pm\,0.11$&$298.7\,\pm\,\phantom{0}33.0$&$0.84\,\pm\,0.02$&$33.34\,\pm\,0.04$&$543.4\,\pm\,\phantom{0}67.2$&\\
    G10&$-24.82\,\pm\,0.10$&$417.2\,\pm\,\phantom{0}25.4$&$0.72\,\pm\,0.02$&$44.88\,\pm\,0.03$&$591.3\,\pm\,\phantom{0}39.6$&\\
    G11&$-24.51\,\pm\,0.13$&$253.9\,\pm\,\phantom{0}24.4$&$0.45\,\pm\,0.02$&$65.85\,\pm\,0.02$&$278.2\,\pm\,\phantom{0}26.9$&\\
    G12&$-26.23\,\pm\,0.19$&$464.0\,\pm\,\phantom{0}21.3$&$0.85\,\pm\,0.02$&$33.08\,\pm\,0.03$&$850.2\,\pm\,\phantom{0}53.6$&\\
    G13&$-24.62\,\pm\,0.09$&$373.5\,\pm\,\phantom{0}50.9$&$0.52\,\pm\,0.01$&$60.98\,\pm\,0.02$&$427.2\,\pm\,\phantom{0}58.4$&\\
    G14&$-25.47\,\pm\,0.14$&$535.0\,\pm\,\phantom{0}35.1$&$0.54\,\pm\,0.02$&$59.26\,\pm\,0.02$&$622.5\,\pm\,\phantom{0}41.9$&\\
    G15&$-26.47\,\pm\,0.19$&$265.1\,\pm\,\phantom{0}16.9$&$0.87\,\pm\,0.02$&$30.27\,\pm\,0.04$&$525.8\,\pm\,\phantom{0}47.0$&\\
    G16&$-25.49\,\pm\,0.04$&$408.3\,\pm\,\phantom{0}14.1$&$0.62\,\pm\,0.02$&$53.12\,\pm\,0.02$&$510.4\,\pm\,\phantom{0}20.0$&\\ 
    G17&$-24.66\,\pm\,0.15$&$421.7\,\pm\,\phantom{00}2.9$&$0.76\,\pm\,0.02$&$41.03\,\pm\,0.03$&$642.4\,\pm\,\phantom{0}22.5$&\\ 
    G18&$-23.85\,\pm\,0.05$&$425.2\,\pm\,\phantom{0}12.9$&$0.46\,\pm\,0.02$&$65.05\,\pm\,0.02$&$467.0\,\pm\,\phantom{0}15.0$&\\ 
    G19&$-23.19\,\pm\,0.21$&$372.2\,\pm\,\phantom{0}28.2$&$0.69\,\pm\,0.02$&$48.03\,\pm\,0.02$&$500.6\,\pm\,\phantom{0}39.6$&\\
    G20&$-23.44\,\pm\,0.19$&$336.1\,\pm\,\phantom{0}25.5$&$0.51\,\pm\,0.02$&$61.39\,\pm\,0.03$&$382.9\,\pm\,\phantom{0}29.5$&\\ 
    G21&$-24.55\,\pm\,0.08$&$424.1\,\pm\,\phantom{00}6.9$&$0.50\,\pm\,0.05$&$62.11\,\pm\,0.06$&$479.8\,\pm\,\phantom{0}17.8$&\\
    G22&$-24.80\,\pm\,0.07$&$256.3\,\pm\,\phantom{0}16.1$&$0.81\,\pm\,0.01$&$36.90\,\pm\,0.01$&$426.8\,\pm\,\phantom{0}27.1$&\\
    G23&$-25.70\,\pm\,0.05$&$572.4\,\pm\,\phantom{0}16.0$&$0.50\,\pm\,0.05$&$62.11\,\pm\,0.06$&$647.6\,\pm\,\phantom{0}28.2$&\\
    G24&$-23.60\,\pm\,0.16$&$163.0\,\pm\,\phantom{0}11.8$&$0.22\,\pm\,0.01$&$84.42\,\pm\,0.01$&$163.0\,\pm\,\phantom{0}11.8$&\\
    G25&$-25.25\,\pm\,0.11$&$684.4\,\pm\,\phantom{0}87.5$&$0.73\,\pm\,0.01$&$43.85\,\pm\,0.01$&$987.8\,\pm\,126.6$&\\ 
    \multicolumn{7}{c}{Remaining initial sub-sample galaxies}\\  
    G26&$-23.86\,\pm\,0.12$&$130.5\,\pm\,\phantom{0}29.7$&$0.59\,\pm\,0.01$&$55.79\,\pm\,0.01$&$157.8\,\pm\,\phantom{0}35.9$&s\\
    G27&$-22.74\,\pm\,0.21$&$299.9\,\pm\,\phantom{0}55.1$&$0.76\,\pm\,0.02$&$41.45\,\pm\,0.03$&$453.0\,\pm\,\phantom{0}84.7$&s\\
    G28&$-24.07\,\pm\,0.11$&$329.1\,\pm\,\phantom{0}38.7$&$0.76\,\pm\,0.02$&$41.74\,\pm\,0.02$&$494.3\,\pm\,\phantom{0}59.7$&s\\
    G29&$-24.21\,\pm\,0.15$&$140.8\,\pm\,\phantom{0}26.5$&$0.81\,\pm\,0.02$&$36.30\,\pm\,0.04$&$237.8\,\pm\,\phantom{0}46.2$&s\\
    G30&$-25.18\,\pm\,0.04$&$524.1\,\pm\,254.5$&$0.52\,\pm\,0.02$&$60.58\,\pm\,0.02$&$601.6\,\pm\,292.2$&s\\
    G31&$-24.64\,\pm\,0.17$&$135.8\,\pm\,\phantom{0}42.4$&$0.63\,\pm\,0.01$&$52.67\,\pm\,0.01$&$170.9\,\pm\,\phantom{0}53.4$&s\\
    G32&$-25.07\,\pm\,0.12$&$164.8\,\pm\,\phantom{0}72.7$&$0.50\,\pm\,0.01$&$62.16\,\pm\,0.01$&$186.4\,\pm\,\phantom{0}82.2$&s\\
    G33&$-25.41\,\pm\,0.11$&$239.7\,\pm\,\phantom{00}8.7$&$0.84\,\pm\,0.08$&$33.63\,\pm\,0.15$&$432.9\,\pm\,100.1$&s\\
    G34&$-24.54\,\pm\,0.12$&$402.9\,\pm\,\phantom{0}93.3$&$0.70\,\pm\,0.01$&$46.57\,\pm\,0.01$&$554.8\,\pm\,128.5$&s\\
    G35&$-24.68\,\pm\,0.09$&$164.8\,\pm\,\phantom{0}24.4$&$0.27\,\pm\,0.01$&$79.34\,\pm\,0.01$&$167.7\,\pm\,\phantom{0}24.8$&s\\
    G36&$-24.73\,\pm\,0.14$&$248.0\,\pm\,\phantom{0}40.3$&$0.57\,\pm\,0.01$&$56.88\,\pm\,0.01$&$296.2\,\pm\,\phantom{0}48.2$&s\\
    G37&$-26.54\,\pm\,0.16$&$398.0\,\pm\,\phantom{0}37.5$&$0.54\,\pm\,0.02$&$59.21\,\pm\,0.02$&$463.3\,\pm\,\phantom{0}44.0$&s\\
    G38&$-25.32\,\pm\,0.14$&$354.8\,\pm\,\phantom{0}52.8$&$0.67\,\pm\,0.02$&$49.08\,\pm\,0.02$&$469.6\,\pm\,\phantom{0}70.6$&s\\
    G39&$-25.14\,\pm\,0.08$&$194.0\,\pm\,\phantom{0}25.6$&$0.76\,\pm\,0.02$&$41.40\,\pm\,0.03$&$293.4\,\pm\,\phantom{0}39.7$&s\\
    G40&$-25.22\,\pm\,0.15$&$139.4\,\pm\,\phantom{0}12.1$&$0.77\,\pm\,0.02$&$40.53\,\pm\,0.04$&$214.6\,\pm\,\phantom{0}20.6$&s\\
    \hline
  \end{tabular}
  \label{tab:TFR}
  \parbox[t]{0.85\textwidth}{\textit{Notes:} Column~1 lists the
    galaxies as in Table~\ref{tab:general}. Column~2: corrected total Vega
    absolute magnitudes, calculated as described in
    Section~\ref{subsec:kcor}. Column~3: velocity widths, calculated as
    described in Section~\ref{subsec:W50}. Column~4: axial ratios,
    taken from the SDSS $r$-band images (see
    Section~\ref{subsec:imeasure}). Column~5: inclinations, calculated
    as described in Section~\ref{subsec:imeasure}. Column~7: reasons 
    why galaxies were excluded from the final sample (``s'' stands for a single Gaussian
    integrated line profile).}
\end{table*}
%
%
\section{Sample selection}
\label{sec:samples}
To draw a more homogenous initial sub-sample of
  galaxies from the parent sample, we applied the further selection
  criteria described below. We then drew a final sub-sample for which
  all galaxies have a double-horned or boxy CO integrated line
  profile, i.e.\ galaxies where the gas likely reaches the flat part
  of the rotation curve \citep[e.g.][]{la97,d11}.
\subsection{Initial sub-sample}
\label{subsec:initsamp}
\subsubsection{Galaxy interactions}
\label{subsec:interactions}
The TFR is only meaningful if the kinematic tracer used is
rotationally-supported and in equilibrium. It is thus important to
remove from our sample galaxies that are strongly interacting. For
the additional literature data, all galaxies showing sings of
interactions in SDSS images were excluded, as done by construction for
the EGNoG sample. We also excluded all galaxies described as
interacting in the original papers or in the NASA/IPAC Extragalactic
Database\footnote{https://ned.ipac.caltech.edu} (NED), as well as
objects showing signs of interactions in archival {\it Hubble Space
  Telescope} ({\it HST}) images (minor disturbances, bridges, tails
and mergers). These checks resulted in the rejection of $22$ objects
from the parent sample (none from EGNoG), leaving all $24$ EGNoG
galaxies and $21$ additional objects from the literature.
\subsubsection{AGN}
\label{subsec:agn}
As the emission from active galactic nuclei (AGN) can contaminate the
measured stellar luminosities, we also removed galaxies with a strong
AGN from our TFR sample. We did this using the MPA-JHU classifications
of the galaxies' optical emission line ratios from the SDSS spectra
\citep{bri04}, inspired by the \citet{bald81} diagnostic diagrams (BPT
diagrams). According to this, galaxies are divided
  into ``Unclassifiable" (BPT class $-1$), ``Not Used" (BPT class
  $0$), ``Star Forming" (BPT class $1$), ``Low S/N Star Forming" (BPT
  class $2$), ``Composite" (BPT class $3$), ``AGN" (BPT class $4$) and
  ``Low S/N AGN" (BPT class $5$) categories.
 
None of the EGNoG galaxies (by construction) or remaining additional
galaxies from the literature harbours a strong AGN (BPT class $4$),
thus leaving $45$ mostly star-forming or low $S/N$ star-forming
galaxies (i.e. \ BPT class 1 or 2 objects) for our TFR analyses,
including all $24$ ENGoG galaxies and $21$ additional objects from the
literature. (see Table~\ref{tab:general}).
\subsubsection{Inclination}
\label{subsec:icut}
As a galaxy approaches a face-on orientation ($i=0\degr$), the
uncertainty in the inclination increases. This is particularly
problematic as the inclination correction to the velocity width
measured is then also large ($\sin^{-1}i$). We therefore only retain
galaxies with $i\geq30\degr$ (a standard cutoff; see e.g.\
\citealt{tp88,ti16}), leading us to exclude $5$ of the $45$ remaining
galaxies ($3$ from EGNoG and $2$ from the literature). This results in
an initial sub-sample of $40$ galaxies, $21$ from EGNoG and $19$ from
the literature. They are listed in Table~\ref{tab:general}.

This stark reduction in the size of the literature sample (from $43$
to $19$ galaxies) is unfortunate but perhaps unsurprising, as LIRGs
and ULIRGs are often disturbed and the fraction of active galaxies
increases with increasing infrared luminosity
\citep{veil99,vei02,wang06}. Nevertheless, we increase the EGNoG
sample by about $90\%$ by including the remaining literature galaxies.
\subsubsection{ULIRGs}
\label{subsec:starform}
Although the parent sample has many ULIRGs, there is no remaining
ULIRG in the initial sub-sample, as a result of excluding all galaxies
showing signs of mergers/interactions. This supports the idea that
ULIRGs at $z\lesssim0.3$ are dominated by mergers/interactions
\citep{arm87, mel90, sur00, bus02}.

We also checked the SFRs of our initial sub-sample galaxies, to verify
whether any ULIRG remained. The SFRs of our initial sub-sample
galaxies were again obtained from the work of the MPA-JHU
group\footnote{https://www.sdss3.org/dr10/spectro/galaxy\_mpajhu.php}.
The SFR range of LIRGs and ULIRGs is $10$--$170$ and
$170$--$1700$~$M_\odot$~yr$^{-1}$, respectively
\citep[e.g.][]{ken98,al13,car15}. None of the $40$ galaxies in our
initial sub-sample has a SFR greater than $75$~$M_\odot$~yr$^{-1}$,
$15$ have $10<\textrm{SFR}<75$~$M_\odot$~yr$^{-1}$, and the rest
($25$) have SFR$<10$~$M_\odot$~yr$^{-1}$. The average SFR of our
initial sub-sample and the final sub-sample (see the sub-section
below) is $18$ and $15$~$M_\odot$~yr$^{-1}$, respectively, again
strengthening the suggestion that our sub-samples do not include any
ULIRG.

We also note that about $60\%$ of the galaxies in the
  initial sub-sample have a SFR smaller than the typical SFR of
  LIRGs. This indicates that the resulting CO TFRs are not dominated
  by LIRGs.

In summary, none of the galaxies in our initialsub-sample
  show any sign of interactions, and they mostly are purely
  star-forming galaxies with an inclination angle $i\geq30\degr$.
\subsection{Final sub-sample}
\label{subsec:shape}
A kinematic tracer that extends significantly past the turnover of 
the galaxy circular velocity curve into its ``flat'' velocity regions will usually 
yield a double-horned or boxy integrated line profile (see e.g.\ \citealt{d11}). Our line
profile analysis in Section~\ref{subsec:W50} indicates that $25$ of
the $40$ galaxies in our initial sub-sample have a double-horned or
boxy profile and are thus likely to yield reliable velocity width
measurements. The remaining $15$ galaxies have line profiles best
represented by a single Gaussian.

There are several reasons for a galaxy's integrated CO profile to exhibit a 
single-Gaussian shape (see Section~\ref{sec:intro}). The most obvious, 
however, is that the CO-emitting gas does not have sufficient radial extent to 
probe beyond the galaxy's circular velocity turnover. In these cases, such galaxies 
clearly warrant exclusion from our TFR analysis in order to avoid biasing our best fit 
relation to higher intercepts (lower velocities), or artificially increasing our measure of 
the TFR scatter. As clearly seen from Figure 1 panels (a), (c) and (e), those systems 
exhibiting single-Gaussian integrated CO line profiles tend to have lower rotational 
velocities ($W_{\rm 50}/\sin i < 10^{2.5}$~km~s$^{-1}$ ) than those with double-horned 
or boxy profiles, lending significance to the postulate that the CO line profile width 
underestimates the total rotation velocity for these systems. We thus exclude them 
from our final sub-sample. The excluded systems are labeled as such in 
column~$7$ of Table~\ref{tab:TFR}.

Overall, our final sub-sample of inactive galaxies with inclination
$i\geq30\degr$ and a double-horned or boxy integrated line profile is
thus composed of $25$ galaxies (from an initial sub-sample of $40$),
$12$ from EGNoG and $13$ additional galaxies from the literature. For
our analyses, we construct TFR relations and report the results for
both the final and initial sub-samples. However, we base our
discussion on the higher quality final sub-sample only.
%
%
\section{Velocity measurements and Tully-Fisher relations}
\label{sec:tfr}	
\subsection{Velocity widths}
\label{subsec:W50}
The second observable required to construct a TFR is a measurement of
the circular velocity, or in the case of integrated spectra (half) the
width of the line profile. The line widths at $20\%$ ($W_{20}$; e.g.\
\citealt{tu77,tp00,d11}) and $50\%$ ($W_{50}$; e.g.\ \citealt{sch94,
  la98, ti16}) of the peak intensity are commonly used measures of the
maximum rotational velocity, but \citet{la97} found that $W_{50}$ has
smaller uncertainties and suffers the least bias. We thus adopt
$W_{50}$ as our measure of (twice) the rotation velocity.

\citet{ti16} tested four fitting functions when using CO integrated
spectra in the context of TFR studies. Using simulated spectra
generated from modelled galaxies, they found that the Gaussian Double
Peak function (a quadratic function bordered by a half-Gaussian on either
side; see below) was the most appropriate, in the sense that it
yielded the most consistent velocity width measures as a function of
both amplitude-to-noise ratio ($A/N$) and inclination, this for a wide
range of maximum circular velocities. The only exceptions were at very
low inclinations and circular velocities, where the single-peaked
Gaussian function was unsurprisingly better suited (as even
intrinsically double-horned spectra appear single-peaked when spread
over only few velocity channels).

Here, we thus fit all our integrated spectra with both the Gaussian
Double Peak function and the single Gaussian function, and we adopt
the fit with the lowest reduced $\chi^2$ ($\chi^2_{\textrm{red}}$,
defined in the standard manner; but see below). The fits were carried
out with the package {\sc MPFIT} \citep{ma09}, that employs a
Levenberg-Marquardt minimisation algorithm. To avoid local minima, in
each case we ran {\sc MPFIT} several times with different initial
guesses. The fitting parameters with the smallest
$|1-\chi_{\textrm{red}}^2|$ value were taken as the best fit.

The form of the Gaussian Double Peak function is
\begin{equation}
  \label{eq:GDPfunction}
  f(v) =
  \begin{cases}
    A_{\text{G}}\times\,{\rm e}^{\frac{-[v-(v_{0}-w)]^{2}}{2\sigma^{2}}} & v<v_{0
    }-w \\
    A_{\text{C}}+a\,(v-v_{0})^{2} & v_{0}-w\leq v\leq v_{0}+w\,\,\,, \\
    A_{\text{G}}\times\,{\rm e}^{\frac{-[v-(v_{0}+w)]^{2}}{2\sigma^{2}}} & v>v_{0
    }+w \\
  \end{cases}
\end{equation}
where $v$ is the velocity, $A_{\textrm{C}}>0$ is the flux at the
central velocity $v_{0}$, $A_{\textrm{G}}>0$ is the peak flux of the
half Gaussians on both sides (centred at velocities $v_{0}\pm w$),
$w>0$ is the half-width of the central parabola, $\sigma>0$ is the
width of the profile edges, and
$a\equiv(A_{\textrm{G}}-A_{\textrm{C}})/w^{2}$.

The velocity width $W_{50}$ can then be easily calculated
analytically. Defining
\begin{equation}
  A_{\textrm{max}} \equiv
  \begin{cases}
    A_{\textrm{C}} & \textrm{ if } A_{\textrm{C}}\geq A_{\textrm{G}} \\
    A_{\textrm{G}} & \textrm{ if } A_{\textrm{C}}<A_{\textrm{G}}\,\,\,, \\
  \end{cases}
\end{equation}
then if $A_{\textrm{G}}\ge A_{\textrm{max}}/2$ (i.e.\ the central
parabola is either concave, as expected for a standard double-horned
profile, or slightly convex), the profile width is determined by the
two half-Gaussians and $W_{50}$ is given by
\begin{equation}
  \label{eq:GDPW50}
  \begin{array}{r@{}l}
    W_{50} &{} =\,2\,(w\,+\,\sqrt{2\ln2}\,\sigma)\,\,\,. \\
    \end{array}
\end{equation}
If $A_{\textrm{G}}<A_{\textrm{max}}/2$ (i.e.\ the central parabola is
strongly convex), the profile width is determined by the central
parabola but the profile is in fact not really double-horned and it is
preferable to adopt a single Gaussian fit irrespective of the
$\chi_{\textrm{red}}^2$ value.

The single Gaussian function is given by
\begin{equation}
  \label{eq:SGfunction}
  f(v) = A\,{\rm e}^{\frac{-(v-v_{0})^{2}}{2\sigma^{2}}}\,\,\,,
\end{equation} 
where $A>0$ is the flux of the peak at the central (and mean) velocity
$v_{0}$, and $\sigma>0$ is the width of the profile (root mean square
velocity). The velocity width $W_{50}$ is then given by
\begin{equation}
  \label{eq:SGW50}
  \begin{array}{r@{}l}
    W_{50} &{} =\,2\,\sqrt{2\ln2}\,\sigma\,\,\,,\\
  \end{array}
\end{equation}
as for the Gaussian Double Peak function with $w=0$.

The uncertainty on the velocity width, $\Delta W_{50}$, is estimated
by generating $150$ realisations of the best-fit model. Random
Gaussian noise (with a root mean square equal to that in line-free
channels of the spectrum) is added to each realisation, which is then
fit as described above. Finally, $\Delta W_{50}$ is taken as the
standard deviation of the measured velocity width distribution.

We further note here that while $A_{\textrm{max}}/2$ is the
mathematically convenient threshold to determine whether one should
use the Gaussian Double Peak function or the single Gaussian function,
\citet{ti16} established that $2A_{\textrm{max}}/3$ is a more
practical threshold to use. We therefore adopt this convention, and
use the Gaussian Double Peak function (Eqs.~\ref{eq:GDPfunction} and
\ref{eq:GDPW50}) when $A_{\textrm{G}}\ge 2A_{\textrm{max}}/3$ and the
single Gaussian function (Eqs.~\ref{eq:SGfunction} and \ref{eq:SGW50})
otherwise.
%
%
\subsection{Tully-Fisher relations}
\label{subsec:tfr}
We constructed $M_{K_{\textrm{s}}}$ TFRs for the galaxies in both our
initial and final sub-samples. We used a standard form for the TFR,
\begin{equation}
  \label{eq:tfr}
  M_{K_{\textrm{s}}} = a\,\left[\,\log\left(\,\frac{W_{\rm 50}/\sin i}{\textrm{km s}^{-1}}\,\right)-2.5\,\right]+b\,\,\,, 
\end{equation}
 where
$a$ is the slope and $b$ the zero-point of the relation. We fit this
linear relationship to the data using the {\sc MPFITEXY} routine
\citep{w10}, that uses the {\sc MPFIT} package. The intrinsic scatter
($\sigma_{\text{int}}$) in the relation was estimated by adjusting its
value to ensure $\chi_{\textrm{red}}^{2}=1$. A fuller description of
the fitting procedure can be found in \citet{w10} and \citet{ti16}.

\begin{figure*}
  \includegraphics[width=9.5cm,clip=]{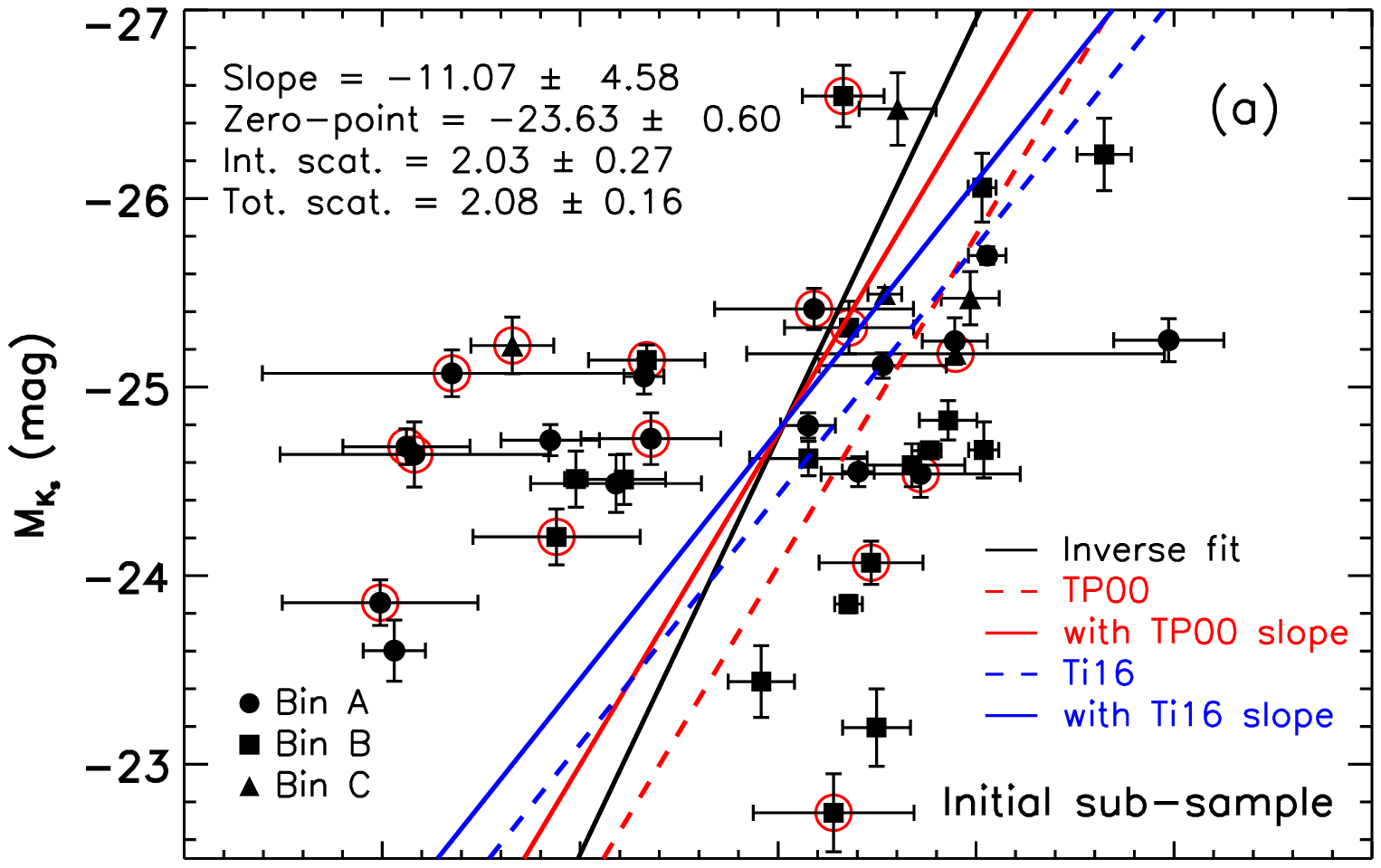}
  \hspace{-50pt}
  \includegraphics[width=9.5cm,clip=]{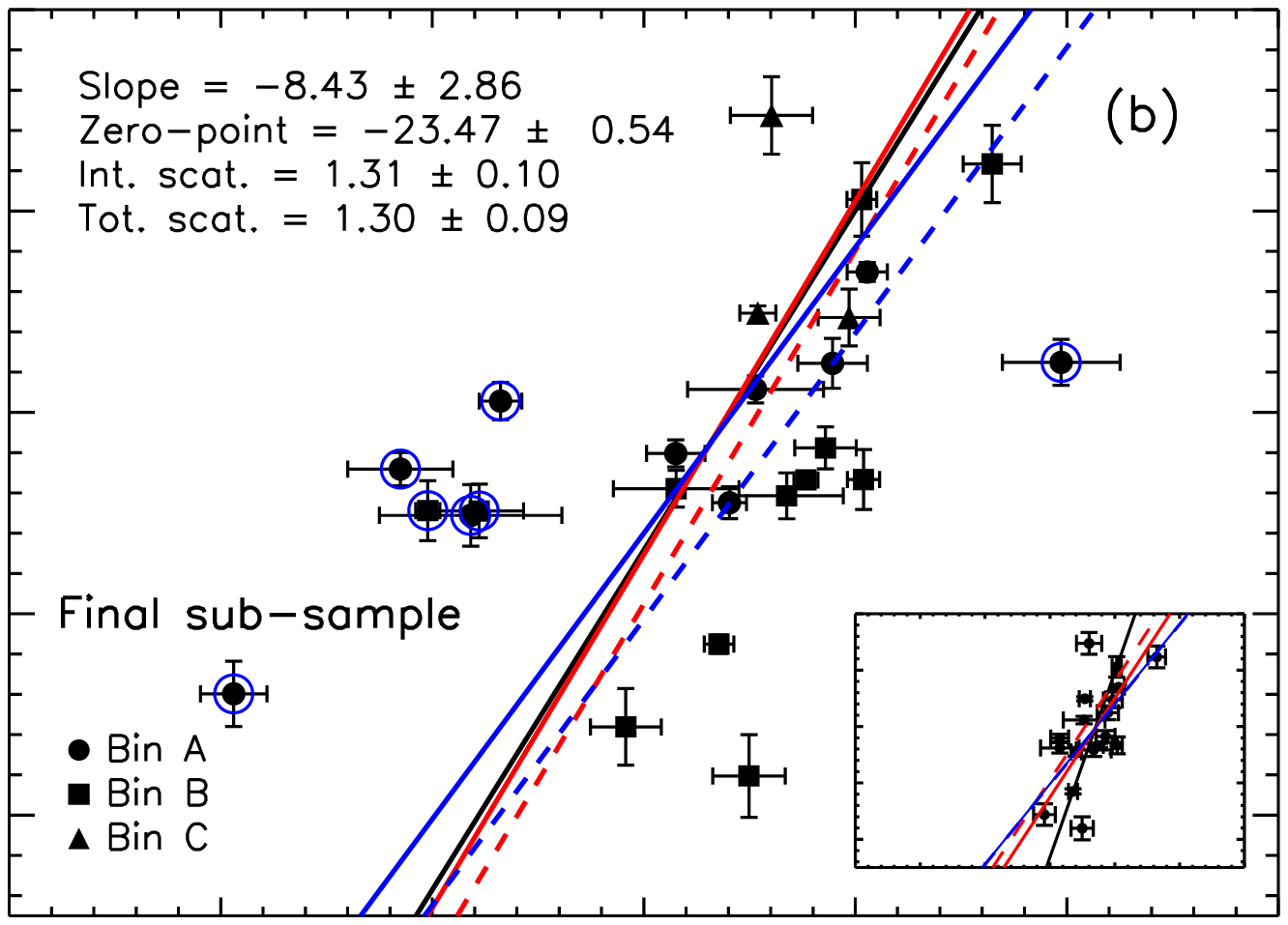}\\
  \vspace{-30pt}
  \includegraphics[width=9.5cm,clip=]{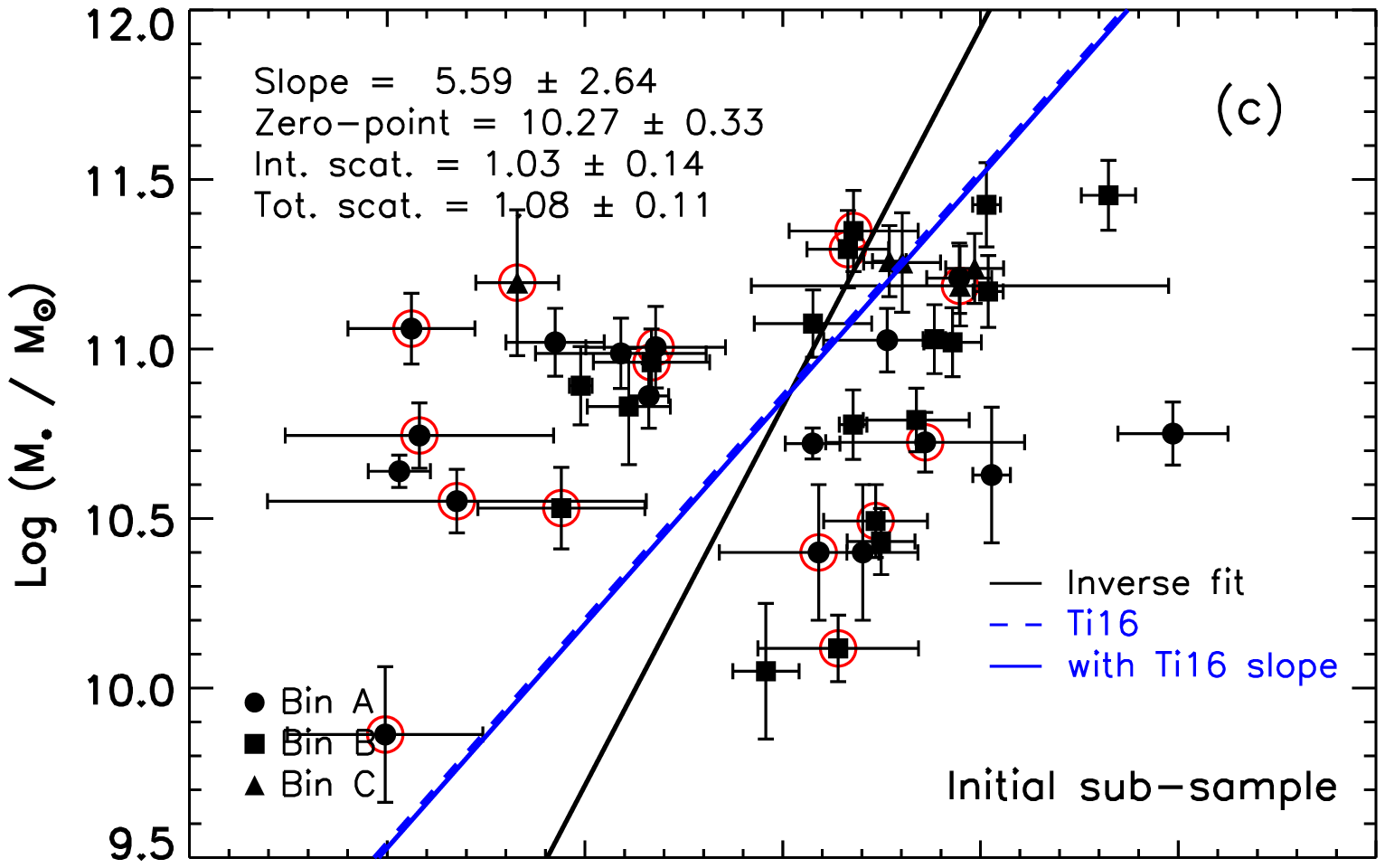}
  \hspace{-50pt}
  \includegraphics[width=9.5cm,clip=]{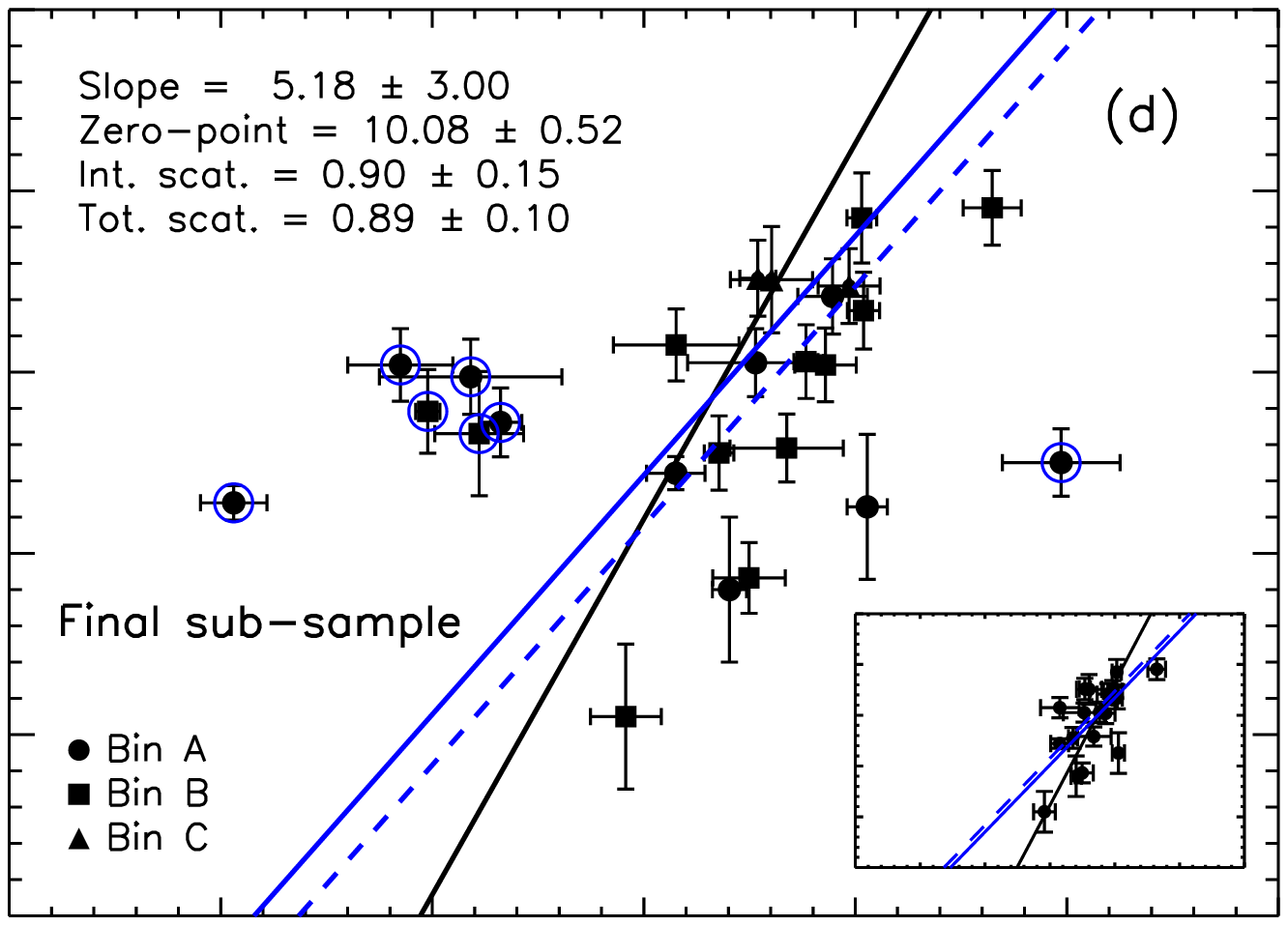} \\
  \vspace{-30pt}
  \includegraphics[width=9.5cm,clip=]{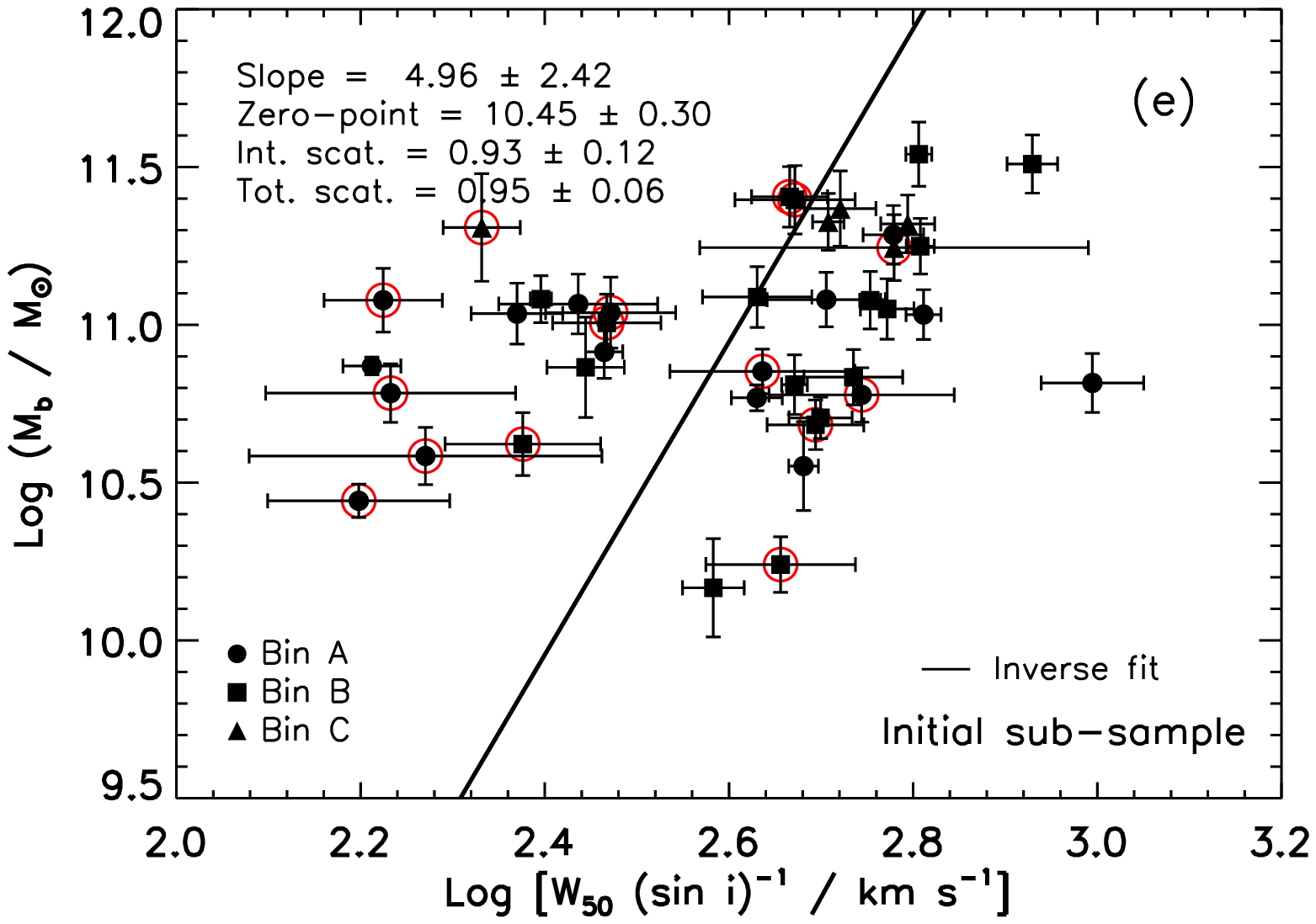}
  \hspace{-50pt}
  \includegraphics[width=9.5cm,clip=]{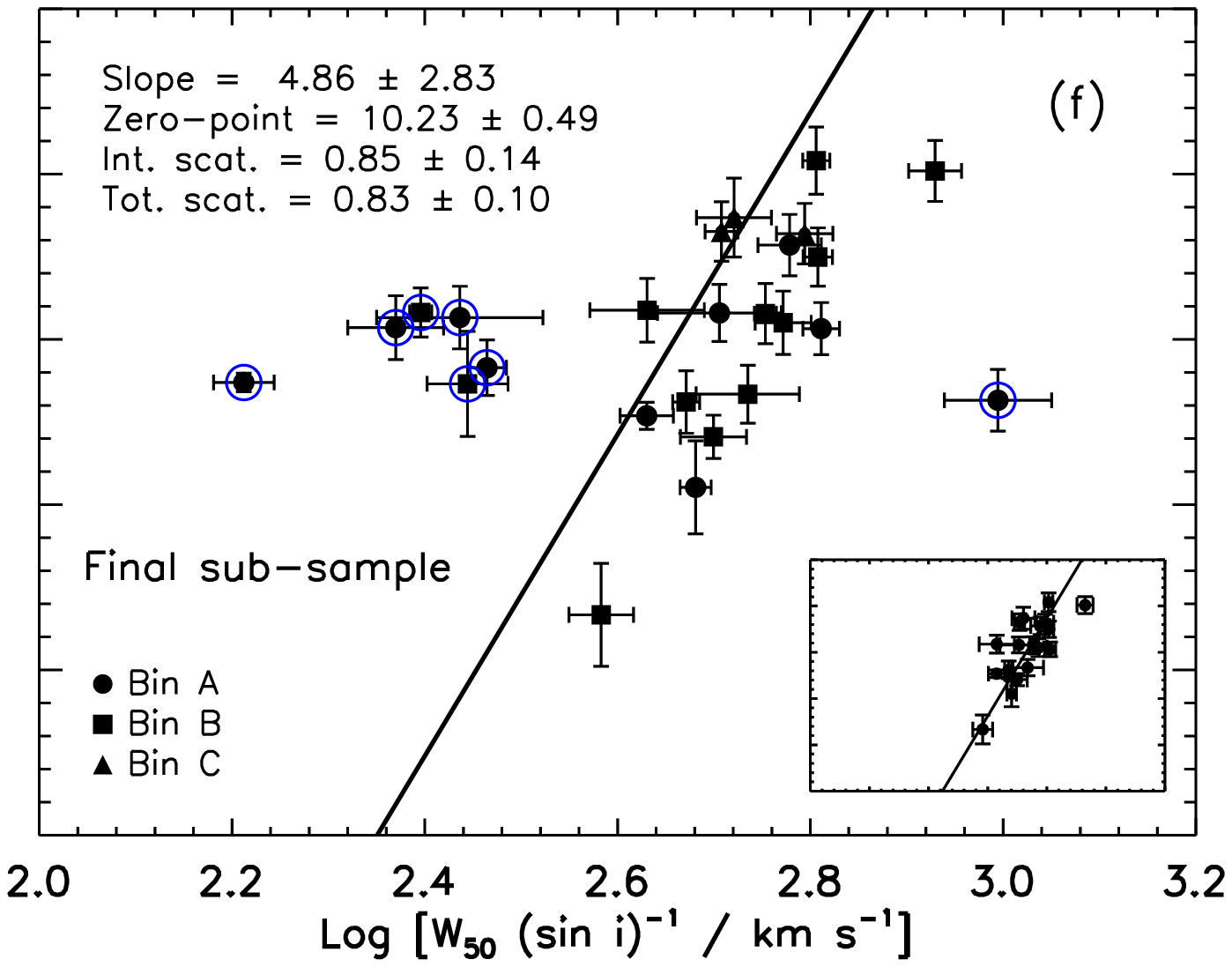} \\
  \caption{{\bf Top}: $K_{\textrm{s}}$-band CO TFR of the initial
    (left) and final (right) galaxy sub-samples. The black solid lines
    show the reverse fits, while the red (respectively blue) lines
    show the reverse fit with slope fixed to that of \citet{tp00}
    (resp.\ \citealt{ti16}). The red dashed lines show the H\,{\small
      I} TFR of local spirals from \citet{tp00}. The blue dashed lines
    show the CO TFR of local spirals \citep{ti16}. {\bf Middle}: As
    for the top panels, but for the $M_{\star}$ CO TFR. {\bf Bottom}:
    As for the top and the middle panels, but for the $M_{\rm b}$ CO
    TFR. In panels (a), (c) and (e), the data points
      with open red circles represent galaxies with a single Gaussian
      line profile, while open blue circles in panels (b), (d) and (f)
      represent the outliers (see Section~\ref{subsec:outliers}). In
      all panels, the black filled circles, black filled squares and
      black filled triangles show the galaxies in bin~A, bin~B and
      bin~C, respectively. All the fits shown were done with all the
      data points. The embedded figures in panels (b), (d) and (f)
      show the TFR for the final galaxy sub-sample after excluding the
      outliers.}
  \label{fig:tfr}
\end{figure*}

Since there is a significant bias in the slope of the forward fit
\citep{w94}, we also fit the inverse of Equation~\ref{eq:tfr}
(similarly to \citealt{w10,d11,ti16}). In addition, we performed a
number of further fits with the slope fixed to that of past studies
(\citealt{tp00},\citealt{ti16} and Torii et al., in prep.). The
$K_{\textrm{s}}$-band TFRs of the initial and final sub-samples are
shown in Figure~\ref{fig:tfr} along with fixed-slope fits. The fit
parameters are listed in Table~\ref{tab:mktfr}. While we list the
results of both the forward and reverse fits in the table, we shall
restrict our discussion to the more robust reverse fits.

As for $M_{K_{\textrm{s}}}$, we also constructed
  stellar and baryonic mass TFRs for both the initial and final galaxy
  sub-samples, where $\log(M_{\star}/M_\odot)$ and
  $\log(M_{\rm b}/M_\odot)$ are, respectively, on the left hand side
  of Equation~\ref{eq:tfr} instead (see Fig.~\ref{fig:tfr} and
  Table~\ref{tab:masstfr}).

The $M_{K_{\textrm{s}}}$, $M_{\star}$ and $M_{\rm b}$
  CO TFRs of the final sub-sample presented in Fig.~\ref{fig:tfr} and
  Tables~\ref{tab:mktfr} and \ref{tab:masstfr} constitute the main
  results of this paper, the first CO TFRs beyond the immediate local
  Universe.
%
%
\begin{table*}
  \caption{Best-fit parameters of the $K_{\textrm{s}}$-band CO TFRs.}
  \begin{tabular}{llccccc}
    \hline
    Sub-sample&Fit type&Slope&Zero-point&$\sigma_{\textrm{int}}$&$\sigma_{\textrm{total}}$&Zero-point offset\\ 
          &&&(mag)&(mag)&(mag)&(Ours - Theirs)\\
    \hline
    Initial&Forward&$-1.49\,\pm\,0.61$&$-24.66\,\pm\,0.14$&$0.76\,\pm\,0.09$&$0.75\,\pm\,0.03$&$-$\\
           &Reverse&$-11.07\,\pm\,4.58\,\,\,$&$-23.63\,\pm\,0.60$&$2.03\,\pm\,0.27$&$2.08\,\pm\,0.16$&$-$\\
           &&($-16.50\,\pm\,9.77$)\phantom{0}&($-22.57\,\pm\,1.46$)&($2.78\,\pm\,0.29$)&($2.85\,\pm\,0.41$)&$-$\\\\
      Final&Forward&$-2.31\,\pm\,0.78$&$-24.47\,\pm\,0.19$&$0.68\,\pm\,0.11$&$0.68\,\pm\,0.03$&$-$\\
           &Reverse&$-8.43\,\pm\,2.86$&$-23.47\,\pm\,0.54$&$1.31\,\pm\,0.10$&$1.30\,\pm\,0.09$&$-$\\
           &&($-16.40\,\pm\,4.76$)\phantom{0}&($-21.04\,\pm\,1.16$)&($0.94\,\pm\,0.10$)&($1.01\,\pm\,0.19$)&$-$\\
           &Fixed (Ti16)&$-7.1$&$-24.26\,\pm\,0.22$&$1.08\,\pm\,0.04$&$1.10\,\pm\,0.04$&$-0.43\,\pm\,0.24$\\
           &&&($-23.84\,\pm\,0.16$)&($0.64\,\pm\,0.00$)&($0.67\,\pm\,0.03$)&($-0.01\,\pm\,0.18$)\\
           &Fixed (TP00)&$-8.8$&$-23.42\,\pm\,0.28$&$1.34\,\pm\,0.10$&$1.35\,\pm\,0.05$&$-0.25\,\pm\,0.52$\\
           &&&($-22.87\,\pm\,0.17$)&($0.63\,\pm\,0.10$)&($0.68\,\pm\,0.04$)&\phantom{0}\,\,($0.30\,\pm\,0.47$)\\\\
          Bin~A Final&Forward&$-1.92\,\pm\,0.55$&$-24.65\,\pm\,0.14$&$0.35\,\pm\,0.11$&$0.34\,\pm\,0.03$&$-$\\
                            &Reverse&$-3.10\,\pm\,0.85$&$-24.52\,\pm\,0.18$&$0.44\,\pm\,0.09$&$0.43\,\pm\,0.04$&$-$\\
                            &&($-7.15\,\pm\,2.12$)&($-23.47\,\pm\,0.50$)&($0.21\,\pm\,0.10$)&($0.23\,\pm\,0.10$)&$-$\\
                            &Fixed (Ti16)&$-7.1$&$-24.64\,\pm\,0.40$&$1.22\,\pm\,0.10$&$1.20\,\pm\,0.08$&$-0.81\,\pm\,0.41$\\
                            &&&($-24.04\,\pm\,0.11$)&($0.15\,\pm\,0.02$)&($0.23\,\pm\,0.03$)&($-0.18\,\pm\,0.14$)\\
                             &Fixed (TP00)&$-8.8$&$-23.89\,\pm\,0.52$&$1.59\,\pm\,0.02$&$1.56\,\pm\,0.14$&$-0.72\,\pm\,0.68$\\
                            &&&($-23.10\,\pm\,0.15$)&($0.21\,\pm\,0.00$)&($0.30\,\pm\,0.03$)&\phantom{0}\,($0.08\,\pm\,0.46$)\\\\
           Bin~B Final&Forward&$-3.03\,\pm\,1.57$&$-24.04\,\pm\,0.37$&$0.78\,\pm\,0.17$&$0.73\,\pm\,0.19$&$-$\\
                            &Reverse&$-10.99\,\pm\,5.69\,\,\,$&$-22.56\,\pm\,1.15$&$1.49\,\pm\,0.10$&$1.40\,\pm\,0.37$&$-$\\
                            &&($-11.95\,\pm\,2.80$)\phantom{0}&($-21.73\,\pm\,0.72$)&($0.59\,\pm\,0.10$)&($0.63\,\pm\,0.25$)&$-$\\
                            &Fixed (Ti16)&$-7.1$&$-23.89\,\pm\,0.31$&$1.03\,\pm\,0.10$&$1.03\,\pm\,0.06$&$-0.05\,\pm\,0.32$\\
                            &&&($-23.48\,\pm\,0.18$)&($0.51\,\pm\,0.01$)&($0.54\,\pm\,0.05$)&\phantom{0}\,($0.35\,\pm\,0.20$)\\
                            &Fixed (TP00)&$-8.8$&$-23.01\,\pm\,0.36$&$1.21\,\pm\,0.10$&$1.20\pm\,0.08$&\phantom{0}\,$0.16\,\pm\,0.56$\\
                            &&&($-22.50\,\pm\,0.18$)&($0.47\,\pm\,0.10$)&($0.53\,\pm\,0.06$)&\phantom{0}\,($0.67\,\pm\,0.48$)\\
                            &Fixed (Bin~A)&$-3.1$&$-24.07\,\pm\,0.23$&$0.79\,\pm\,0.07$&$0.77\,\pm\,0.06$&$\phantom{0}\,\,0.45\,\pm\,0.29$\\
                                               
    \hline	
  \end{tabular}		
  \label{tab:mktfr}
  \parbox[t]{0.9\textwidth}{\textit{Notes:} Ti16: \citet{ti16}, TP00:
    \citet{tp00}. Values in parentheses are for the reverse fits after
    excluding the outliers (see Section~\ref{subsec:outliers}).}
\end{table*}
%
%
\begin{table*}
  \caption{Best-fit parameters of the $M_{\star}$ and $M_{\rm b}$ CO TFRs}
  \begin{tabular}{lllccccc}
    \hline
    &Sub-sample&Fit type&Slope&Zero-point&$\sigma_{\textrm{int}}$&$\sigma_{\textrm{total}}$&Zero-point offset\\ 
          &&&&$\log(M/M_\odot)$&$\log(M/M_\odot)$&$\log(M/M_\odot)$&(Ours - Theirs)\\
    \hline
   $M_{\star}$ CO TFRs  &Initial&Forward&$0.60\,\pm\,0.27$&$10.81\,\pm\,0.06$&$0.33\,\pm\,0.05$&$0.35\,\pm\,0.01$&$$\\
      &     &Reverse&$5.59\,\pm\,2.64$&$10.27\,\pm\,0.33$&$1.03\,\pm\,0.14$&$1.08\,\pm\,0.11$&$$\\
       &     &&($5.60\,\pm\,2.38$)&($10.09\,\pm\,0.38$)&($0.90\,\pm\,0.14$)&($0.94\,\pm\,0.11$)&$$\\
     & Final&Forward&$0.55\,\pm\,0.34$&$10.84\,\pm\,0.08$&$0.29\,\pm\,0.05$&$0.30\,\pm\,0.02$&$$\\
       &    &Reverse&$5.18\,\pm\,3.00$&$10.08\,\pm\,0.52$&$0.90\,\pm\,0.15$&$0.89\,\pm\,0.10$&$$\\
       	&    &&($6.06\,\pm\,1.54$)&\phantom{0}($9.51\,\pm\,0.38$)&($0.31\,\pm\,0.09$)&($0.36\,\pm\,0.06$)&$$\\
        &   &Fixed (Ti16)&$3.3$&$10.65\,\pm\,0.12$&$0.57\,\pm\,0.12$&$0.58\,\pm\,0.02$&$\phantom{0}\,0.14\,\pm\,0.13$\\
        &   &&&($10.44\,\pm\,0.06$)&($0.22\,\pm\,0.09$)&($0.27\,\pm\,0.01$)&($-0.07\,\pm\,0.07$)\\\\
        
        &Bin~A Final&Reverse&$-22.13\,\pm\,166.57\phantom{0}$&$\phantom{0}13.23\,\pm\,18.25$&$5.40\,\pm\,0.30$&$4.93\,\pm\,4.34$&$$\\
        &&&($11.37\,\pm\,16.86$)&\phantom{0}($8.25\,\pm\,3.78$)&($0.86\,\pm\,0.27$)&($0.74\,\pm\,1.24$)&$$\\
        &   &Fixed (Ti16)&$3.3$&$10.73\,\pm\,0.26$&$0.79\,\pm\,0.18$&$0.78\,\pm\,0.08$&$\phantom{0}\,0.22\,\pm\,0.26$\\
        &   &&&($10.35\,\pm\,0.15$)&($0.28\,\pm\,0.12$)&($0.30\,\pm\,0.04$)&($-0.15\,\pm\,0.16$)\\
        &Bin~B Final&Reverse&$4.35\,\pm\,2.04$&$10.11\,\pm\,0.42$&$0.57\,\pm\,0.15$&$0.55\,\pm\,0.13$&$$\\
        &&&($4.91\,\pm\,1.19$)&\phantom{0}($9.75\,\pm\,0.31$)&($0.23\,\pm\,0.11$)&($0.26\,\pm\,0.10$)&$$\\
        &   &Fixed (Ti16)&$3.3$&$10.57\,\pm\,0.13$&$0.42\,\pm\,0.14$&$0.43\,\pm\,0.03$&$\phantom{0}\,0.06\,\pm\,0.14$\\
        &   &&&($10.41\,\pm\,0.08$)&($0.19\,\pm\,0.10$)&($0.23\,\pm\,0.02$)&($-0.09\,\pm\,0.10$)\\\\
    
   $M_{\rm b}$ CO TFRs&Initial&Forward&$0.55\,\pm\,0.23$&$10.92\,\pm\,0.05$&$0.29\,\pm\,0.04$&$0.30\,\pm\,0.01$&$$\\
      &     &Reverse&$4.96\,\pm\,2.42$&$10.45\,\pm\,0.30$&$0.92\,\pm\,0.12$&$0.95\,\pm\,0.06$&$$\\
      &     &&($4.78\,\pm\,1.98$)&($10.32\,\pm\,0.31$)&($0.76\,\pm\,0.10$)&($0.80\,\pm\,0.08$)&$$\\
     & Final&Forward&$0.51\,\pm\,0.31$&$10.93\,\pm\,0.08$&$0.27\,\pm\,0.05$&$0.28\,\pm\,0.02$&$$\\
       &    &Reverse&$4.86\,\pm\,2.83$&$10.23\,\pm\,0.50$&$0.85\,\pm\,0.14$&$0.83\,\pm\,0.12$&$$\\
       &    &&($5.30\,\pm\,1.11$)&\phantom{0}($9.78\,\pm\,0.28$)&($0.23\,\pm\,0.05$)&($0.28\,\pm\,0.06$)&$$\\
       \hline	
  \end{tabular}
  \label{tab:masstfr}		
  \parbox[t]{0.9\textwidth}{\textit{Notes:} Ti16: \citet{ti16}. Values
    in parentheses are for the reverse fits after excluding the
    outliers (see Section~\ref{subsec:outliers}).}
\end{table*}
%
%
%
%
\begin{figure*}
\includegraphics[width=9.28cm,clip=]{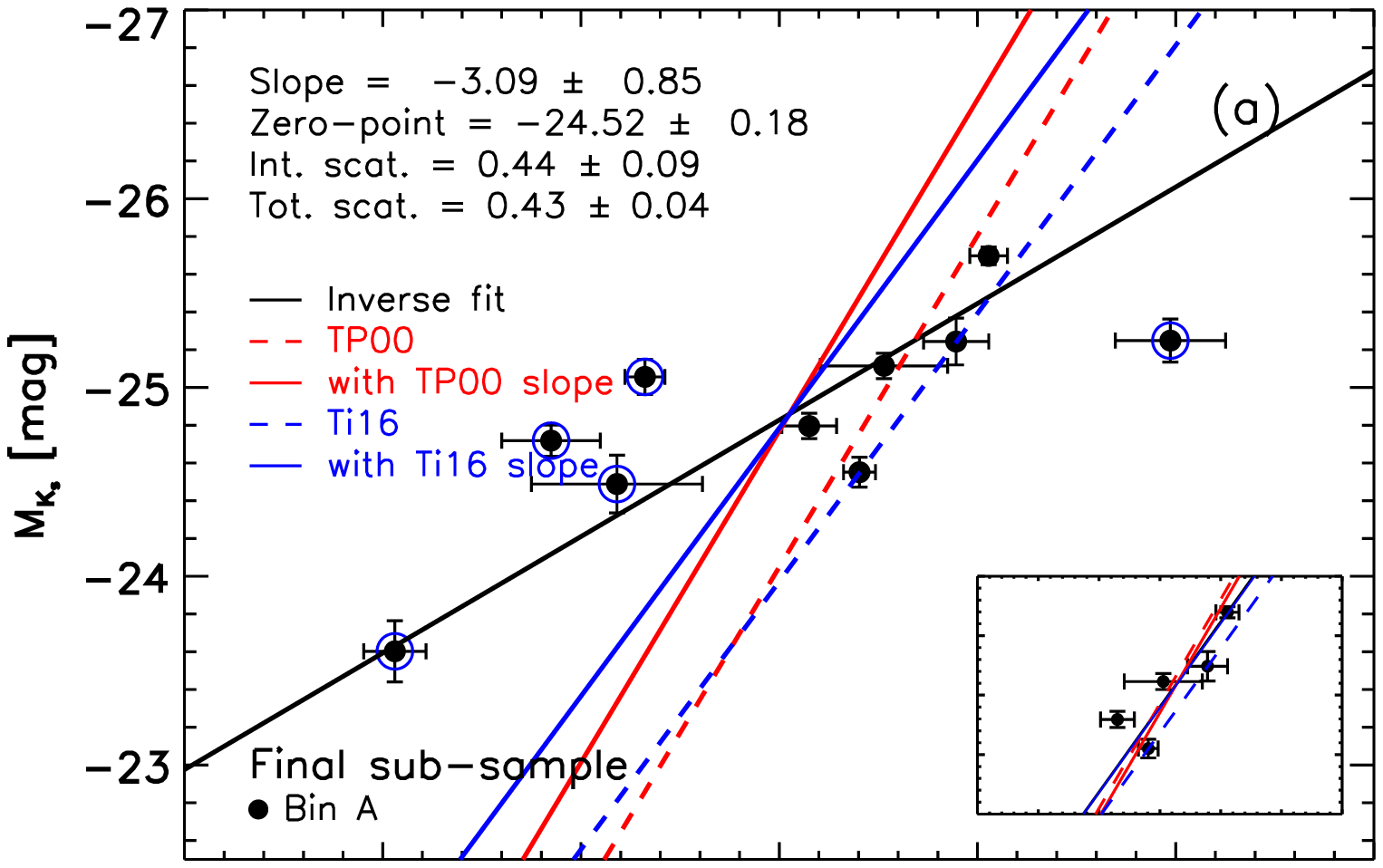}
  \hspace{-40pt}
  \includegraphics[width=9.28cm,clip=]{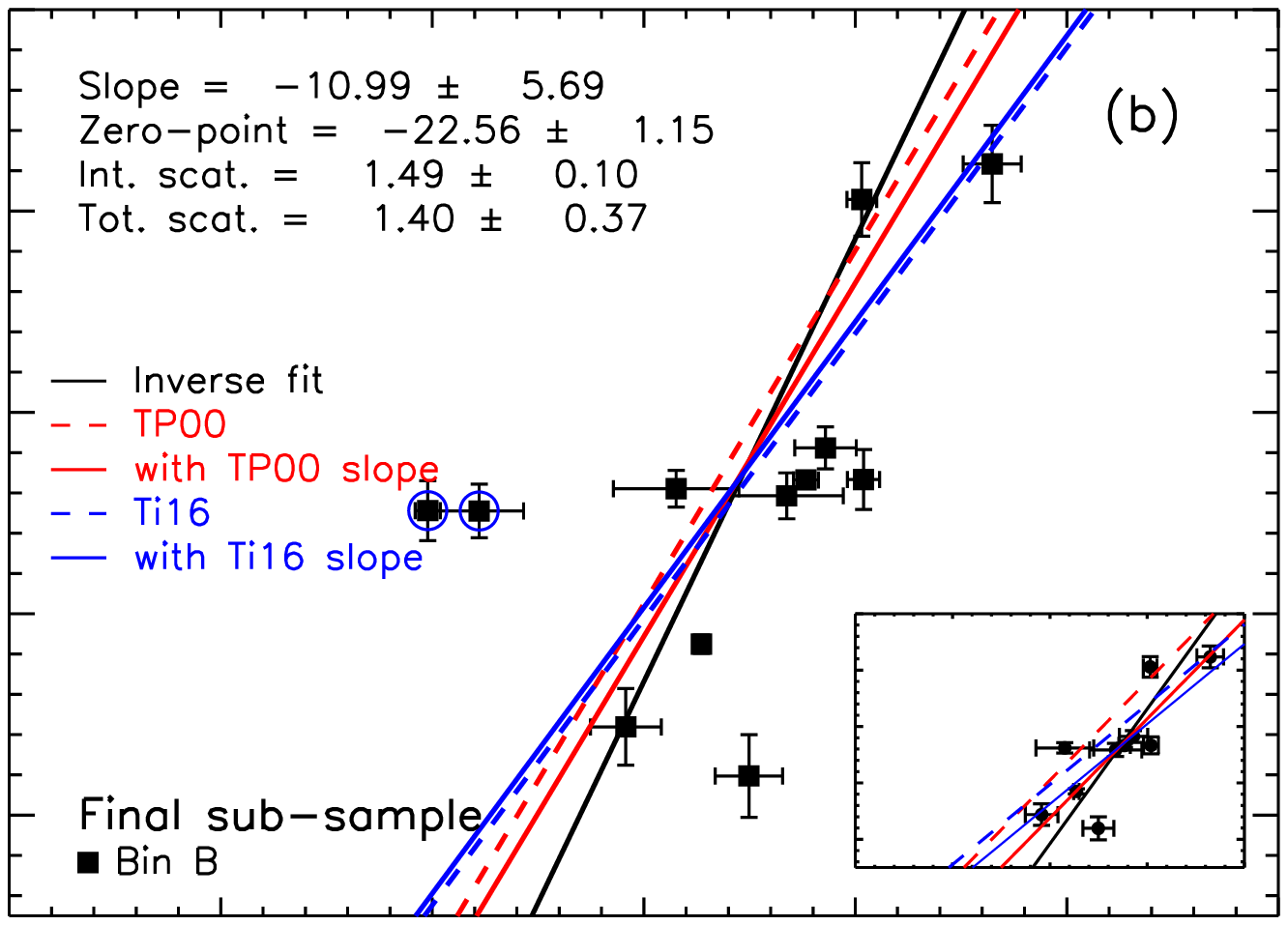}\\
  \vspace{-30pt}
  \includegraphics[width=9.28cm,clip=]{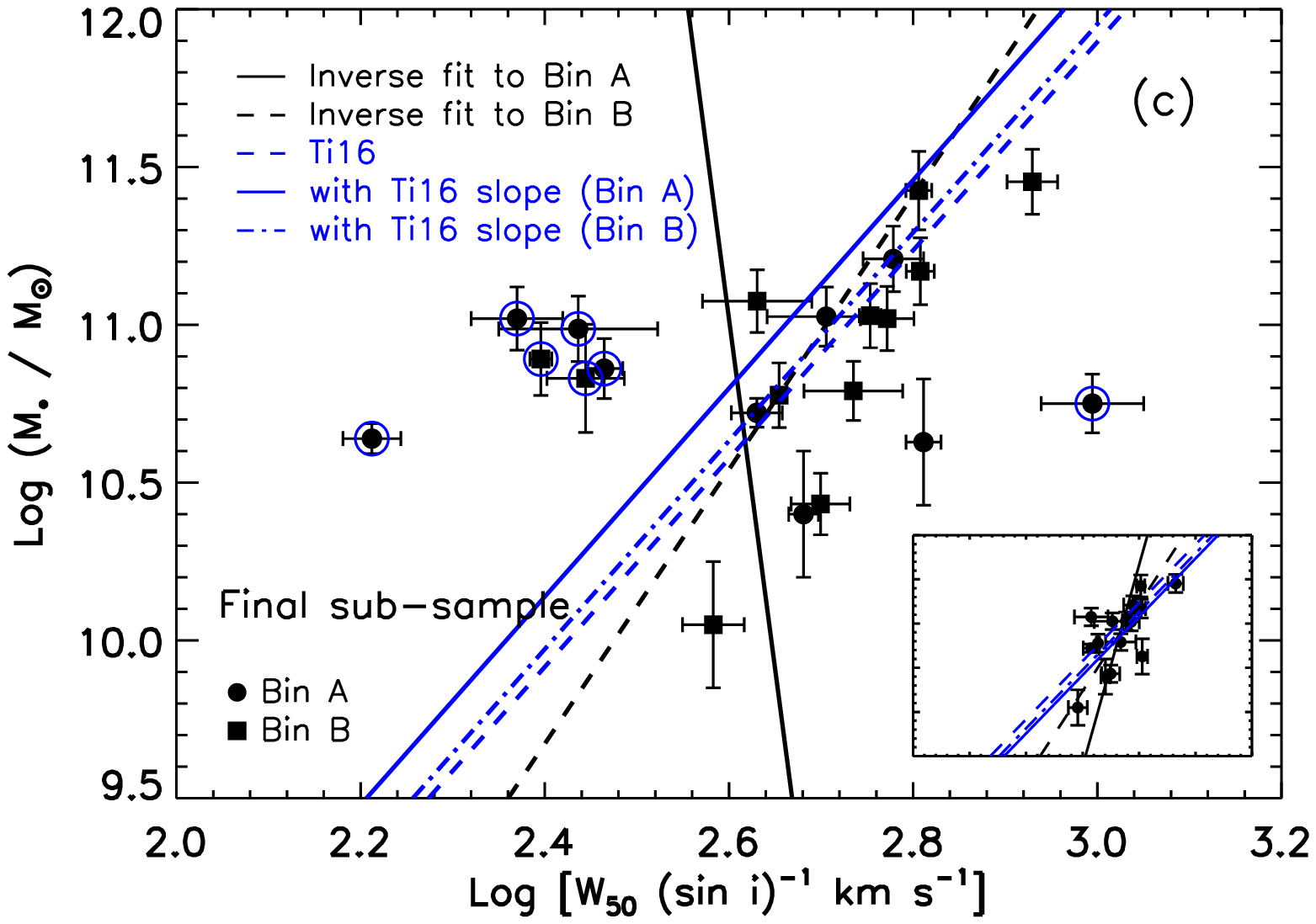}
   \hspace{-40pt}
  \includegraphics[width=9.28cm,clip=]{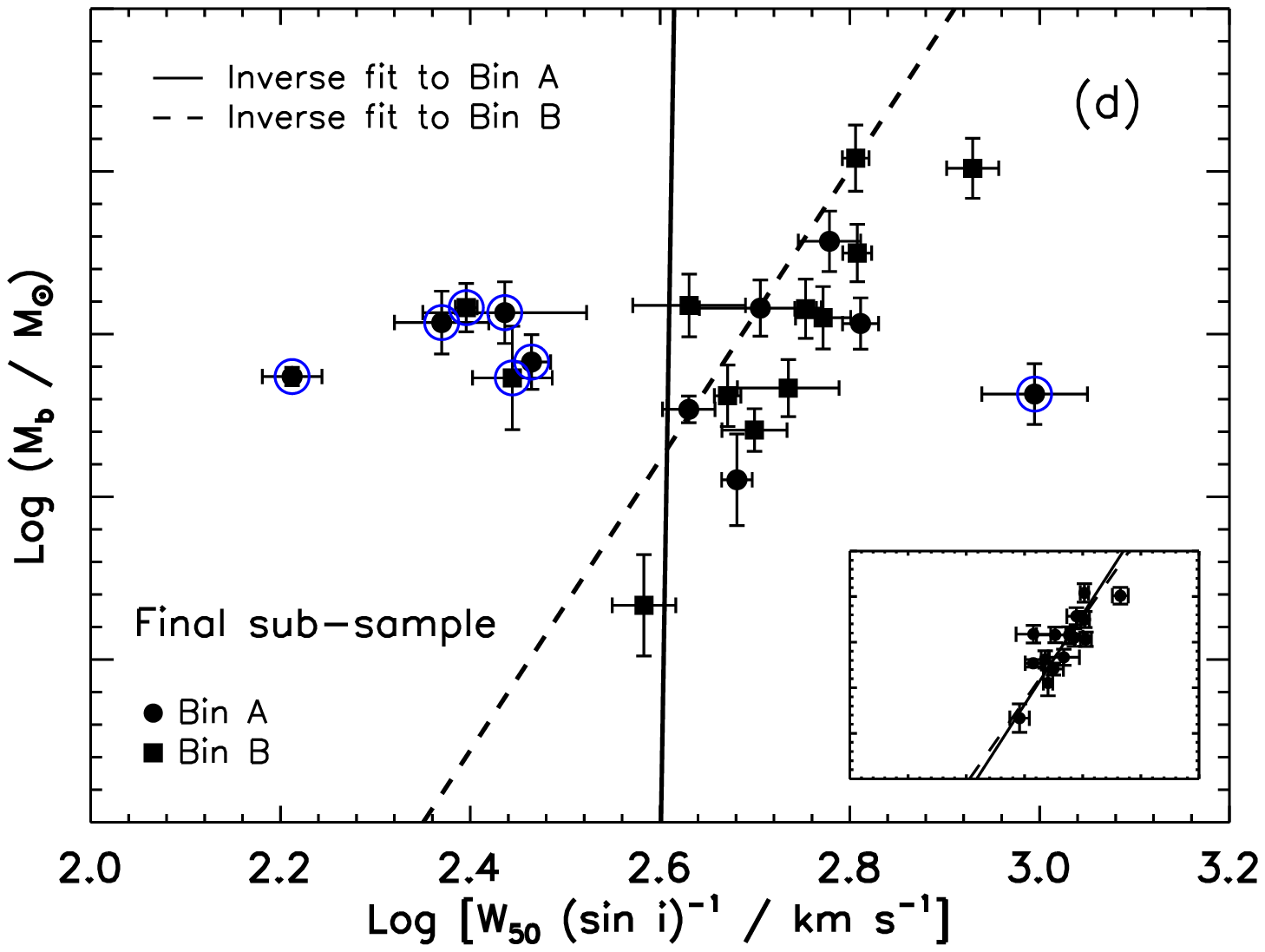} \\
  \caption{{\bf Top}: $K_{\textrm{s}}$-band CO TFR of
      bin~A (left) and bin~B (right) galaxies only in the final
      sub-sample. The black solid lines show the reverse fits, while
      the red (respectively blue) lines show the reverse fit with
      slope fixed to that of \citet{tp00} (resp.\ \citealt{ti16}). The
      red dashed lines show the H\,{\small I} TFR of local spirals
      from \citet{tp00}. The blue dashed lines show the CO TFR of
      local spirals \citep{ti16}. {\bf Bottom}: In panel (c), the
      $M_{\star}$ CO TFRs of the final sub-sample galaxies in bin~A
      and bin~B are shown in the same plot. Similarly, in panel (d)
      the $M_{\rm b}$ CO TFRs of the final sub-sample galaxies in
      bin~A and bin~B are shown in the same plot. In both panels (c)
      and (d), the black solid lines and black dashed lines show the
      reverse fits for the final sub-sample galaxies in bin~A and
      bin~B, respectively. In panel (c), the blue solid line and the
      blue dot-dashed line shows the reverse fit with slope fixed to
      that of \citet{ti16} for the final sub-sample galaxies in bin~A
      and bin~B, respectively. The blue dashed line in panel (c) shows
      the CO TFR of \citet{ti16}, as for the top panels. In all
      panels, the black filled circles and black filled squares show
      the final sub-sample galaxies in bin~A and bin~B,
      respectively. In all panels, the data points shown as open blue
      circles represent the outliers (see
      Section~\ref{subsec:outliers}). The embedded figures in all
      panels show the TFR of the final sub-sample galaxies in bin~A
      and bin~B after excluding the outliers.}
  \label{fig:bintfr}
\end{figure*}
%
%
\section{Discussion}
\label{sec:discussion}
A relation between luminous and dynamical (i.e.\ total) mass, the TFR
informs us about both the structural and the dynamical properties of
galaxies, particularly the total mass surface density and total
mass-to-light ratio (and thus all properties affecting this ratio,
including the stellar $M/L$, gas content and dark matter
content). Comparing the TFRs measured for different galaxy samples
thus reveals differences in those properties. These differences are,
however, also tightly connected to the way the samples were
selected. For example, comparing the TFRs of galaxies of different
morphological types at a given redshift will inform on differences
between those types \citep[e.g.][]{rus04,shen09,lag13}, while
comparing the TFRs of galaxies of a given type at different redshifts
will inform on the evolution of those galaxies
\citep[e.g.][]{cons05,flo06,pu08}. Similarly, luminosities for a given
sample measured in different bands will inform on the stellar
$M/L$. However, when comparing different TFRs, one must make sure that
all the parameters used (e.g.\ luminosity, rotation velocity,
inclination, and any corrections to those) are measured or calculated
in an identical manner, as otherwise any difference between the
zero-points and/or slopes of different TFRs could be due to different
systematics between the methods used rather than any intrinsic
physical differences between the samples. With those caveats in mind,
we compare our results to others in the literature below.
%
%
\subsection{Previous studies}
\label{subsec:previous}
The literature on the TFR is vast, but the number of studies using CO
as the kinematic tracer is small. The most recent studies are those of
\citet{d11}, \citet{ti16} and Torii et al.\ (in prep.). \citet{d11}
studied ETGs, however, so we will refrain from a comparison here as we
would be unable to assign any difference to a redshift evolution
rather than structural differences, or vice-versa. \citet{ti16} and
Torii et al.\ (in prep.) also targeted disc galaxies and measured the
velocity widths in a manner identical to us. They are thus best suited
for comparison. The sample of \citet{ti16} is composed of $\approx300$
disc galaxies from the CO Legacy Database for the GALEX Arecibo SDSS
survey (GASS, \citealt{cat10}; COLD GASS, \citealt{sain11}) and they
used the Wide-Field Infrared Survey Explorer (WISE) Band~1 ($W1$,
$\approx3.4\micron$) to construct their TFR. Although the $W1$ band is
not identical to the 2MASS $K_{\textrm{s}}$ band
($\approx2.4\micron$), both filters trace similar stellar populations
and $K-W1\approx0.0\pm0.2$~mag for late-type galaxies \citep{lag13},
so a direct comparison is appropriate. Torii et al.\ (in prep.)
studied $\approx50$ late-type nearby galaxies and also used 2MASS
$K_{\textrm{s}}$-band magnitudes, but their study is
  not finalised yet and so will not be considered further.

At a fixed stellar mass, galaxies are generally
  smaller at higher redshifts. The COLD GASS galaxies of \citet{ti16}
  are all located at $z\approx0.03$ and have an average effective
  (half-light) radius ($R_{\rm e}$) of $2\farcs6$ (based on SDSS
  $r$-band photometry), corresponding to a linear size of $1.6$~kpc at
  an average distance of $\approx130$~Mpc (based on the cosmology
  calculator of \citealt{wr06}). On the other hand, all galaxies in
  our sample, located $z=0.05$--$0.3$, have $R_{\rm e}>1.6$~kpc. This
  suggests that our galaxies are both CO bright and larger on average
  than \citeauthor{ti16}'s (\citeyear{ti16}) galaxies.

Despite their use of H\,{\small I} rather than CO, we also discuss the
work of \citet{tp00} below, as it is generally considered the standard
reference on the subject. \citeauthor{tp00} measured the velocity
width at $20\%$ of the peak (rather than $50\%$ as done here). While
the difference is generally small, the velocity width at $20\%$ of the
peak is systematically larger.
%
%
\subsection{Outliers}
\label{subsec:outliers}
As seen from Figure~\ref{fig:tfr}, there are a few low-velocity
galaxies and one high-velocity galaxy that appear to constitute clear
outliers with respect to the general trend in the data.
Those galaxies with velocities
  $2.5\ge\log(W_{\rm 50}\sin^{-1}i$~/~km~s$^{-1})\ge3.0$ are listed in
  Tables~\ref{tab:general} and \ref{tab:TFR} (i.e.\ galaxies labeled
  as G1, G3, G4, G5, G11, G24 and G25). We examine the properties of
  those $7$ galaxies and then the TFR results with/without them, to
  better understand whether they are intrinsically different.

Five out of seven galaxies are located at $z\le0.1$, while the
remaining two galaxies are located at $z\approx0.2$. The SFRs of the
galaxies range from $5$ to $56$~$M_{\odot}$~yr$^{-1}$, except one
galaxy (G24) with a very low SFR of $0.02$~$M_{\odot}$~yr$^{-1}$.
Very high inclinations can bias photometric
  measurements due to internal absorption, but the outlier galaxies
  have inclinations ranging from $43$ to $65\degr$, except one galaxy
  (G24) with a relatively high inclination of $84\degr$. Furthermore,
  the stellar and baryonic masses of the outliers are similar to those
  of the other galaxies in the final sub-sample (see
  Table~\ref{tab:TFR}). Overall, the outliers thus seem unremarkable.

The embedded panels in Figures~\ref{fig:tfr} and \ref{fig:bintfr}
represent the TFR fits to the data after excluding the $7$
outliers. The results based on those fits are also listed in
Tables~\ref{tab:mktfr} and \ref{tab:masstfr} (values shown in
parentheses) and are discussed in the following sub-sections.
%
%
\subsection{Evolution with redshift}
\label{subsec:redshift}
Some early works suggest an evolution (i.e.\ a different zero-point
and/or slope) of the $B$-band TFR at intermediate redshifts, in the
sense that higher redshift galaxies are brighter at a given rotational
velocity \citep[e.g.][]{vo96,si98,zi02,mil03,bo04,bam06}, although it
may be that only low-mass systems show such an offset
\citep{zi02,bo04}. Furthermore, \citet{zi02} studied the $B$-band TFR
of $60$ late-type galaxies at $z=0.1$--$1.0$, and found that the
slope is flatter for distant galaxies. More recent works however seem
to indicate that there is no redshift evolution in the $K$-band
relation (e.g.\ \citealt{cons05,flo06,ti16}; but see \citealt{pu08}
who found that $z\approx0.6$ galaxies are {\em fainter} than local
galaxies by $0.66\pm0.14$~mag at $K$-band). There is thus clearly some
disagreement in the literature as to whether the TFR evolves with
redshift or not.
\subsubsection{Evolution in slope and luminosity}
\label{subsubsec:slobri}
As the galaxies in our samples cover a reasonable range in redshift
($z=0.3$ corresponds to a $\approx3.5$~Gyr lookback time), it is
possible to probe whether the TFR has evolved during that period by
simply breaking down our galaxy samples by redshift.
We therefore split our samples into three redshift
  bins and constructed the CO TFR for each bin separately
  (Fig.~\ref{fig:bintfr}). Bin~A includes galaxies at $z=0.05$--$0.1$,
  bin~B galaxies at $z=0.1$--$0.2$, and bin~C galaxies at
  $z=0.2$--$0.3$ (the black filled circles, black filled squares and
  black filled triangles in Fig.~\ref{fig:tfr}, respectively). Bin~A
  includes $17$ galaxies ($10$ galaxies with a double-horned profile),
  bin~B $18$ galaxies ($12$ galaxies with a double-horned profile),
  but in bin~C only $5$ galaxies ($3$ galaxies with a double-horned
  profile), not enough for a reliable fit. The average galaxy SFRs are
  also different from each other, $14$, $18$ and
  $31$~$M_\odot$~yr$^{-1}$ for bin~A, B and C, respectively, while
  COLD GASS galaxies have an average SFR of
  $\approx4$~$M_\odot$~yr$^{-1}$. Interestingly, the galaxies in bin~A
  have masses similar to each other (both stellar and baryonic)
  despite a wide range of rotational velocities, causing the rather
  flat distributions in the $M_{K_{\textrm{s}}}$, $M_{\star}$ and
  $M_{\rm b}$ CO TFRs (Fig.~\ref{fig:bintfr}). The galaxies in bin~B
  have a wider range of masses but a relatively narrow range of
  rotational velocities (Fig.~\ref{fig:bintfr}).

The limitations described above lead to unreliable
  fits for individual bins (see Tables~\ref{tab:mktfr} and
  \ref{tab:masstfr}), and the slopes are essentially unconstrained
  (very large uncertainties). Nevertheless, it is possible to
  constrain zero-point offsets by using a unique slope across all
  bins. Fixing the slope of the final sub-sample reverse fit in bin~B
  to that of bin~A, we found a zero-point offset of $0.45\pm0.29$~mag
  (bin~B - bin~A), indicating that the galaxies in bin~B are on
  average about $1.5$ times fainter than those in bin~A. However, this
  offset is not statistically significant, i.e.\ S/N$\,<3$.

We also examined the galaxies in bin~A and bin~B only after excluding
the outliers. Except for the smaller scatters (as expected by
construction), we again found no significant change in the slopes and
zero-points (see Table~\ref{tab:mktfr}).

The other way to test for redshift evolution is to compare our results
with those of other studies of local galaxies, although we must then
be aware of differences between the samples and/or methods. Comparing
to the TFR studies discussed above (Section~\ref{subsec:previous}),
our results (see Table~\ref{tab:mktfr}) indicate that the slope of the
reverse $K_{\textrm{s}}$-band CO TFR of the final sub-sample is
consistent with that of nearby disc galaxies within the uncertainties
(that are however quite large; e.g.\ \citealt{tp00,ti16}; see also
\citealt{zi02}). This therefore suggests that there is no significant
evolution of the slope of the TFR between local spirals and the
galaxies in our final sub-sample at $z=0.05$--$0.3$.
When different redshift bins are considered, the slope
  of the reverse $K_{\textrm{s}}$-band CO TFR of bin~A is somewhat
  flatter than all the other samples, but this difference disappears
  when excluding the outliers from bin~A and is thus doubtful (see
  Fig.~\ref{fig:bintfr} and Table~\ref{tab:mktfr}).

Fixing the slope of the reverse fit to that of the spirals in
\citet{ti16} and \citet{tp00}, we found a zero-point offset (our fit
minus theirs) of $-0.43\pm0.24$ and
  $-0.25\pm0.52$~mag, respectively ($-0.01\pm0.18$ and
  $0.30\pm0.47$~mag, respectively, after excluding the outliers; see
  Table~\ref{tab:mktfr}). \citet{ti16} being the study most similar
to ours, this suggests that, at a given rotation velocity, our final
sub-sample galaxies are on average brighter than local galaxies by
$\approx0.43$~mag or a factor of about $1.5$. As
  expected from the comparisons of bin~A and bin~B galaxies above,
  when fixing the slope to that of the spirals of \citet{ti16} and
  repeating the fit for bin~A galaxies only, we find a larger
  zero-point offset of $-0.81\pm0.41$~mag ($-0.18\pm0.14$~mag after
  excluding the outliers), a factor of slightly more than two in
  luminosity, while the offset between the galaxies in bin~B and
  \citet{ti16} is negligible, i.e. $-0.05\pm0.32$~mag
  ($0.35\pm0.20$~mag after excluding the outliers). All these offsets
  are listed in Table~\ref{tab:mktfr}, but none is truly significant.
\subsubsection{Evolution in mass}
\label{subsubsec:evomass}
We now turn our attention to the $M_{\star}$ and $M_{\rm b}$ CO TFRs
(see Fig.~\ref{fig:tfr}).
The slope of the $M_{\star}$ CO TFR for our final
  galaxy sub-sample is similar to that of \citet{ti16} for local
  spiral galaxies. In addition, fixing the slope to that of
  \citet{ti16}, we find a zero-point offset (our fit minus their) of
  $0.14\pm0.13$~dex ($-0.07\pm0.07$~dex after excluding the outliers;
  see Table~\ref{tab:masstfr}), indicating no significant evidence for
  evolution of the $M_{\star}$ TFR zero-point since $z\approx0.3$. If
  we probe the offset between bin~A and \citet{ti16} galaxies, we
  obtain an offset of $0.22\pm0.26$~dex ($-0.15\pm0.16$~dex after
  excluding the outliers). For bin~B galaxies, we obtain an offset of
  $0.06\pm0.14$~dex ($-0.09\pm0.10$~dex after excluding the
  outliers). All results based on the final sub-sample thus arrive the
  same conclusion: no offset in stellar mass.

The results for the $M_{\rm b}$ CO TFRs agree with
  those of the $M_{\star}$ CO TFRs, i.e.\ same slopes and zero-points
  within the uncertainties, but with slightly smaller scatters. The
  results after excluding the outliers also indicate the same slopes
  and zero-points within the uncertainties (see
  Table~\ref{tab:masstfr}). Since, there is no baryonic mass CO TFR in
  the literature, we shall not discuss this relation further.
  Otherwise, any comparison between $M_{\rm b}$ TFR studies exploiting
  other kinematic tracers would introduce too many potential
  systematics.

Overall, when all galaxies in the final sub-sample are considered, and
given the large uncertainties in the offsets found, our results for
the $K_{\textrm{s}}$-band, $M_{\star}$ and $M_{\rm b}$ CO TFRs suggest
no significant redshift evolution in either luminosity or mass (even
after excluding the outliers).

As they were selected to be on the upper envelope of the
star-formation main sequence at their redshifts \citep{ba13}, and as
the SFR of the main sequence increases with redshift
\citep[e.g.][]{noe07}, one would naively have expected the higher
redshift galaxies to have slightly lower stellar (and thus dynamical)
$M/L$. This would naturally explain any brightening in luminosity with
$z$, but would not predict a commensurate increase in mass (as the
stellar $M/L$ of our sample galaxies would then be lower than those of
local galaxies). This effect is not observed here, presumably because
of a combination of the relatively small coverage in redshift and the
relatively large uncertainties in the data (and thus TFRs).
%
%
\subsection{Intrinsic scatter} 
\label{subsec:scatter}
As can be seen from Table~\ref{tab:mktfr}, the intrinsic scatter of
the reverse $K_{\textrm{s}}$-band CO TFR of the final sub-sample
($\sigma_{\textrm{int}}=1.3$) is higher than that of
\citet{ti16} ($\sigma_{\textrm{int}}=0.7$) and \citet{d11}
($\sigma_{\textrm{int}}=0.6$). The reasons for this relatively higher
scatter are unclear, since the final sub-sample only contains galaxies
that should yield robust measurements, but we can
speculate. In fact, the scatter decreases to a
  comparable or even lower value when the outliers are excluded,
  particularly when different redshift bins are considered (see
  Table~\ref{tab:mktfr}).

The TFR is known to have a much greater scatter at higher redshifts
\citep[e.g.][]{til16}, this for a variety of reasons such as greater
variations of the stellar mass fraction (and thus total $M/L$ ratio)
and stellar $M/L$ ratio, and most importantly increased morphological
and dynamical anomalies \citep[e.g.][]{knap02,flo06,kas07}. It could
thus be that some of these effects are already significant at
$z\lesssim0.3$.

Our inclinations derived from the stellar axial ratios could also
introduce more scatter than superior measurements \citep[e.g][]{d11},
although as long as the uncertainties are properly quantified this
should only affect the total scatter ($\sigma_{\textrm{total}}$) and
not the intrinsic scatter ($\sigma_{\textrm{int}}$). In addition, due
to the difficulty of identifying interacting and/or disturbed galaxies
at the modest resolution of SDSS, it is possible that despite our best
efforts to exclude them some interacting galaxies do remain in the
initial and final samples. Overall, however, the main reason behind
the large intrinsic scatter measured remains unclear.
%
%
\subsection{Inclinations}
\label{subsec:idiscuss}
In view of the comments in Section~\ref{subsec:scatter}, it is worth
noting that the accuracy of the TFR fits strongly depends on the
accuracy of the inclination measurements, as the circular velocity
measurements (here the velocity widths) must be corrected for the
inclination of the galaxies. We used here stellar axial ratios to
estimate the inclinations, as is common in the literature
\citep[e.g.][]{tp00,d11,ti16}. Although these inclinations can lead to
a large scatter in the TFR, they do not generally affect its slope
and/or zero-point \citep[e.g.][]{d11}. For our sample, the slope and
zero-point obtained for the initial sub-sample are consistent with
those of the final sub-sample within the uncertainties (see
Tables~\ref{tab:mktfr} and \ref{tab:masstfr}), indicating that our
results are indeed robust and only minimally affected by inclination
uncertainties.

We assumed a value of $0.2$ for $q_{0}$ (i.e.\ we
  assumed late-type systems; \citealt{tu77,tp88}). However, it is
  clear that any variation in $q_{0}$ will affect the inclinations
  inferred, and thus the TFR results. We investigated this effect and
  found that the effect is very small. For example, assuming $q_{0}=0$
  would yield the same zero-point and slope for both the initial and
  final sub-samples within the errors. In particular, the change in
  the zero-points are tiny. Similarly, if we assume $q_{0}=0.34$ (as
  for ETGs; \citealt{d11}), the results for the slopes and zero-points
  are again unchanged within the errors. This indicates that $q_{0}$
  uncertainties have an insignificant effect on the
  inclination-corrected velocities and thus our TFR results.
%
%
\section{Conclusions}
\label{sec:conclusions}
We studied the $K_{\textrm{s}}$-band, stellar mass and baryonic mass
CO TFRs of $25$ carefully selected galaxies at $z\approx0.05$--$0.3$,
and compared our results to those obtained for similar local disc
galaxy samples. This represents the first attempt to construct TFRs
for disc galaxies beyond the local universe using CO as a kinematic
tracer. The principal results are summarised below.

\begin{enumerate}
\item The best-fit reverse $K_{\text{s}}$-band,
    stellar mass and baryonic mass TFRs are
    $M_{K_{\text{s}}}=(-8.4\pm2.9)\left[\log{\left(\frac{W_{50}/\rm{km~s^{-1}}}{\sin{i}}\right)}-2.5\right]\,+\,(-23.5\pm0.5)$,
    $\log{\left(M_{\star} /
        M_\odot\right)}=(5.2\pm3.0)\left[\log{\left(\frac{W_{50}/\rm{km~s^{-1}}}{\sin{i}}\right)}-2.5\right]\,+\,(10.1\pm0.5)$
    and
    $\log{\left(M_{\rm b} /
        M_\odot\right)}=(4.9\pm2.8)\left[\log{\left(\frac{W_{50}/\rm{km~s^{-1}}}{\sin{i}}\right)}-2.5\right]\,+\,(10.2\pm0.5)$,
    respectively, where $M_{K_{\text{s}}}$ is the total absolute
    $K_{\text{s}}$-band magnitude of the objects, $M_{\star}$ and
    $M_{\rm b}$ their total stellar and baryonic mass, respectively,
    and $W_{50}$ the width of their integrated CO line profile at
    $50\%$ of the maximum.

\item When different redshift bins are considered
    within our sample, we find no significant change in the slope or
    zero-point of the TFRs, in either luminosity or mass.

\item When comparing to other TFR studies of local
    ($z=0$) disc galaxies, we again find no significant offset in
    either luminosity or mass.

\item Similarly to galaxies at much higher redshifts, our sample
  galaxies show higher intrinsic scatters around the best-fit TFRs
  than local galaxies. The main drivers of this are also likely
  analogous, i.e.\ higher gas fractions coupled with more intense star
  formation, and morphological as well as dynamical disturbances.
  
\item Although the scatter in the $M_{K_{\text{s}}}$
    TFR is high compared to that of local studies, the scatter
    decreases to comparable and even lower values when obvious
    outliers are excluded, particularly for the case of different
    redshift bins, thus suggesting that the increased scatter is due
    to a few pathological galaxies rather than the general galaxy
    population.
\end{enumerate}  

More generally, our study supports the view that CO is an excellent
kinematical tracer for TFR studies. As CO is relatively easy to detect
even in distant galaxies, our study provides a useful benchmark for
future high-redshift CO TFR studies, themselves a powerful tool to
probe the cosmological evolution of the $M/L$ of galaxies.
%
%
\section*{Acknowledgements}
The authors would like to thank the anonymous referee for his/her
insightful comments and suggestions. ST thanks to Amber Bauermeister
for kindly providing the CO data for the EGNoG sample.  ST was
supported by the Republic of Turkey, Ministry of National Education,
The Philip Wetton Graduate Scholarship at Christ Church, and also
acknowledges support from Van 100. Yil University, project code:
FBA-2017-5874. MB acknowledges support from STFC rolling grant
‘Astrophysics at Oxford’ PP/E001114/1. AT acknowledges support from an
STFC Studentship. This research made use of the NASA/IPAC
Extragalactic Database (NED), which is operated by the Jet Propulsion
Laboratory, California Institute of Technology, under contract with
the National Aeronautics and Space Administration.  This work is based
in part on data obtained as part of the UKIRT Infrared Deep Sky
Survey.  This publication also makes use of data products from the Two
Micron All Sky Survey, which is a joint project of the University of
Massachusetts and the Infrared Processing and Analysis
Center/California Institute of Technology, funded by the National
Aeronautics and Space Administration and the National Science
Foundation. This research also made use of SDSS-III data
release. SDSS-III is managed by the Astrophysical Research Consortium
for the Participating Institutions of the SDSS-III Collaboration
including the University of Arizona, the Brazilian Participation
Group, Brookhaven National Laboratory, Carnegie Mellon University,
University of Florida, the French Participation Group, the German
Participation Group, Harvard University, the Instituto de Astrofisica
de Canarias, the Michigan State/Notre Dame/JINA Participation Group,
Johns Hopkins University, Lawrence Berkeley National Laboratory, Max
Planck Institute for Astrophysics, Max Planck Institute for
Extraterrestrial Physics, New Mexico State University, New York
University, Ohio State University, Pennsylvania State University,
University of Portsmouth, Princeton University, the Spanish
Participation Group, University of Tokyo, University of Utah,
Vanderbilt University, University of Virginia, University of
Washington, and Yale University.
\bibliographystyle{mn2e}
\bibliography{reference}
%
%
\appendix
%
%
\section{Integrated CO profiles and best fits}
\label{sec:Ap1}
%
%
\begin{figure*}
  \hspace{-15pt}
  \includegraphics[width=5.8cm,clip=]{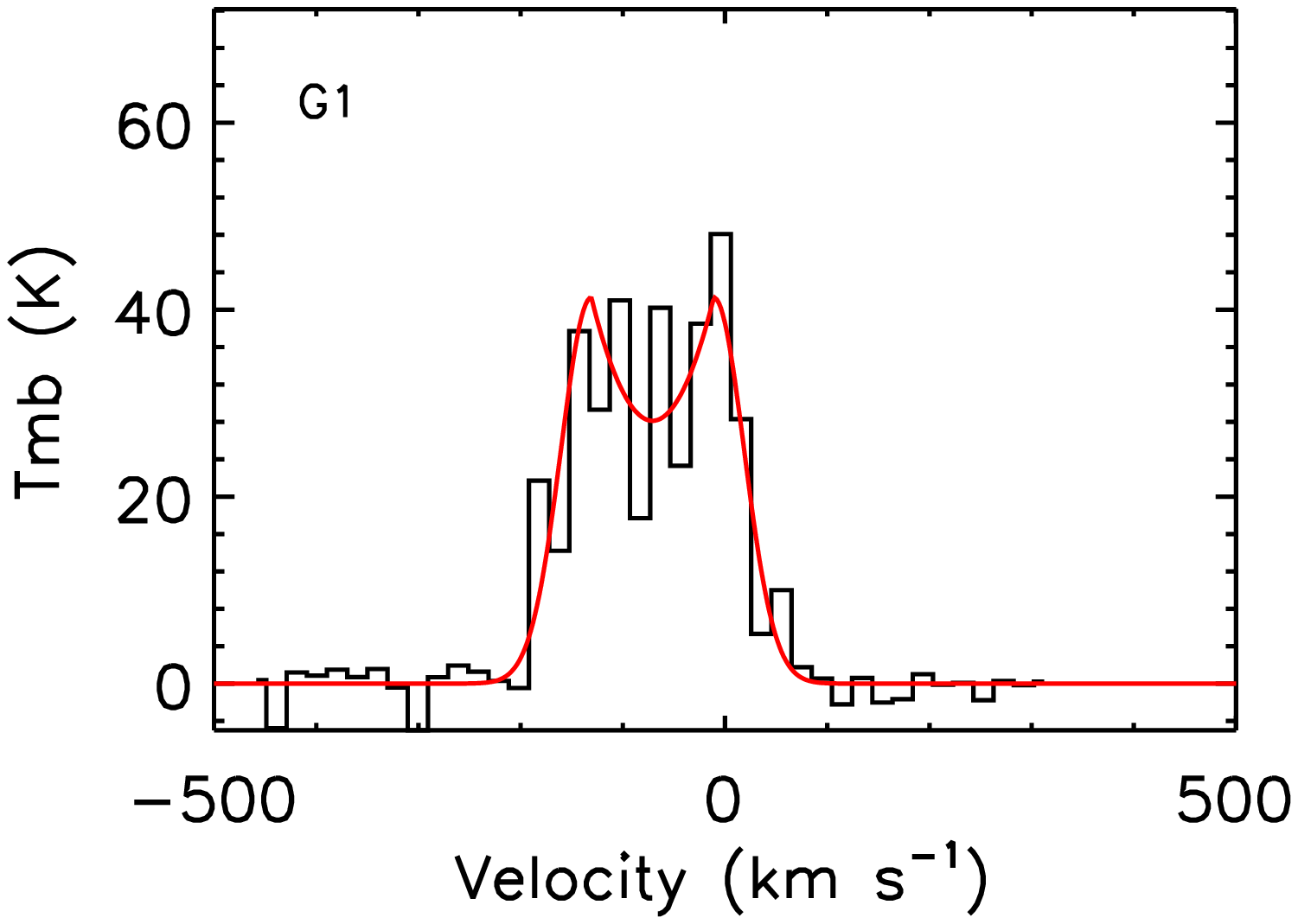}
  \includegraphics[width=5.8cm,clip=]{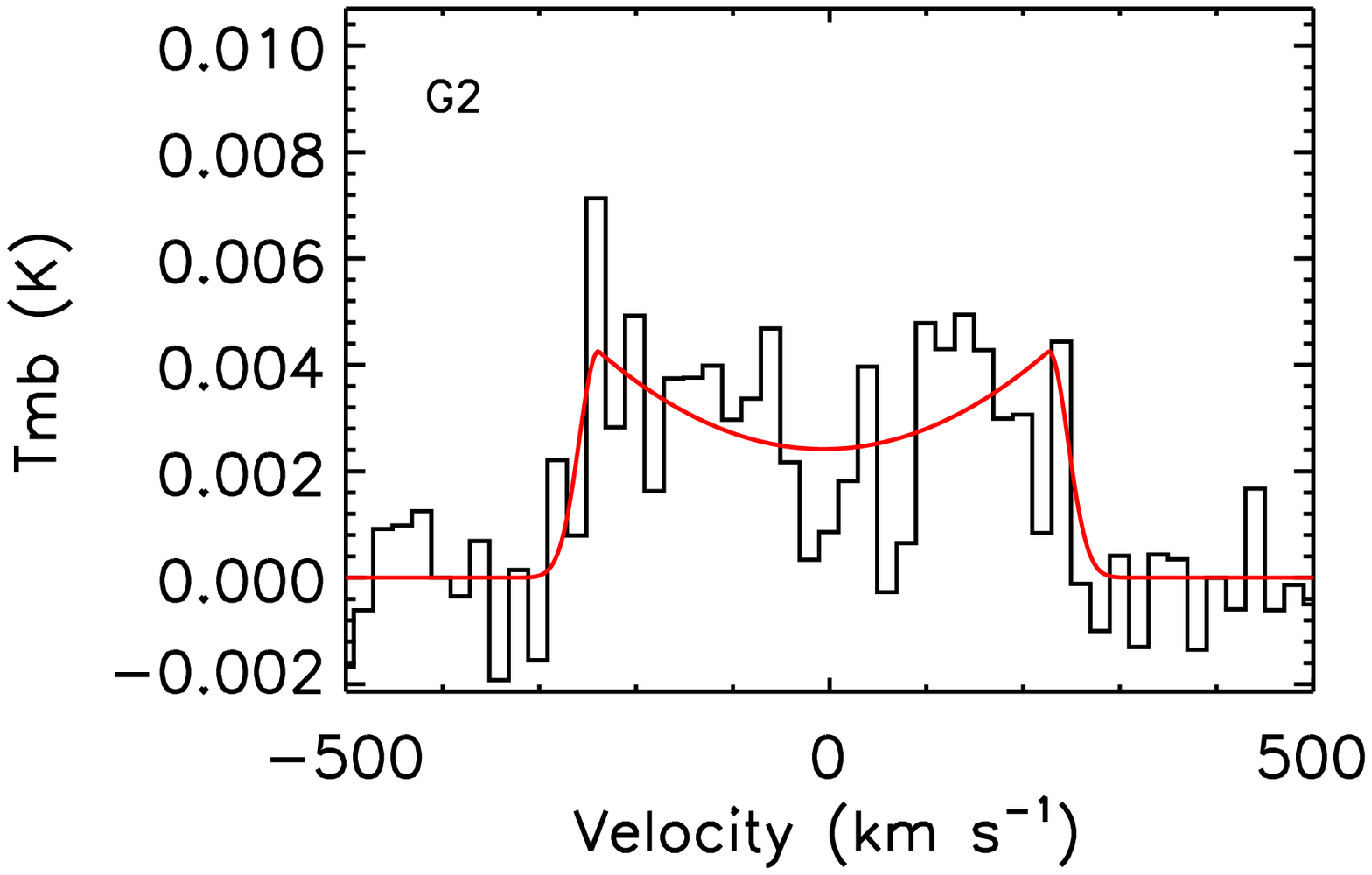}
  \includegraphics[width=5.8cm,clip=]{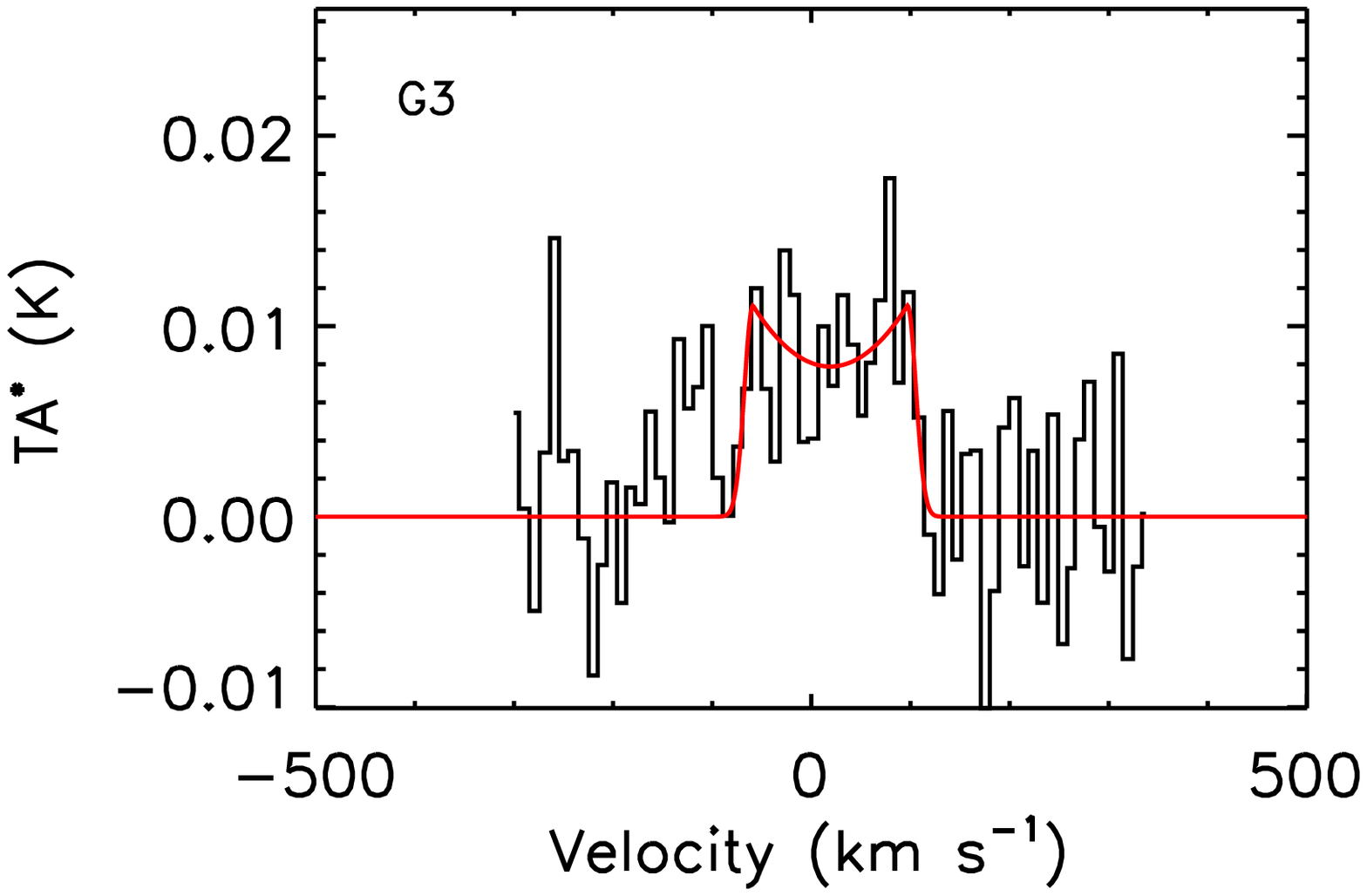}\\
  \vspace{-10pt}
  \hspace{-15pt}
  \includegraphics[width=5.8cm,clip=]{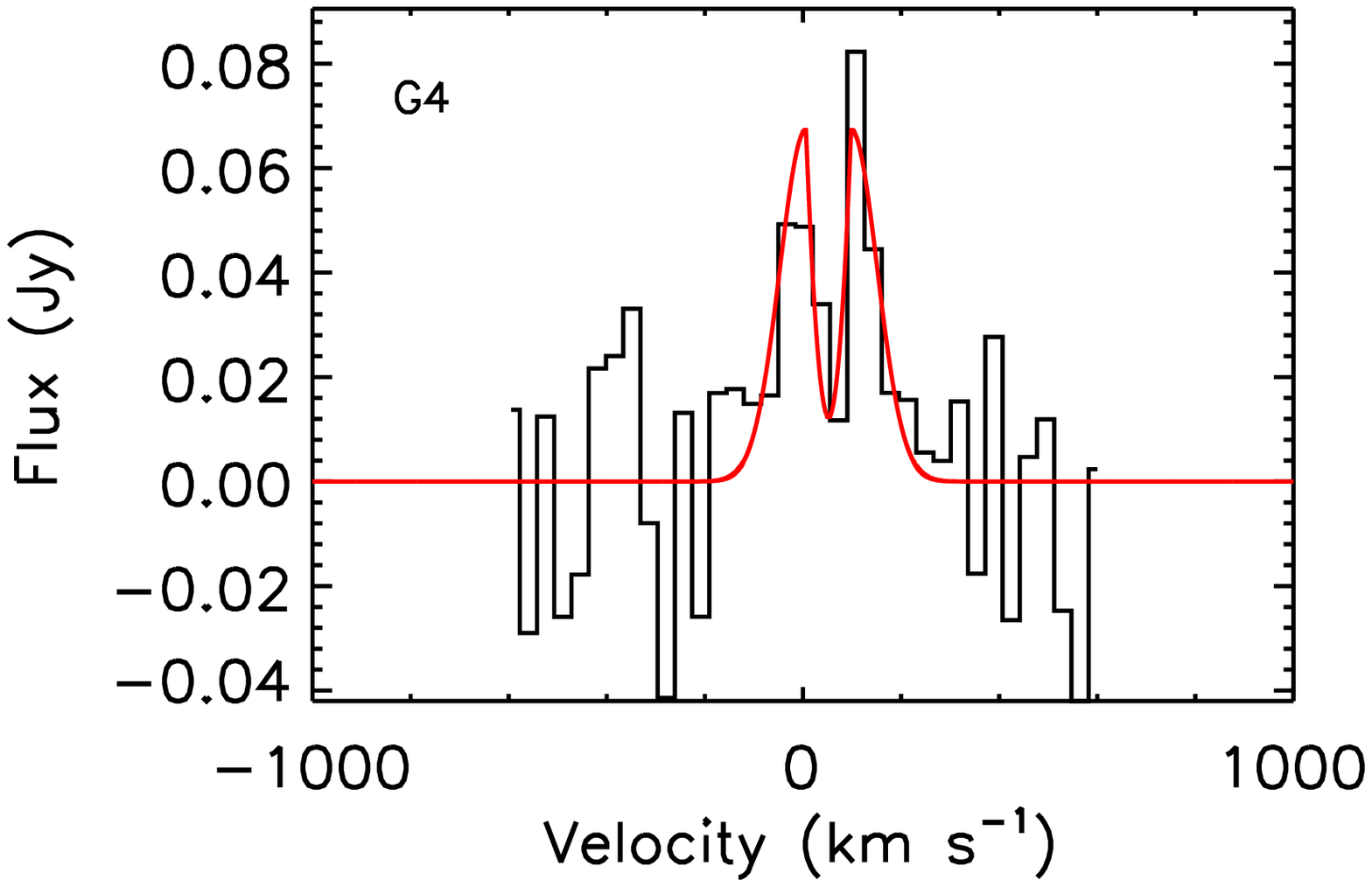}
  \includegraphics[width=5.8cm,clip=]{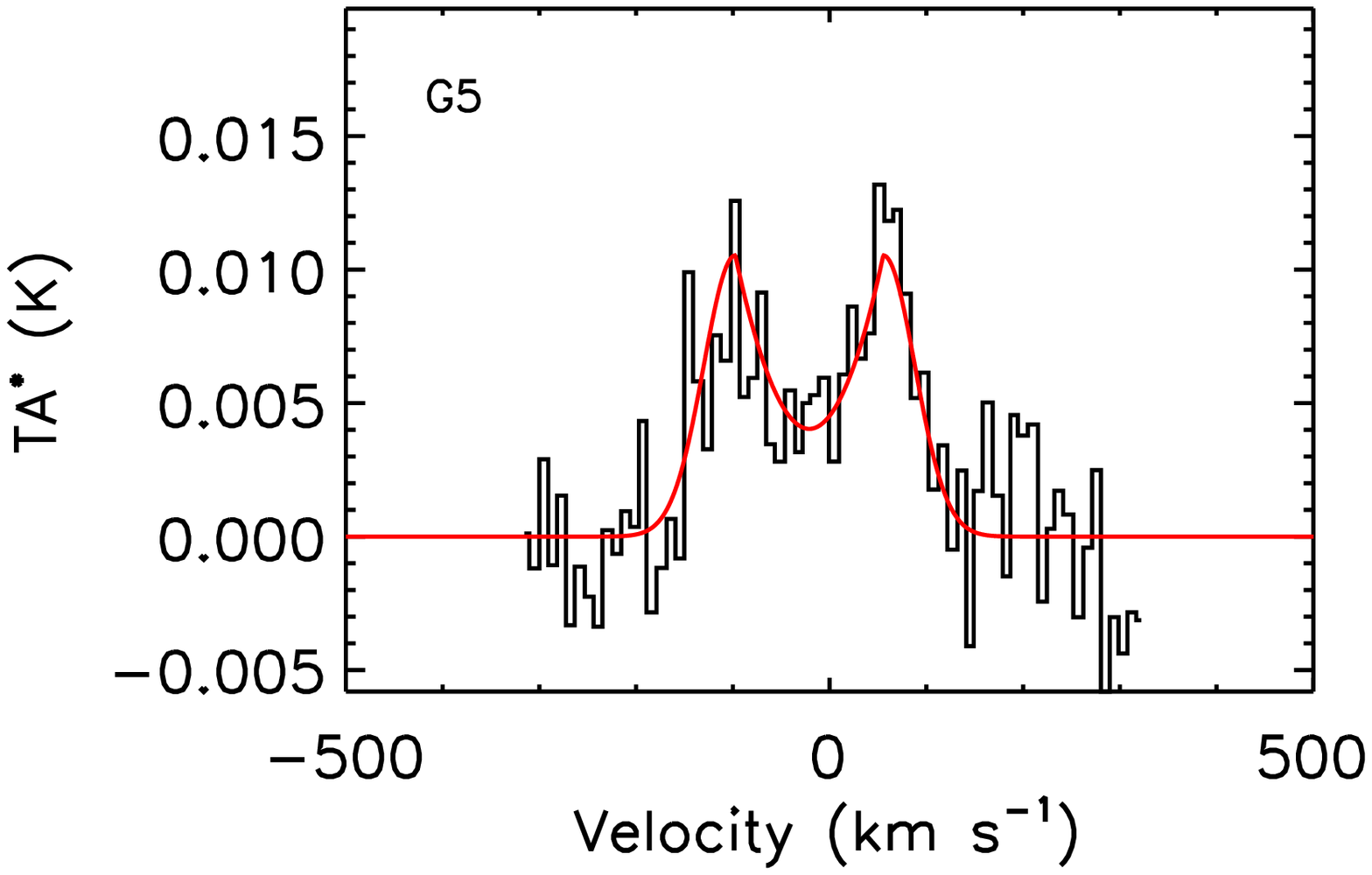}
  \includegraphics[width=5.8cm,clip=]{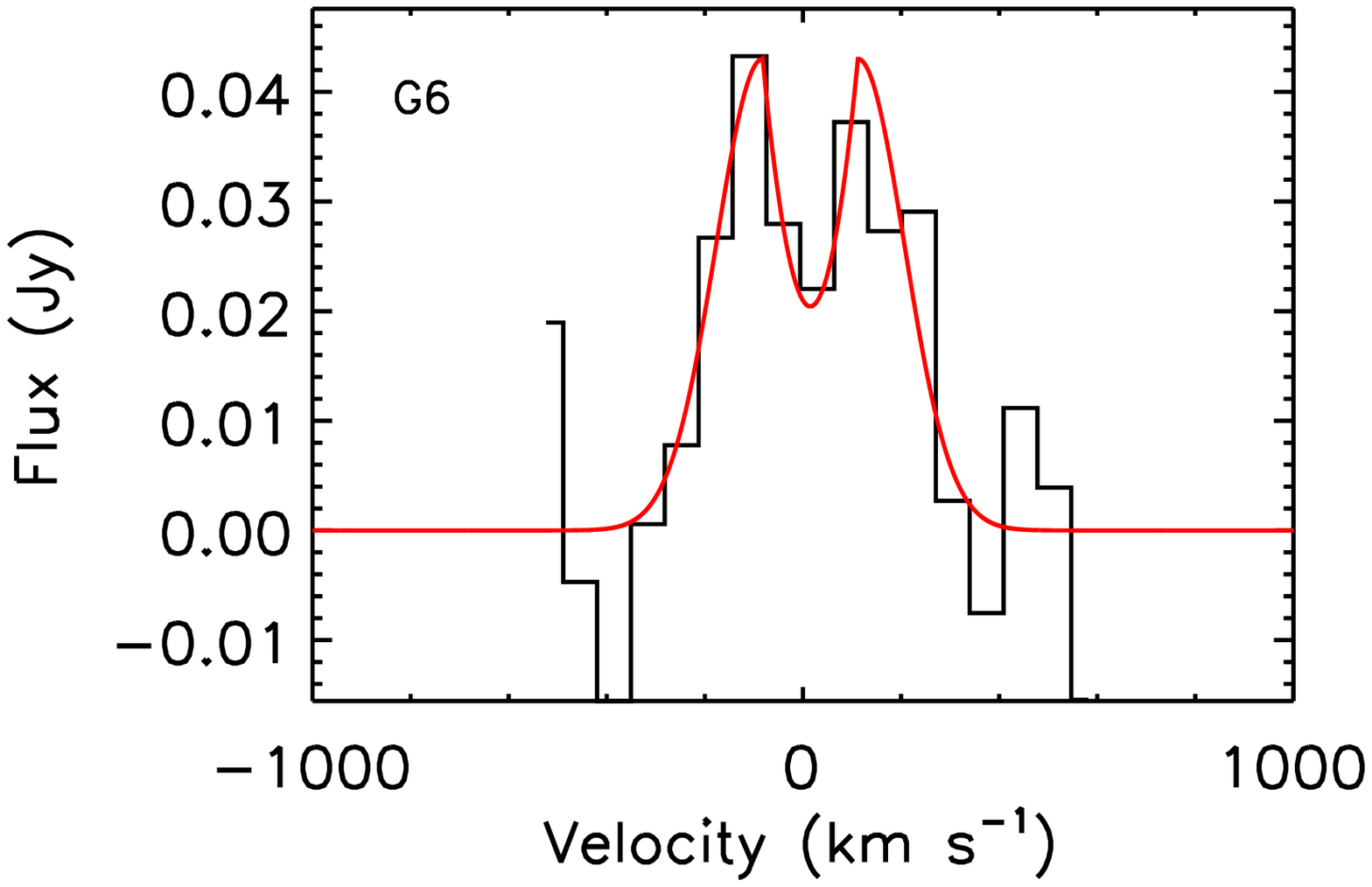}\\
  \vspace{-10pt}
  \hspace{-15pt}
  \includegraphics[width=5.8cm,clip=]{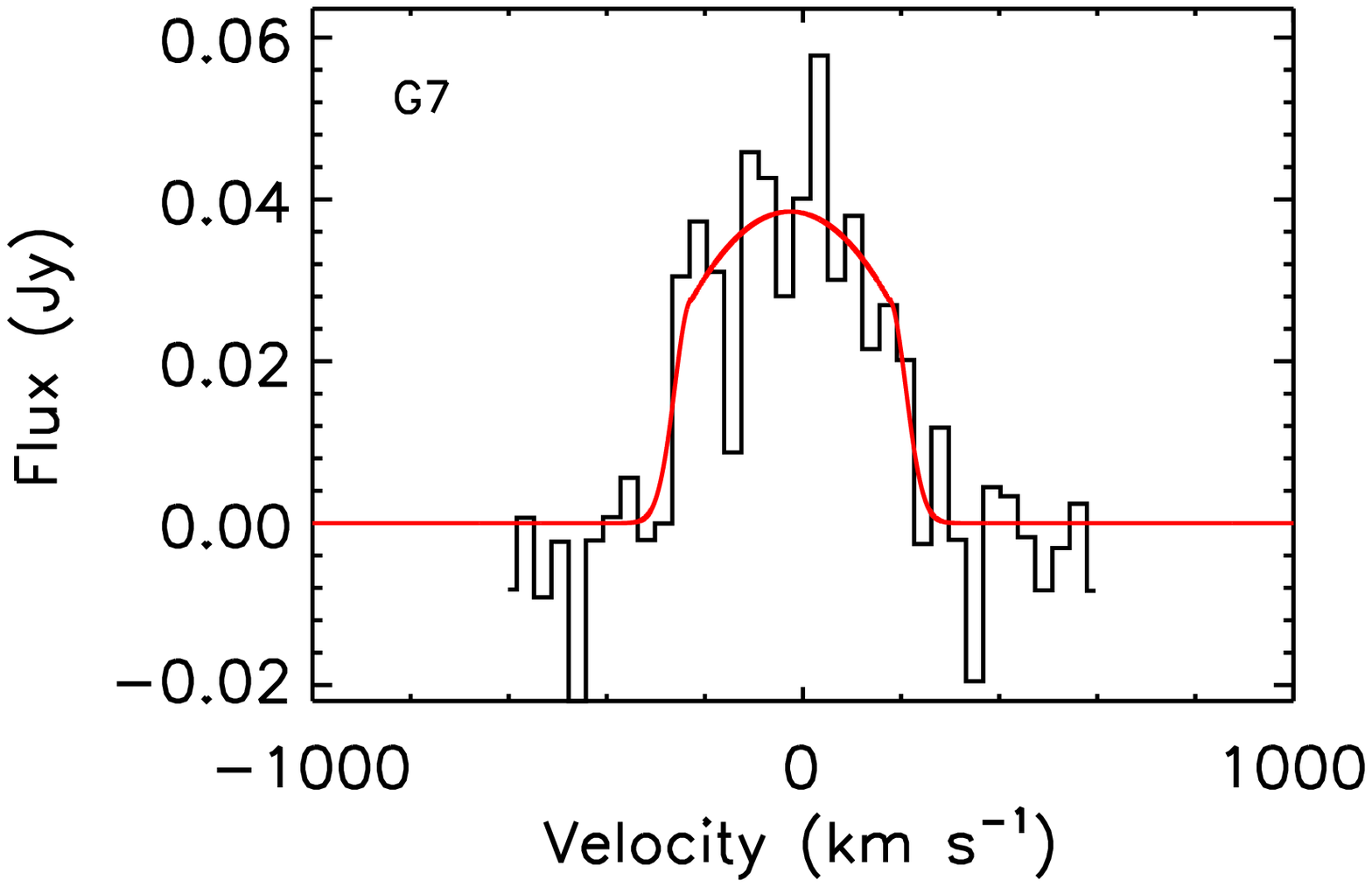}
  \includegraphics[width=5.8cm,clip=]{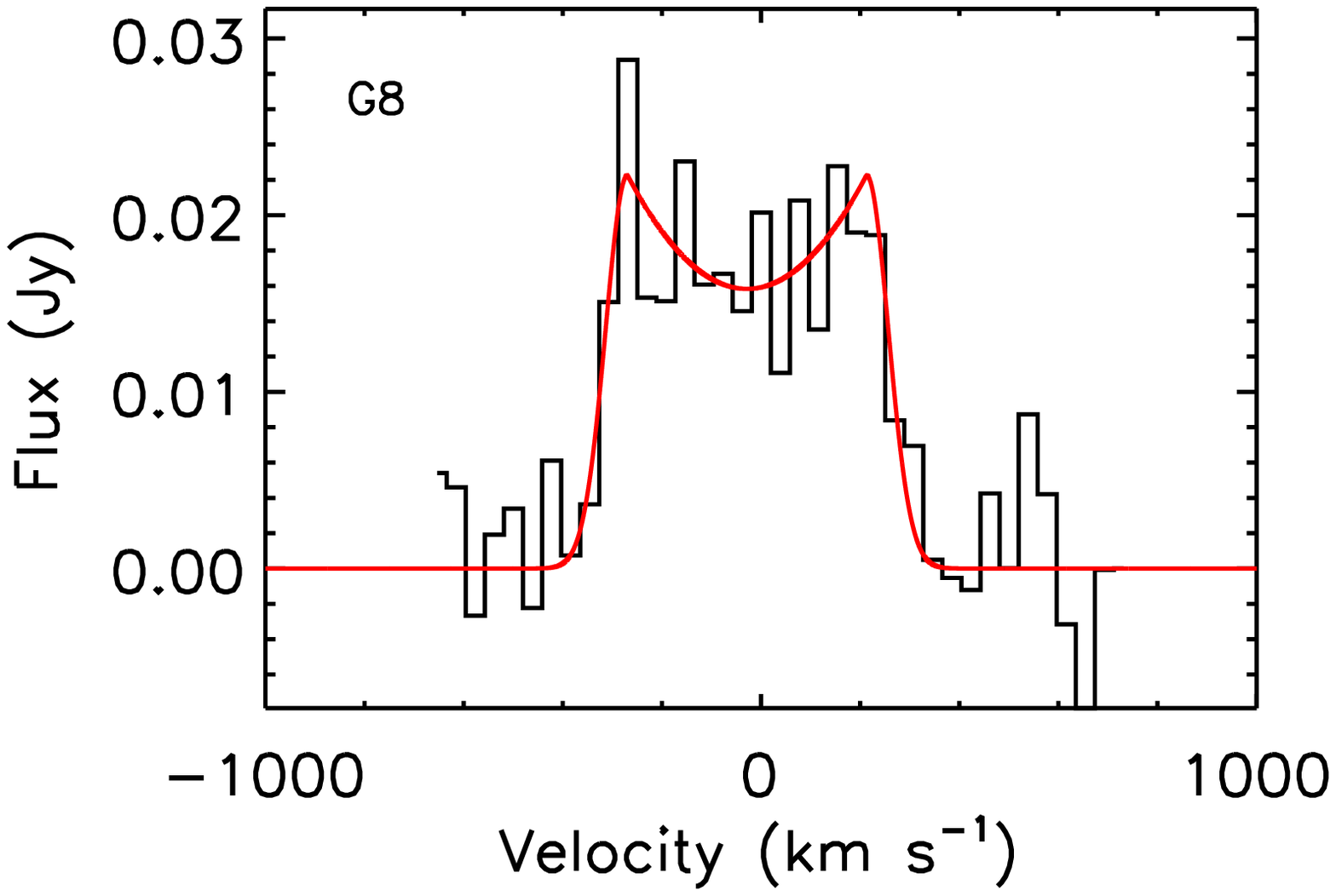}
  \includegraphics[width=5.8cm,clip=]{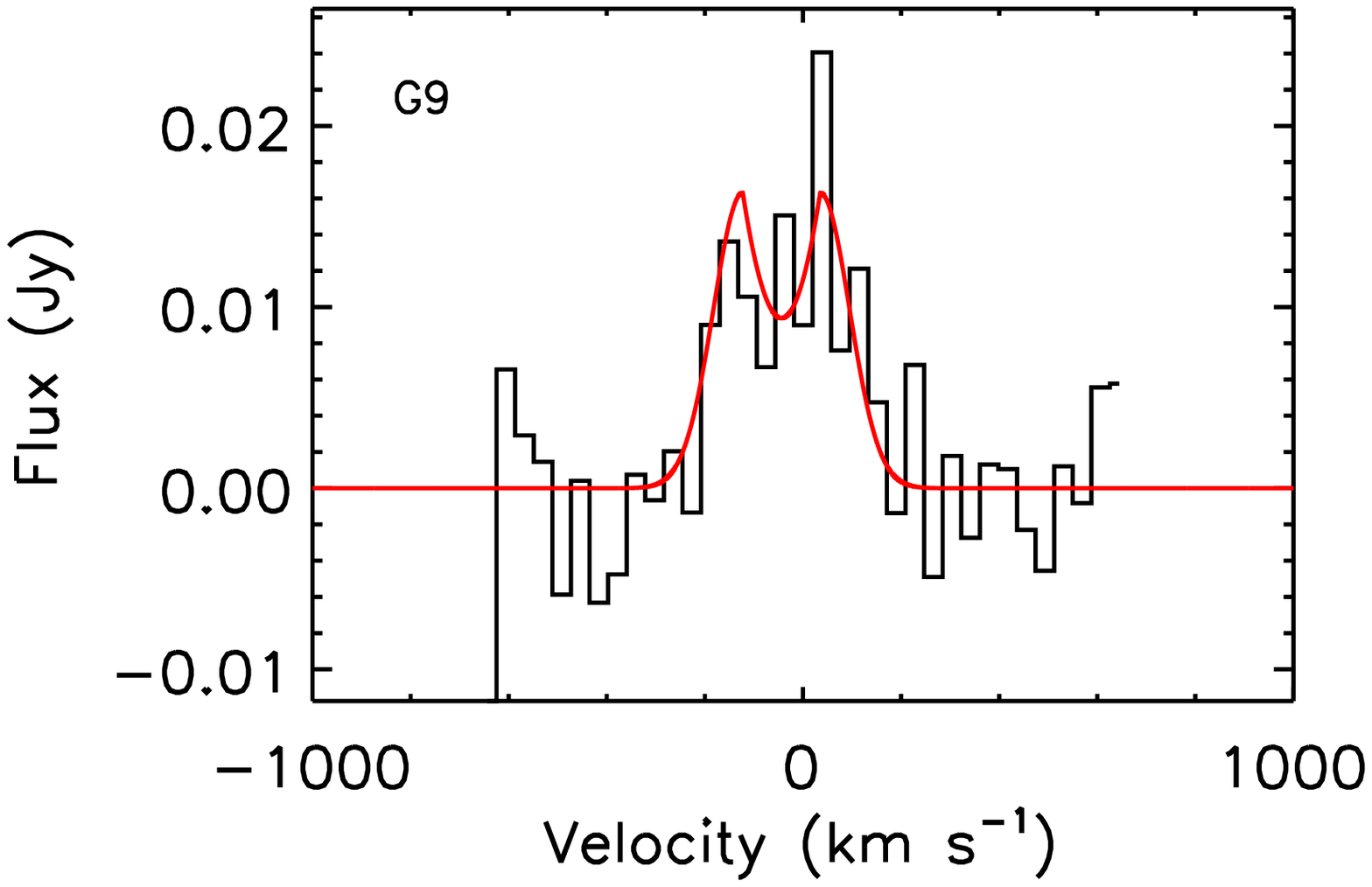}\\
  \vspace{-10pt}
  \hspace{-15pt}
  \includegraphics[width=5.8cm,clip=]{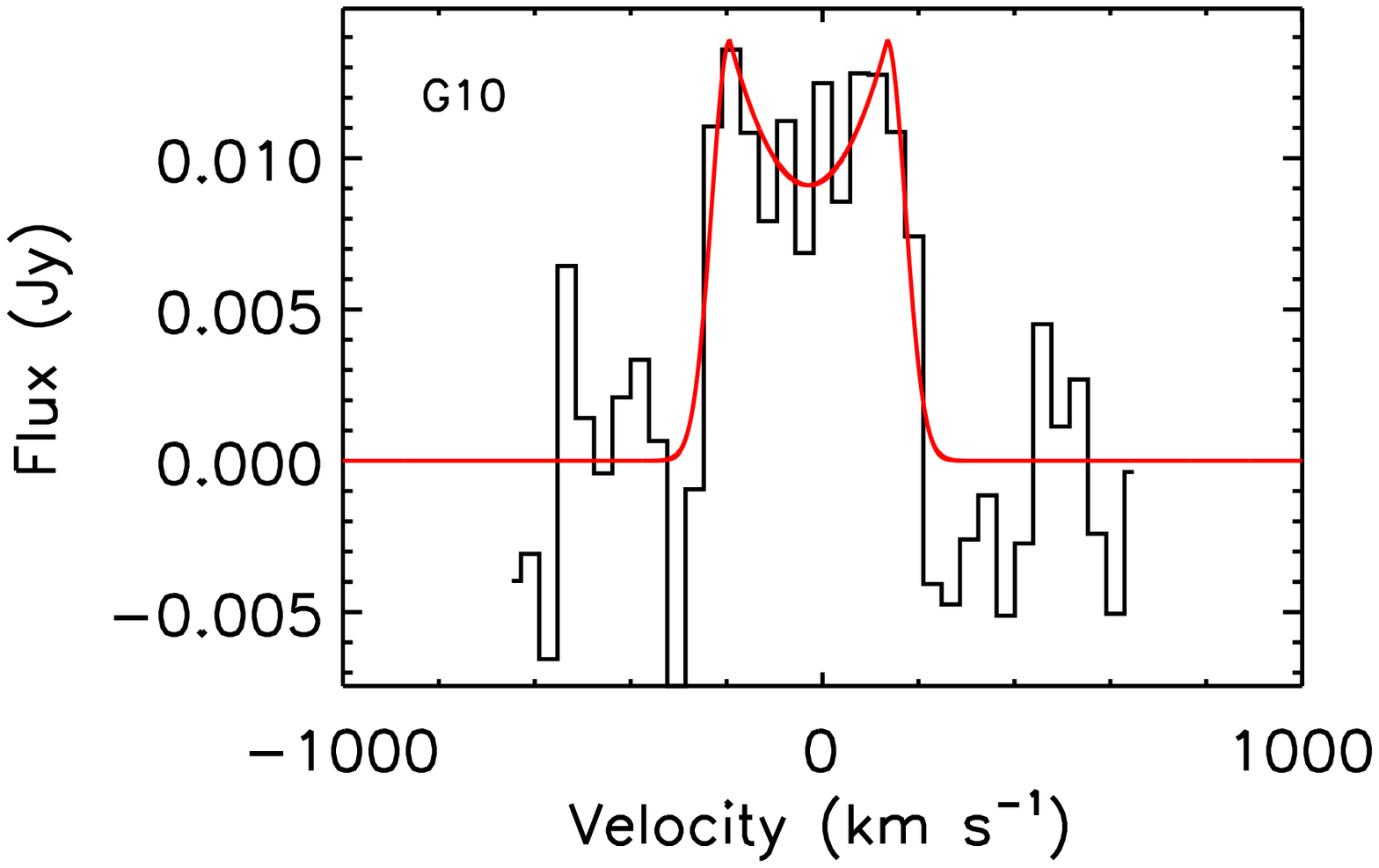}
  \includegraphics[width=5.8cm,clip=]{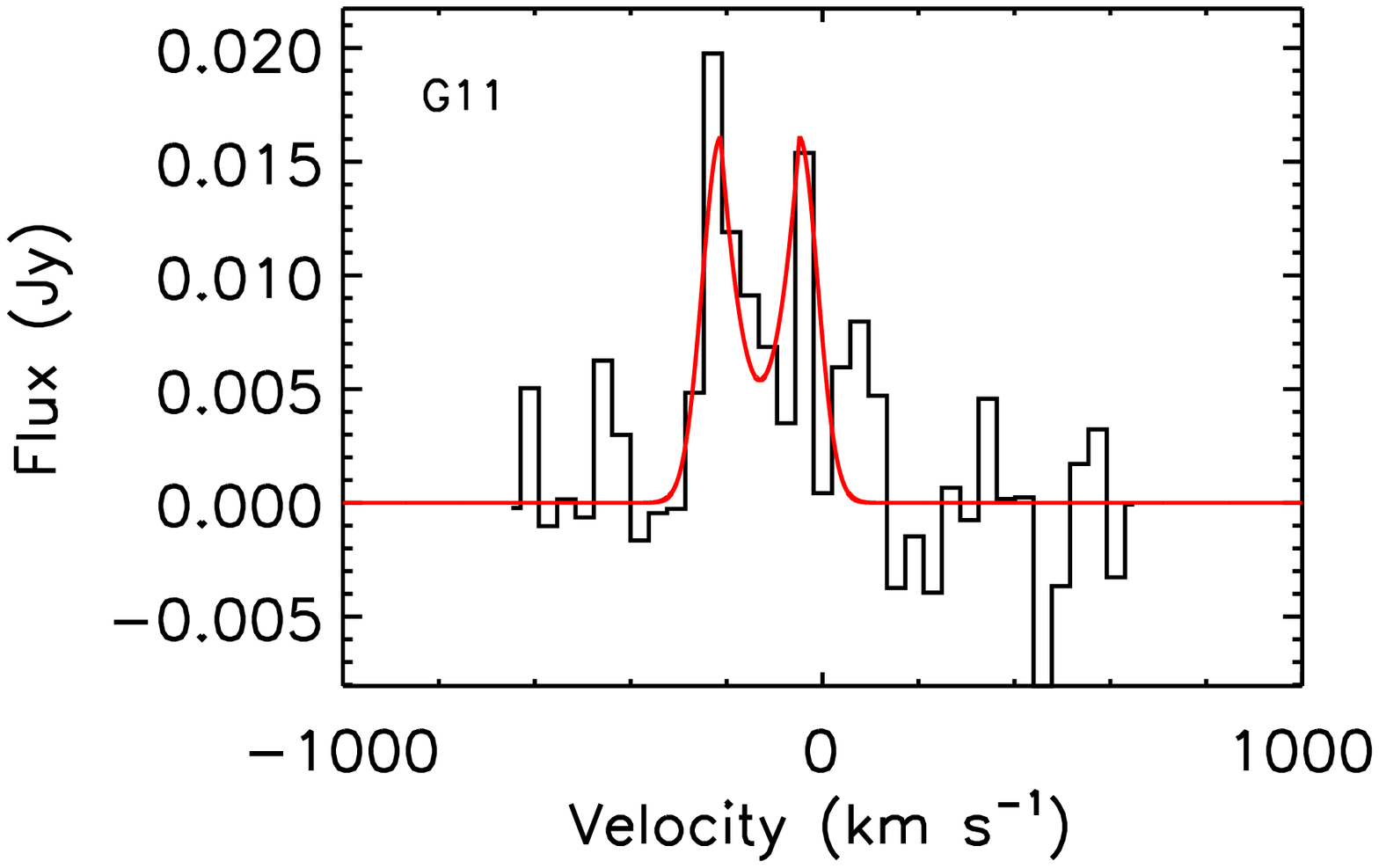}
  \includegraphics[width=5.8cm,clip=]{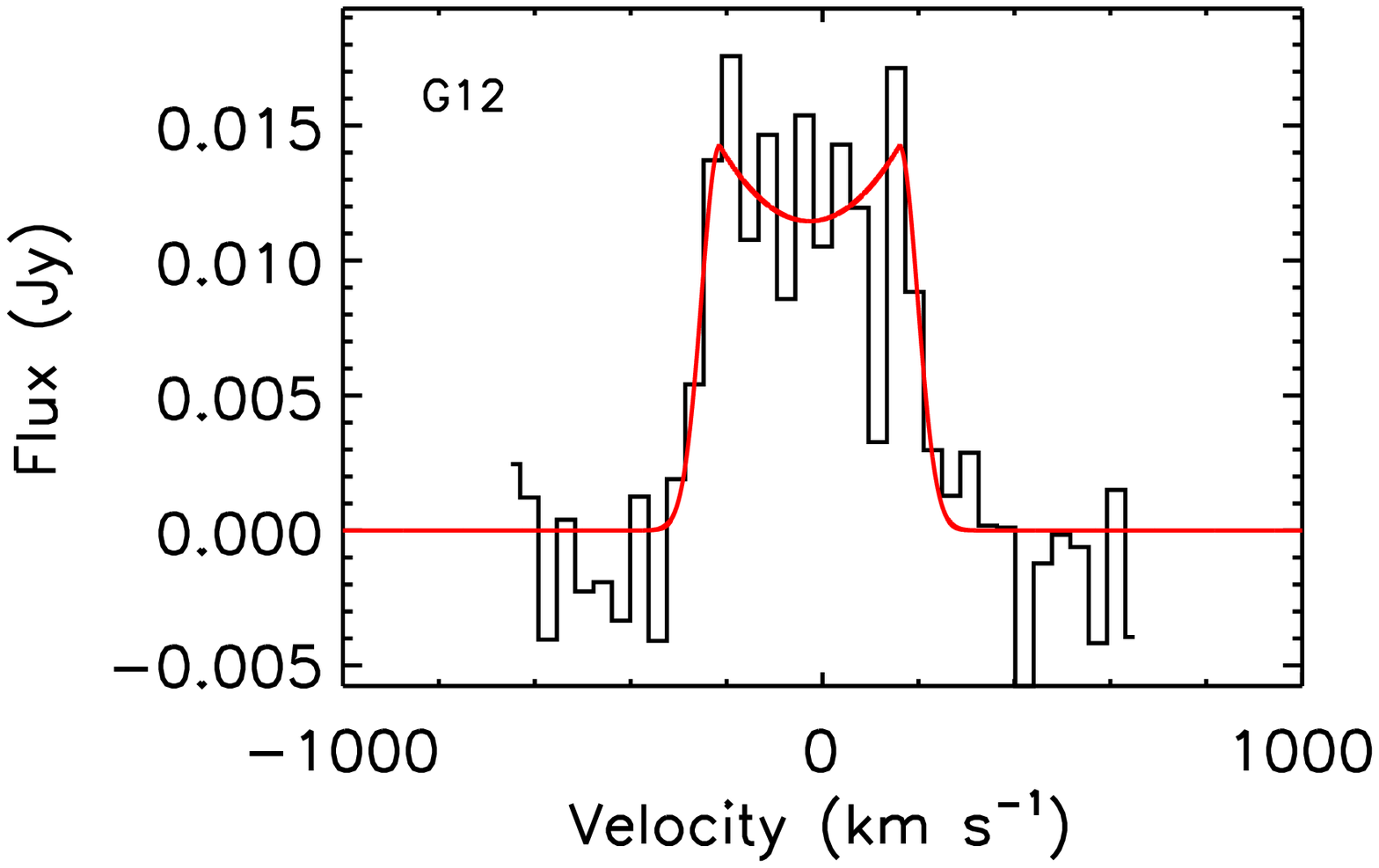}\\
  \vspace{-10pt}
  \hspace{-15pt}
  \includegraphics[width=5.8cm,clip=]{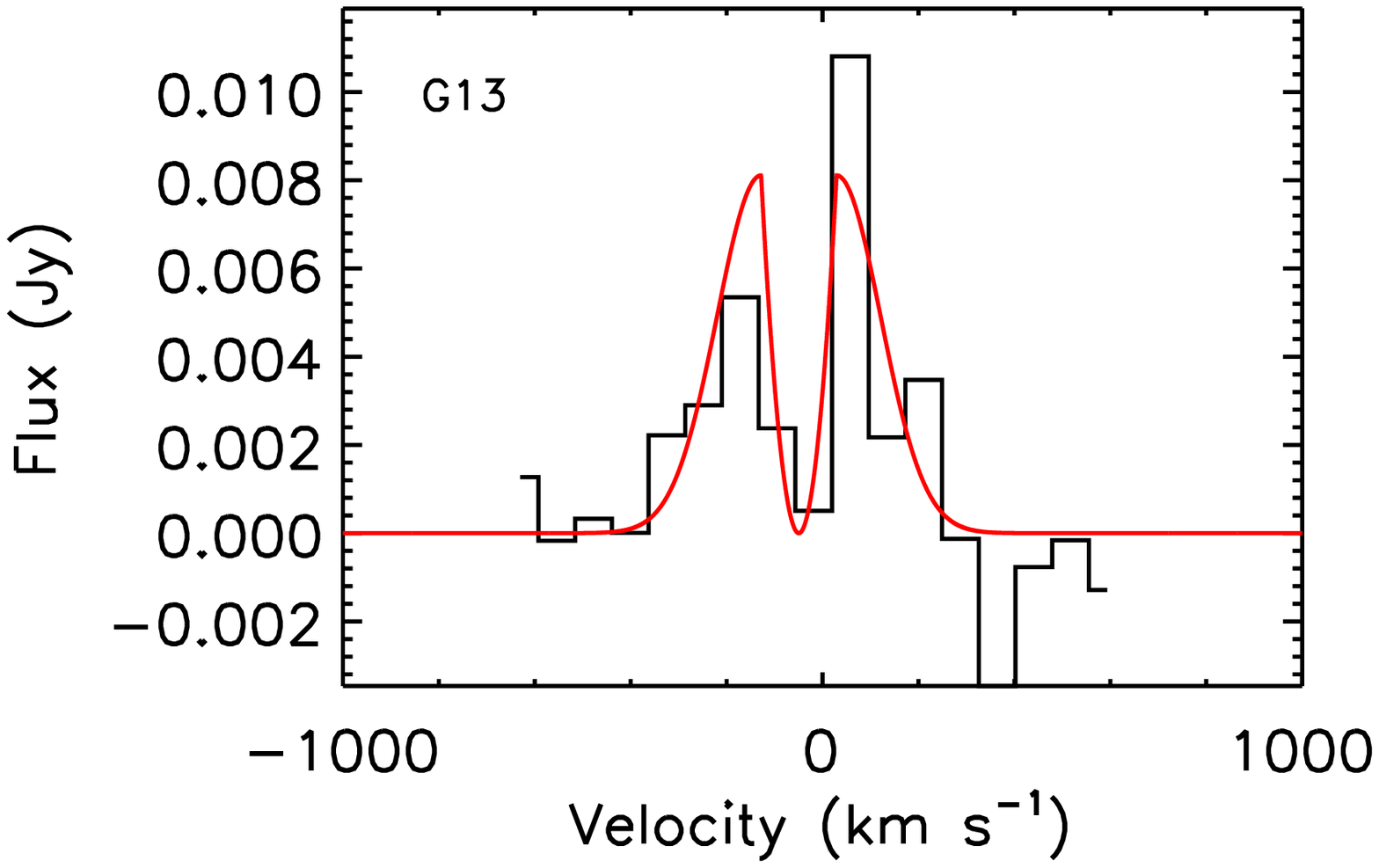}
  \includegraphics[width=5.8cm,clip=]{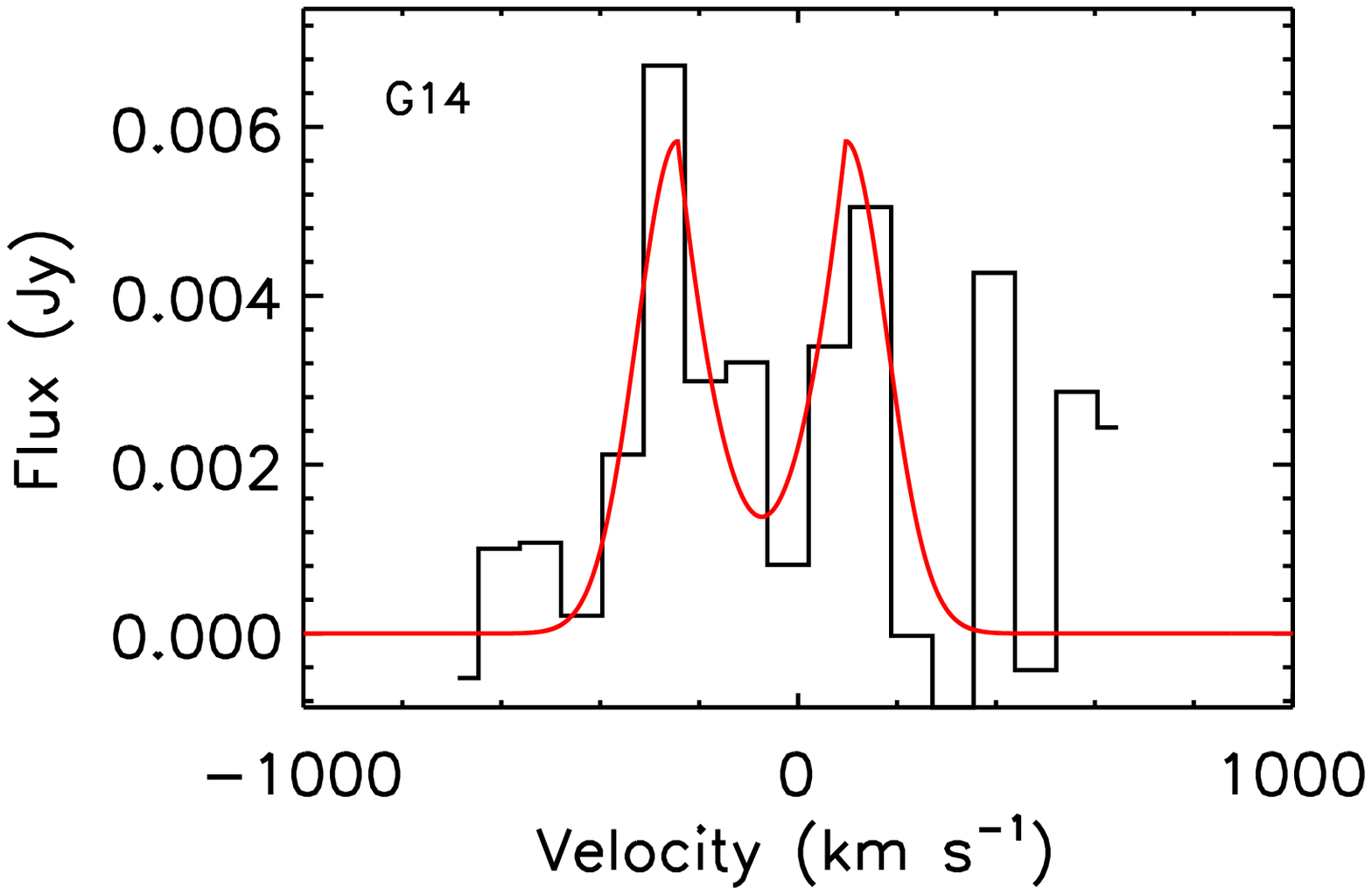}
  \includegraphics[width=5.8cm,clip=]{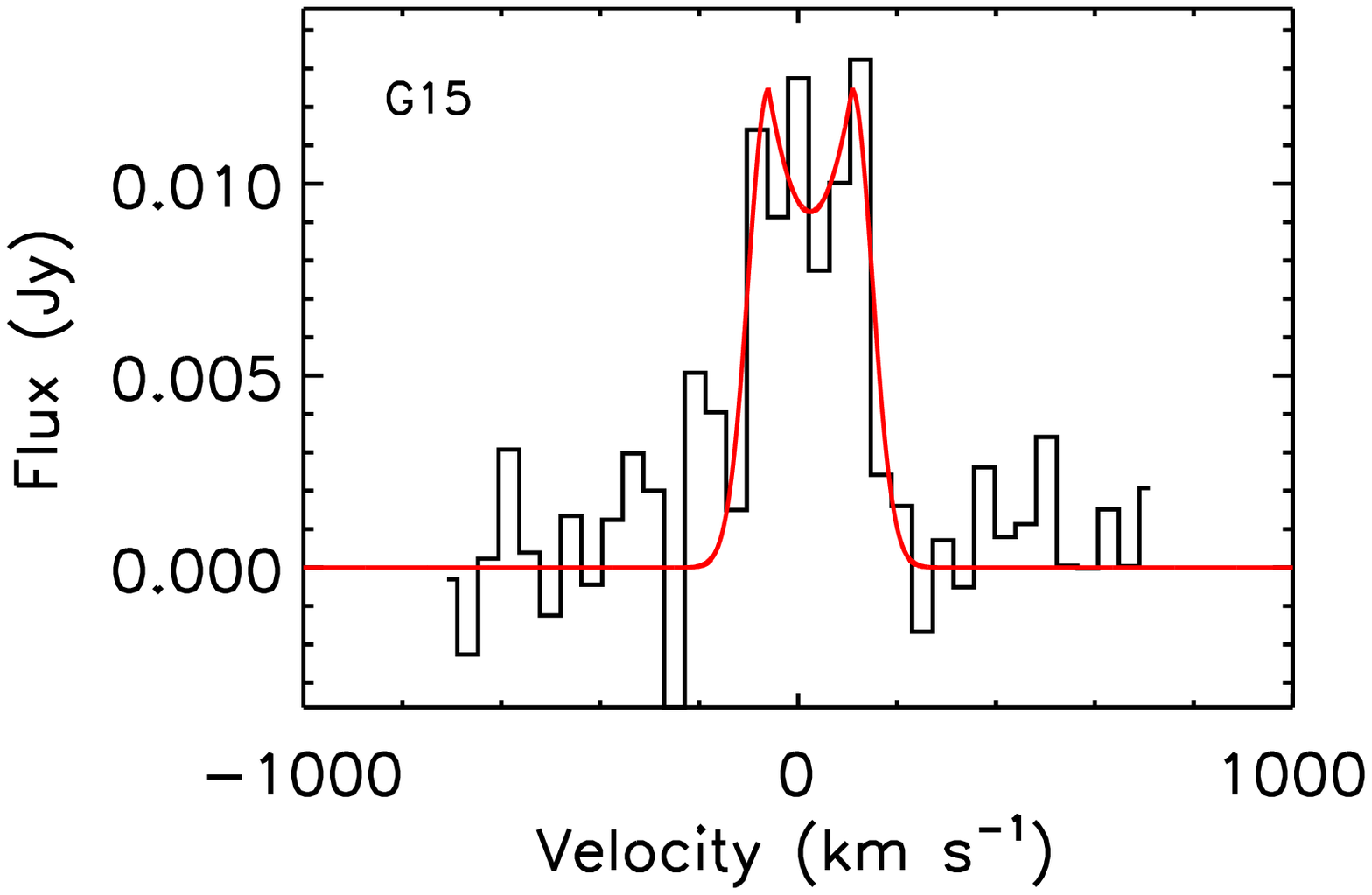}\\
  \caption{Integrated CO profiles of our initial sub-sample galaxies,
    taken from \citet{ba13} and additional literature sources (see
    Section~\ref{subsec:literature}). For each plot, the red line
    shows the best parametric fit to the spectrum (see
    Section~\ref{subsec:W50}). The name of the galaxy as listed in
    Tables~\ref{tab:general} and \ref{tab:TFR} is indicated in the
    top-left corner of each plot. The flux units are as in the
    original publications, but this has no bearing on the derived line
    widths.}
  \label{fig:egnog}
\end{figure*}
\addtocounter{figure}{-1}
\begin{figure*}
  \hspace{-15pt}
  \includegraphics[width=5.7cm,clip=]{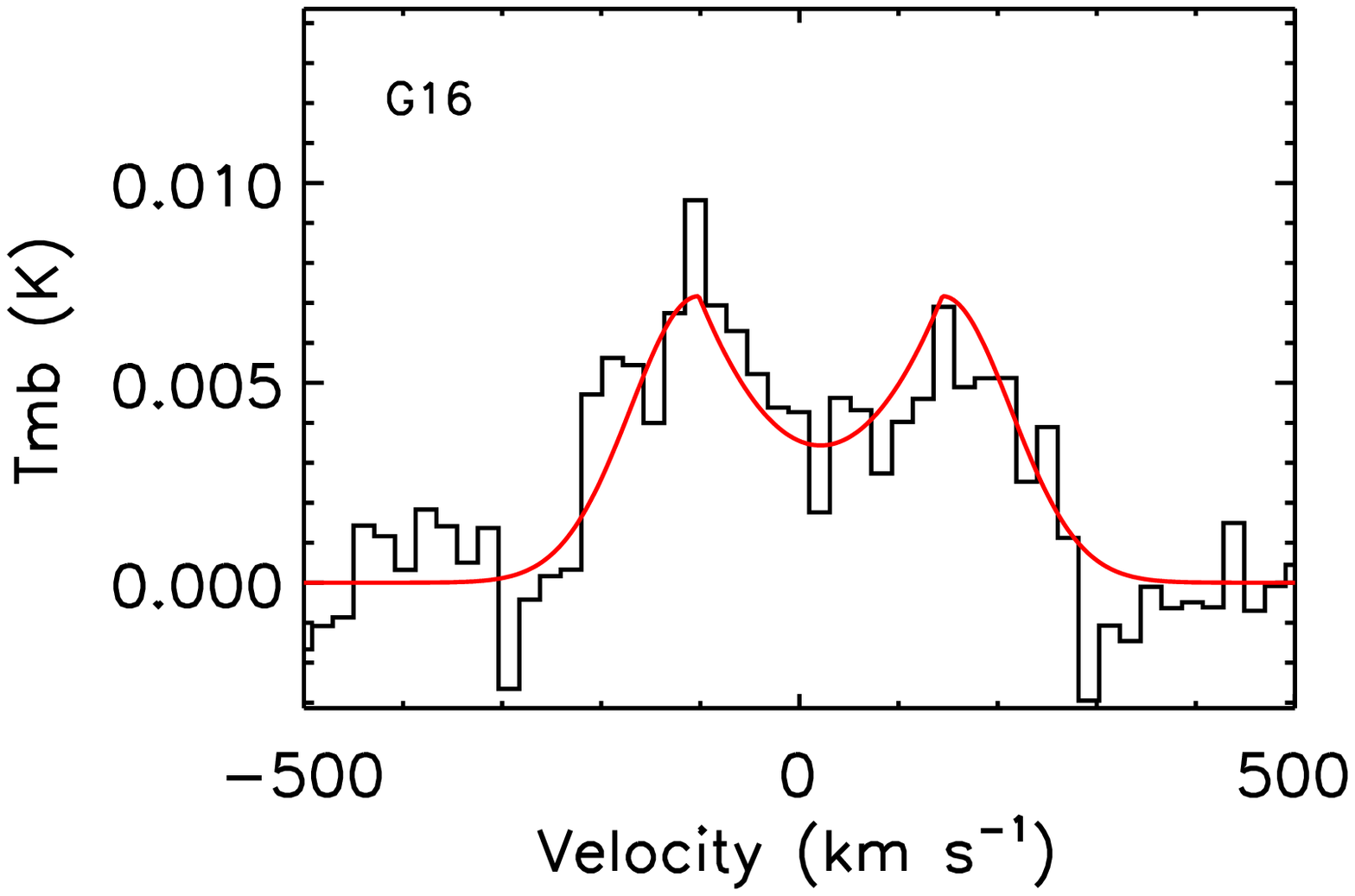}
  \includegraphics[width=5.7cm,clip=]{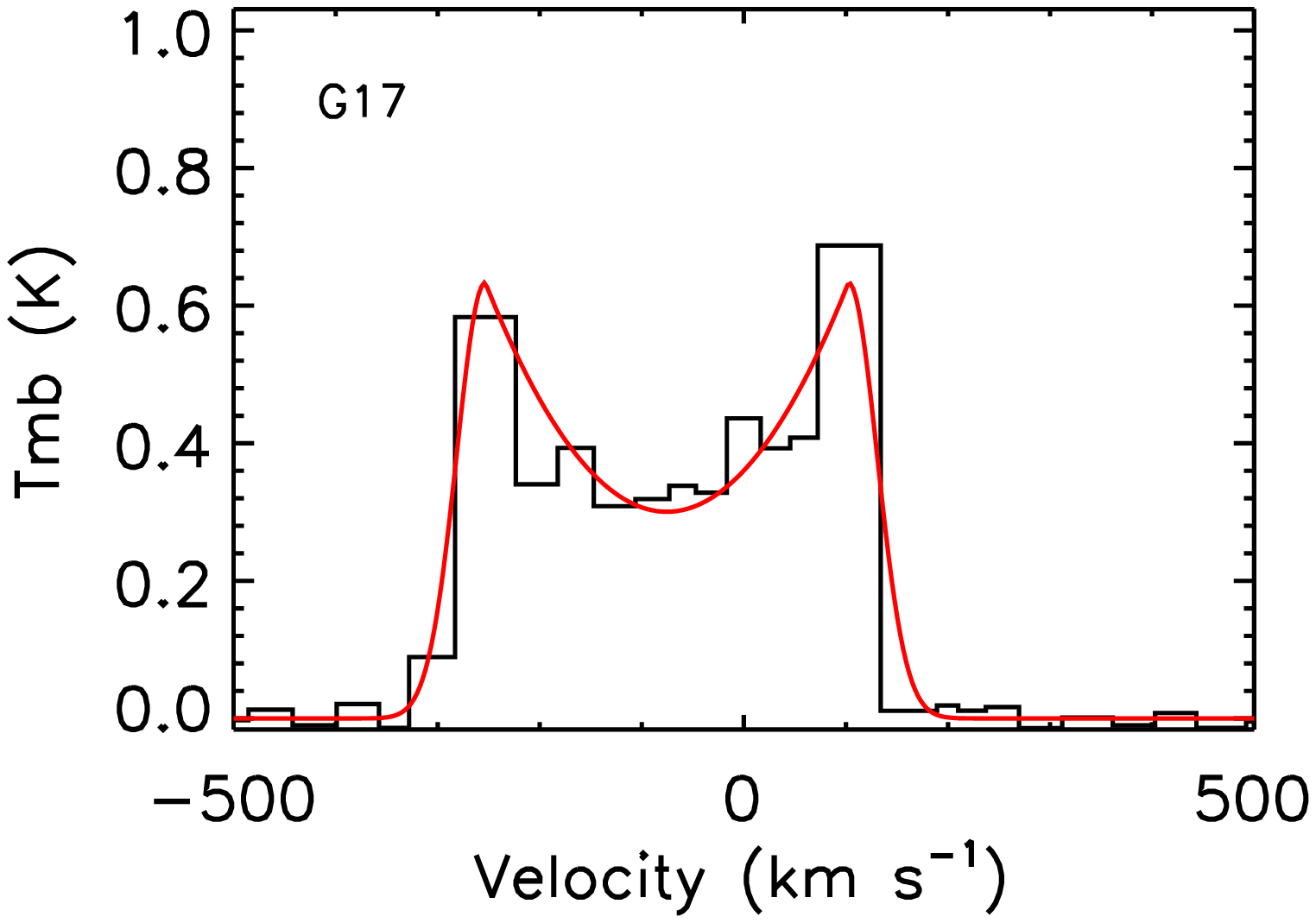}
  \includegraphics[width=5.7cm,clip=]{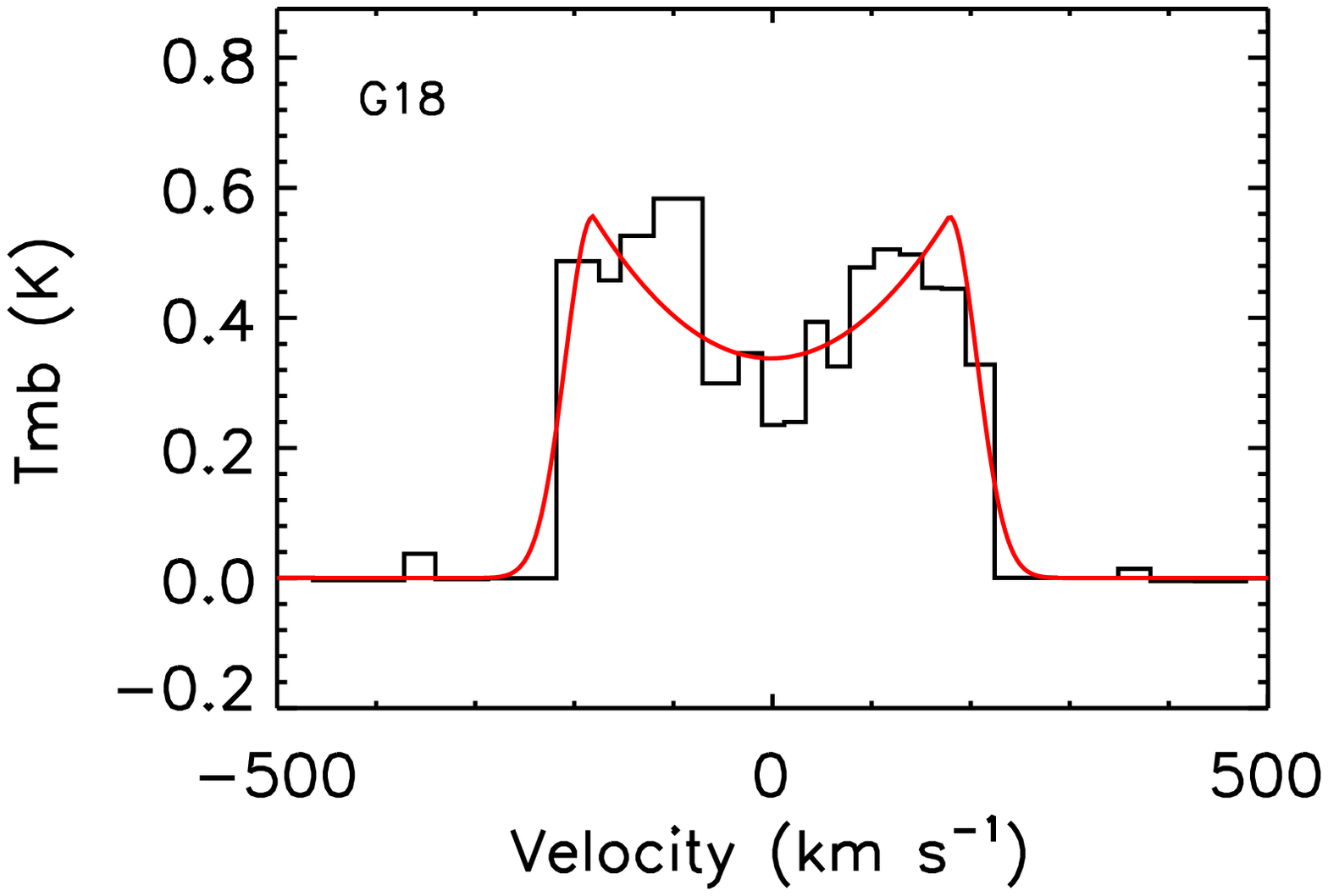}\\
  \vspace{-10pt}
  \hspace{-17pt}
  \includegraphics[width=5.7cm,clip=]{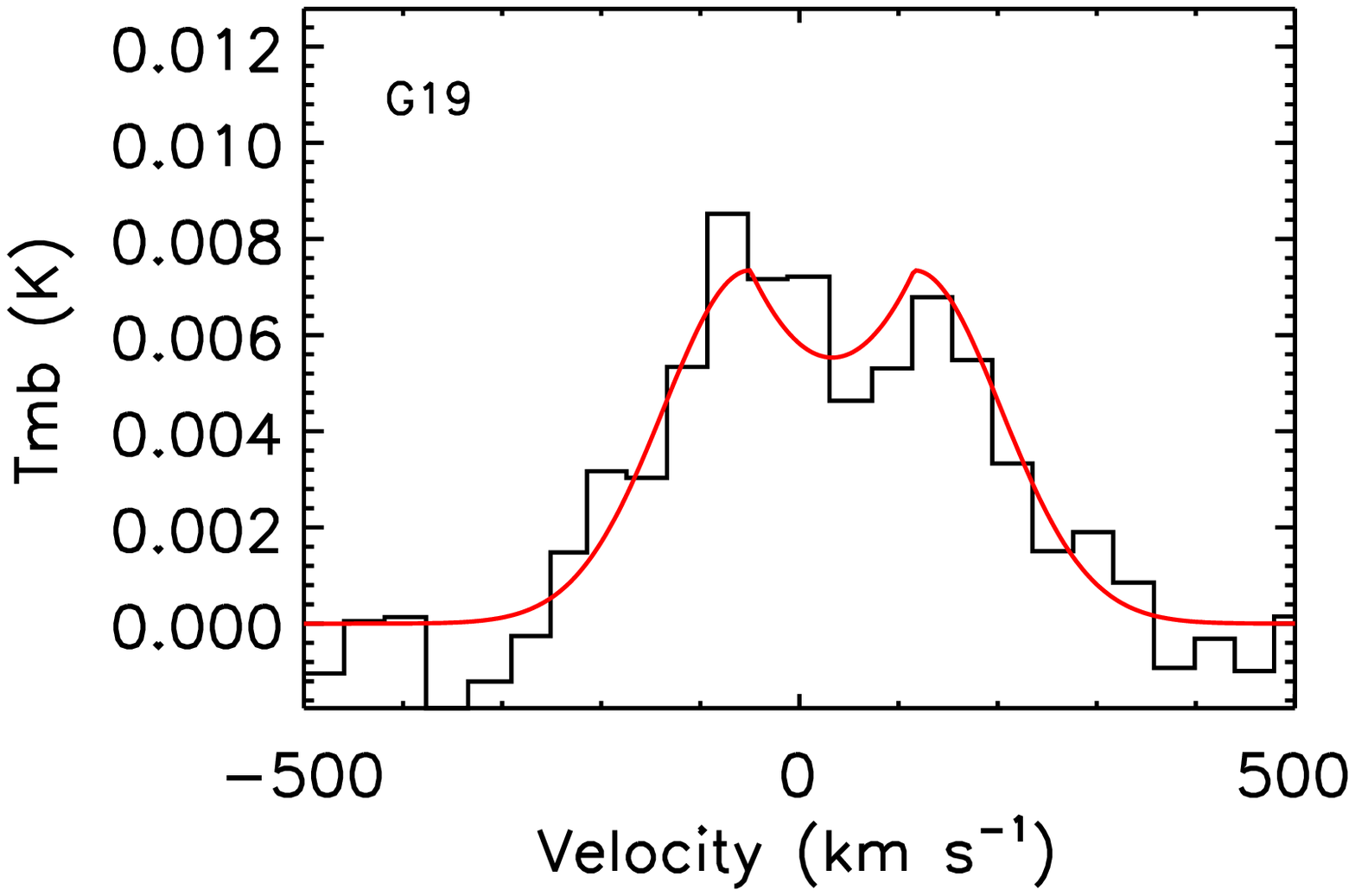}
  \includegraphics[width=5.7cm,clip=]{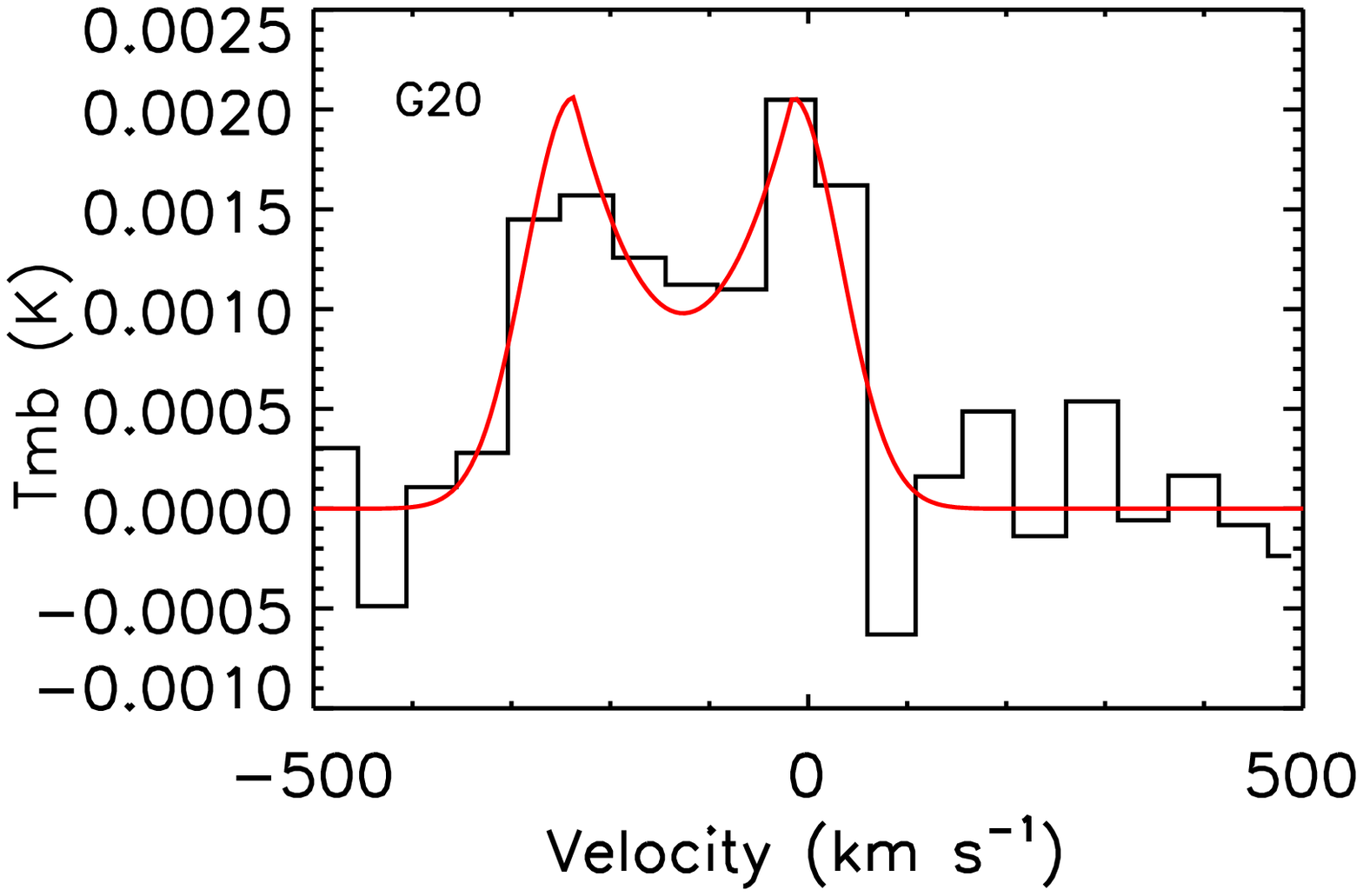}
  \includegraphics[width=5.7cm,clip=]{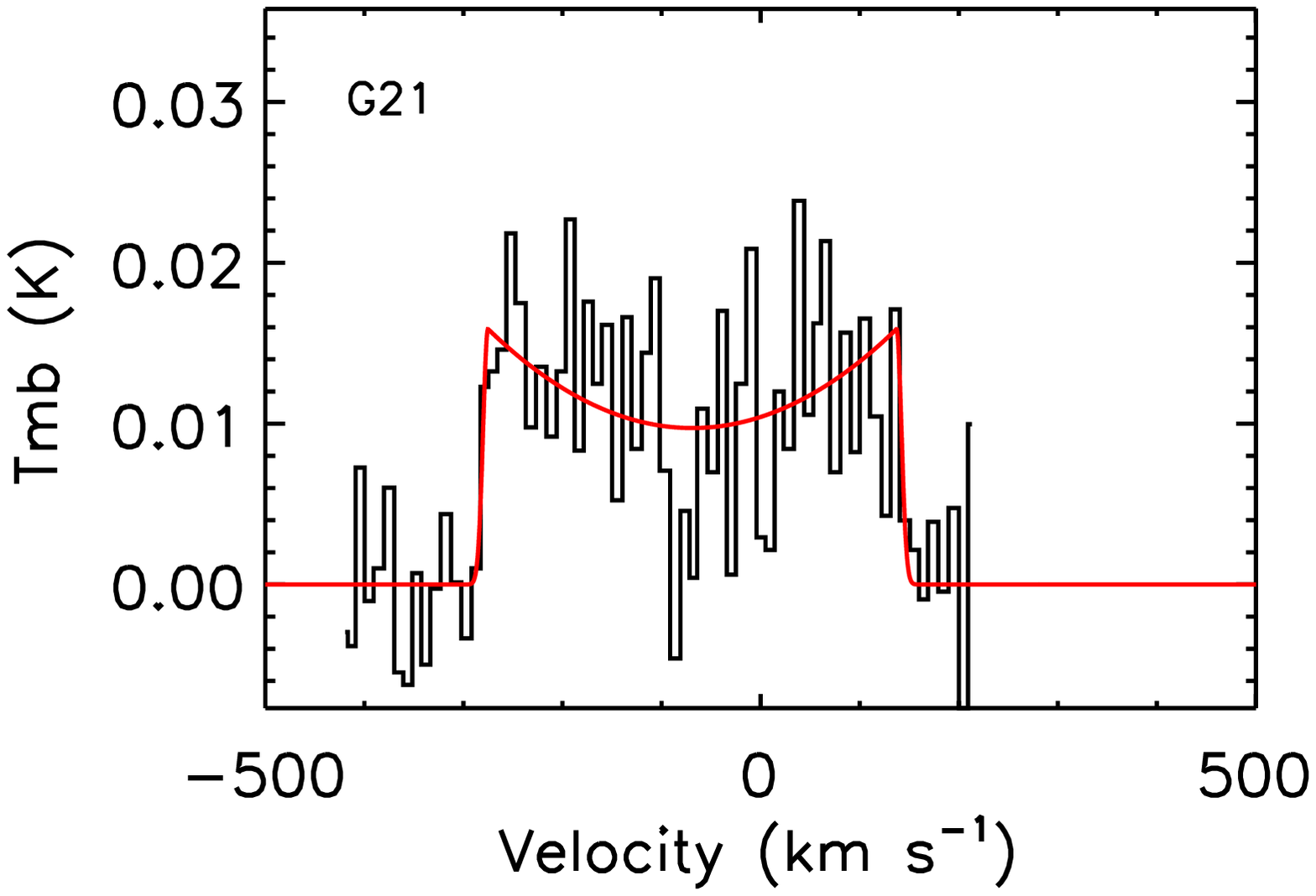}\\
  \vspace{-10pt}
  \hspace{-17pt}
  \includegraphics[width=5.7cm,clip=]{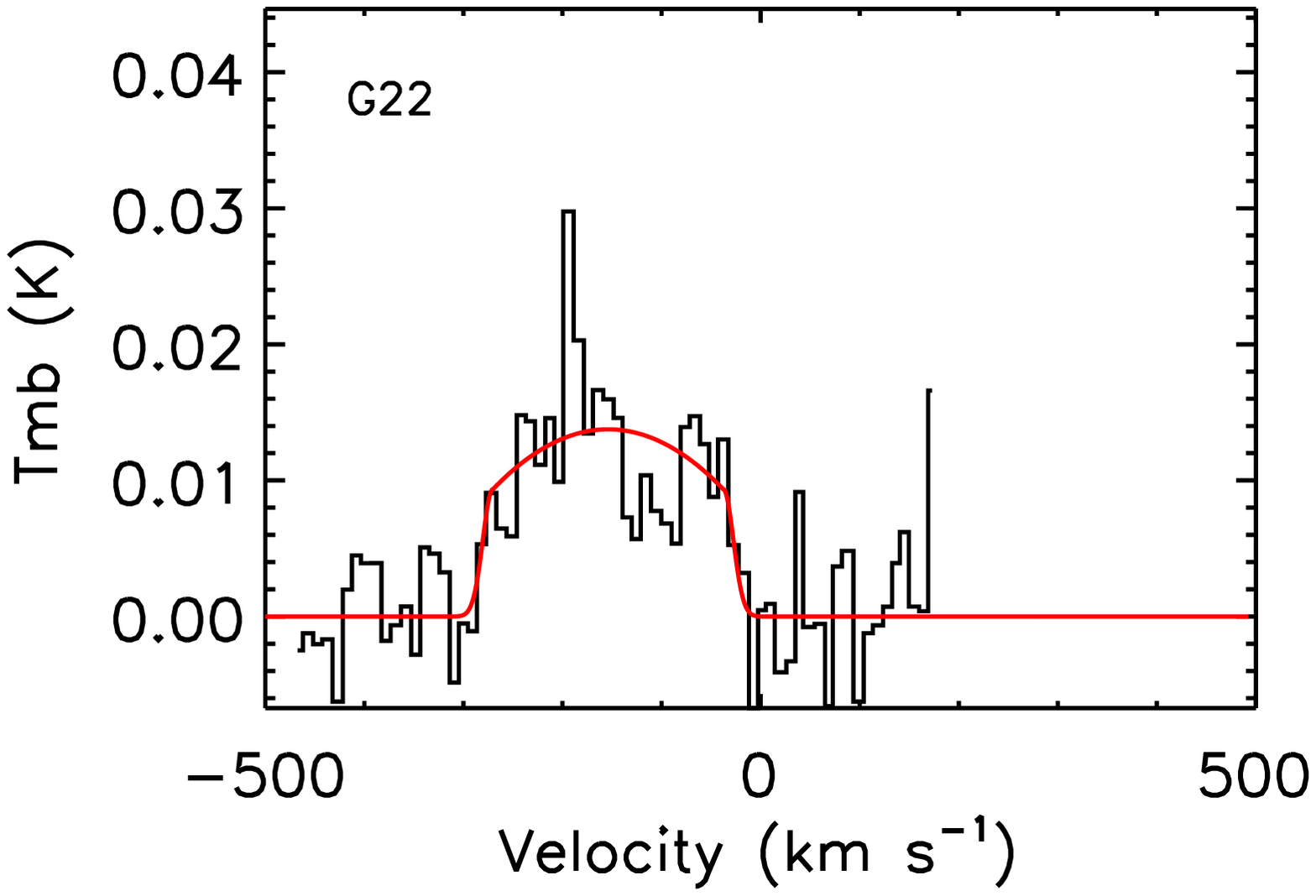}
  \includegraphics[width=5.7cm,clip=]{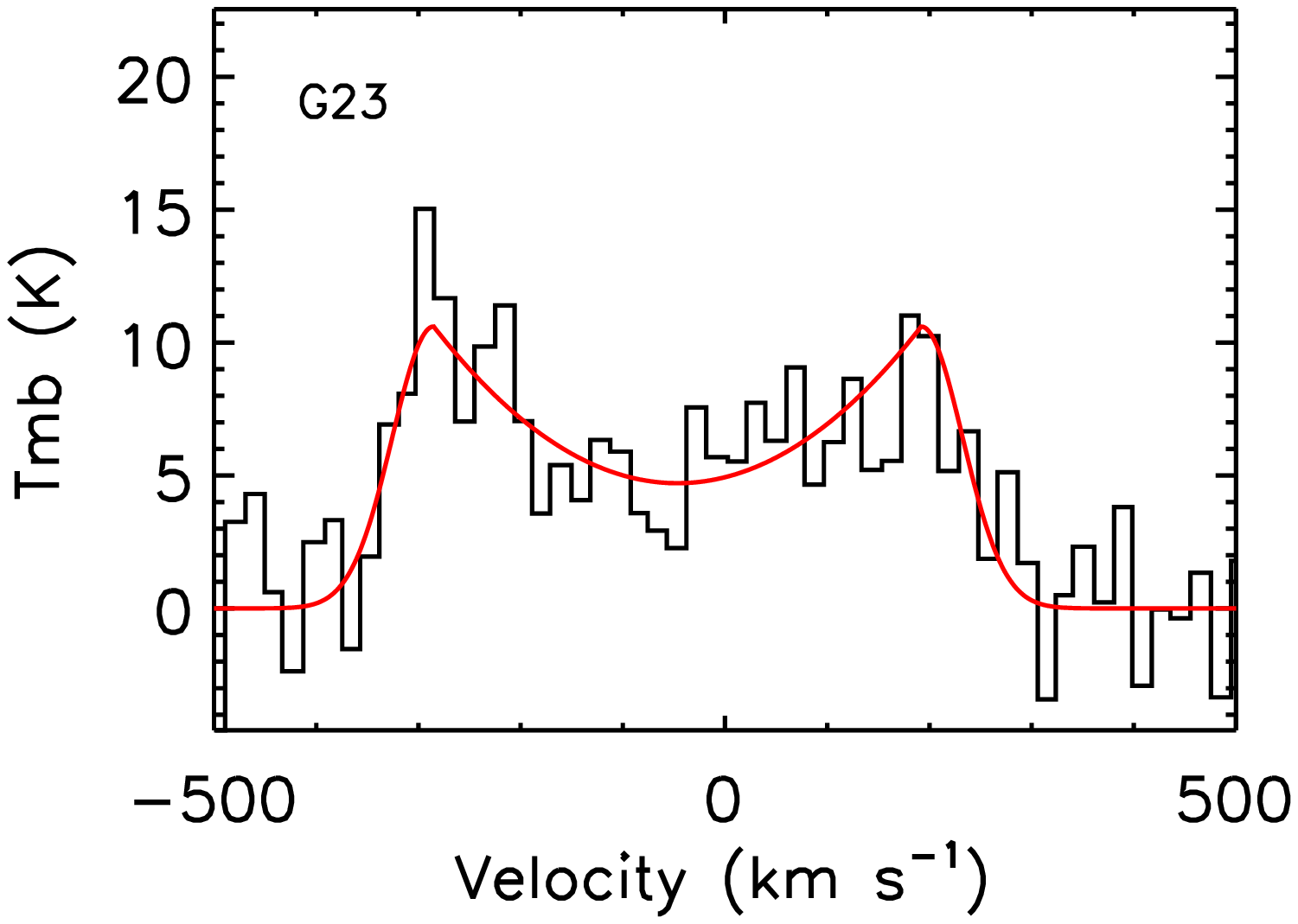}
  \includegraphics[width=5.7cm,clip=]{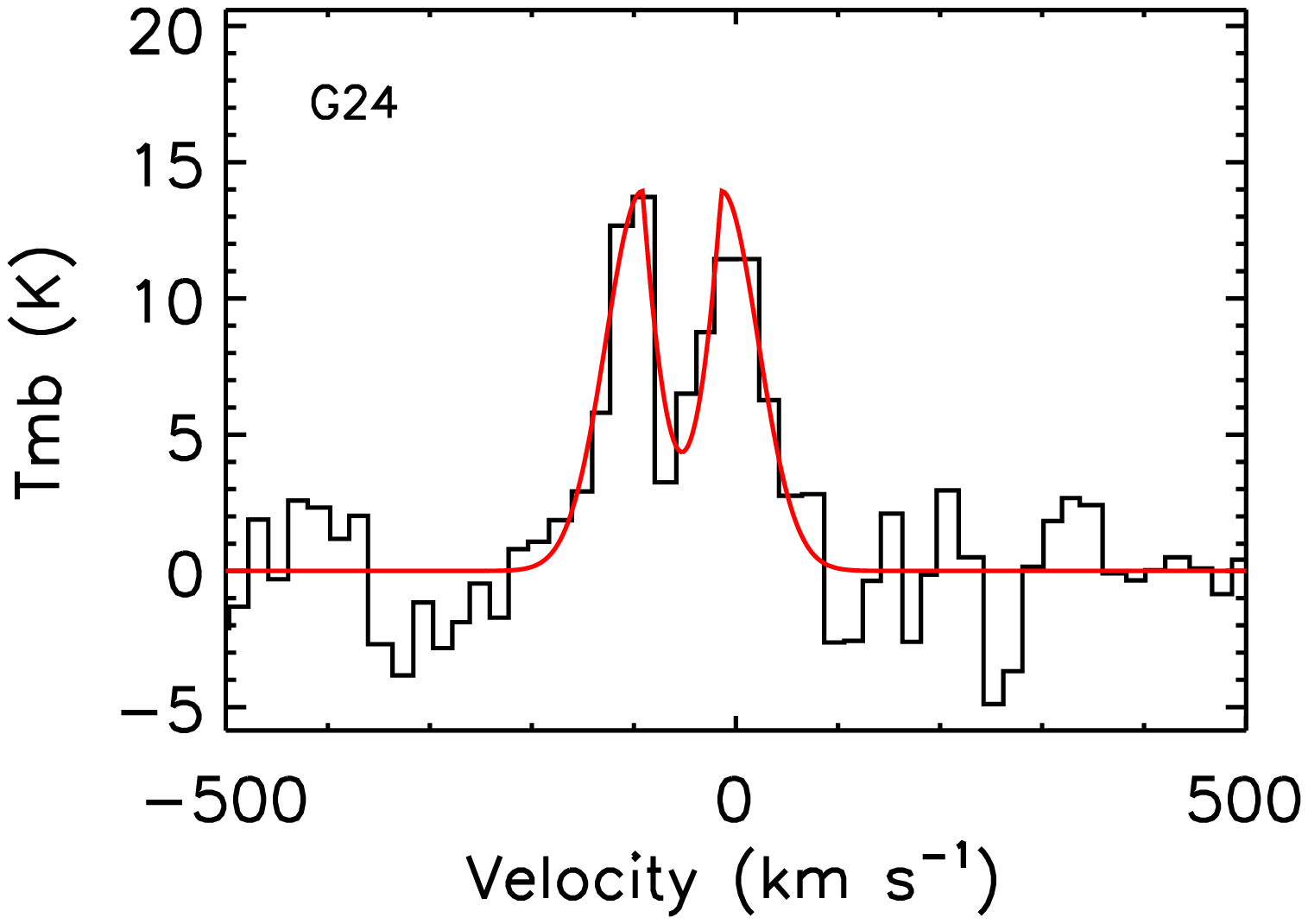}\\
  \vspace{-10pt}
  \hspace{-17pt}
  \includegraphics[width=5.7cm,clip=]{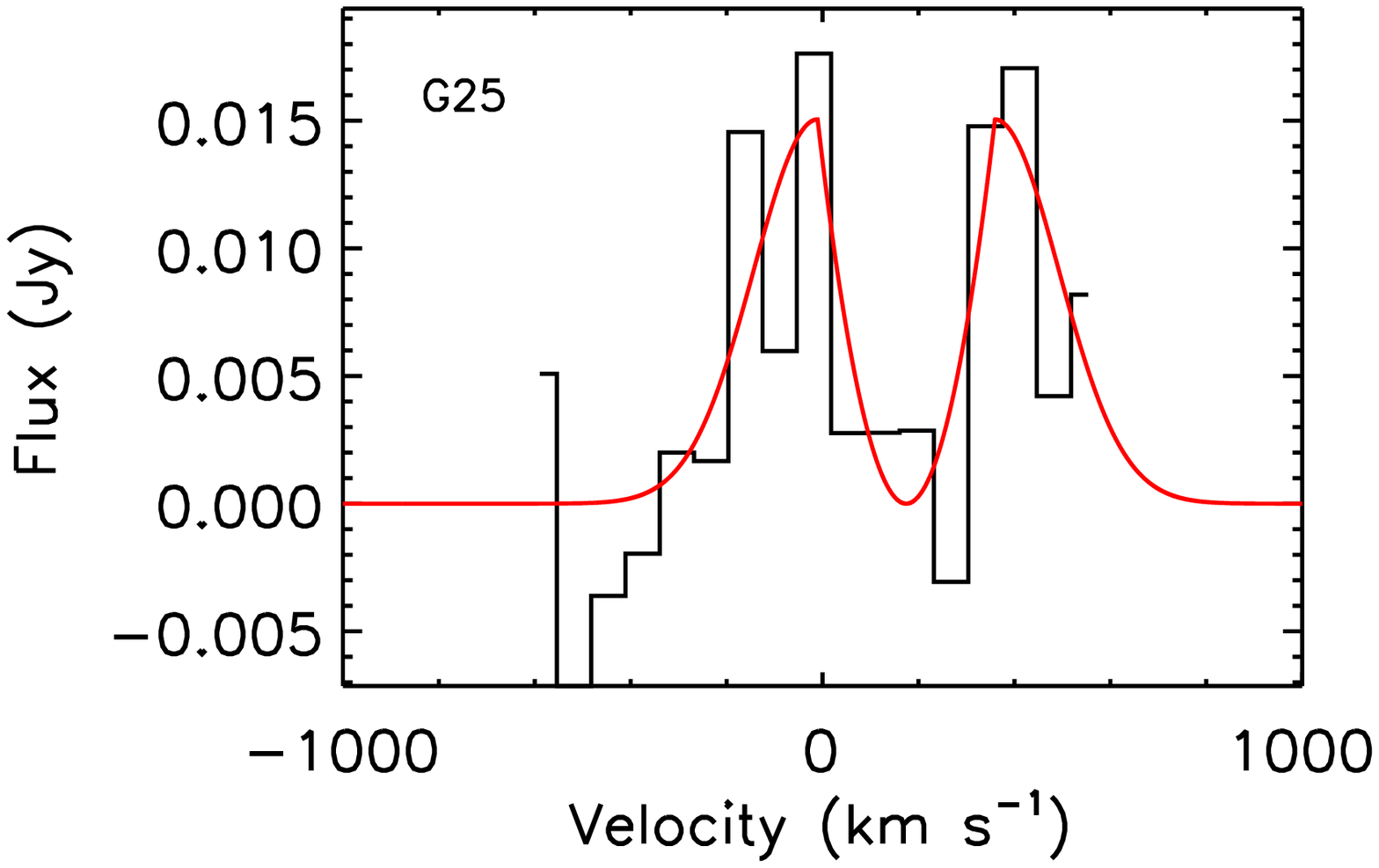}
  \includegraphics[width=5.7cm,clip=]{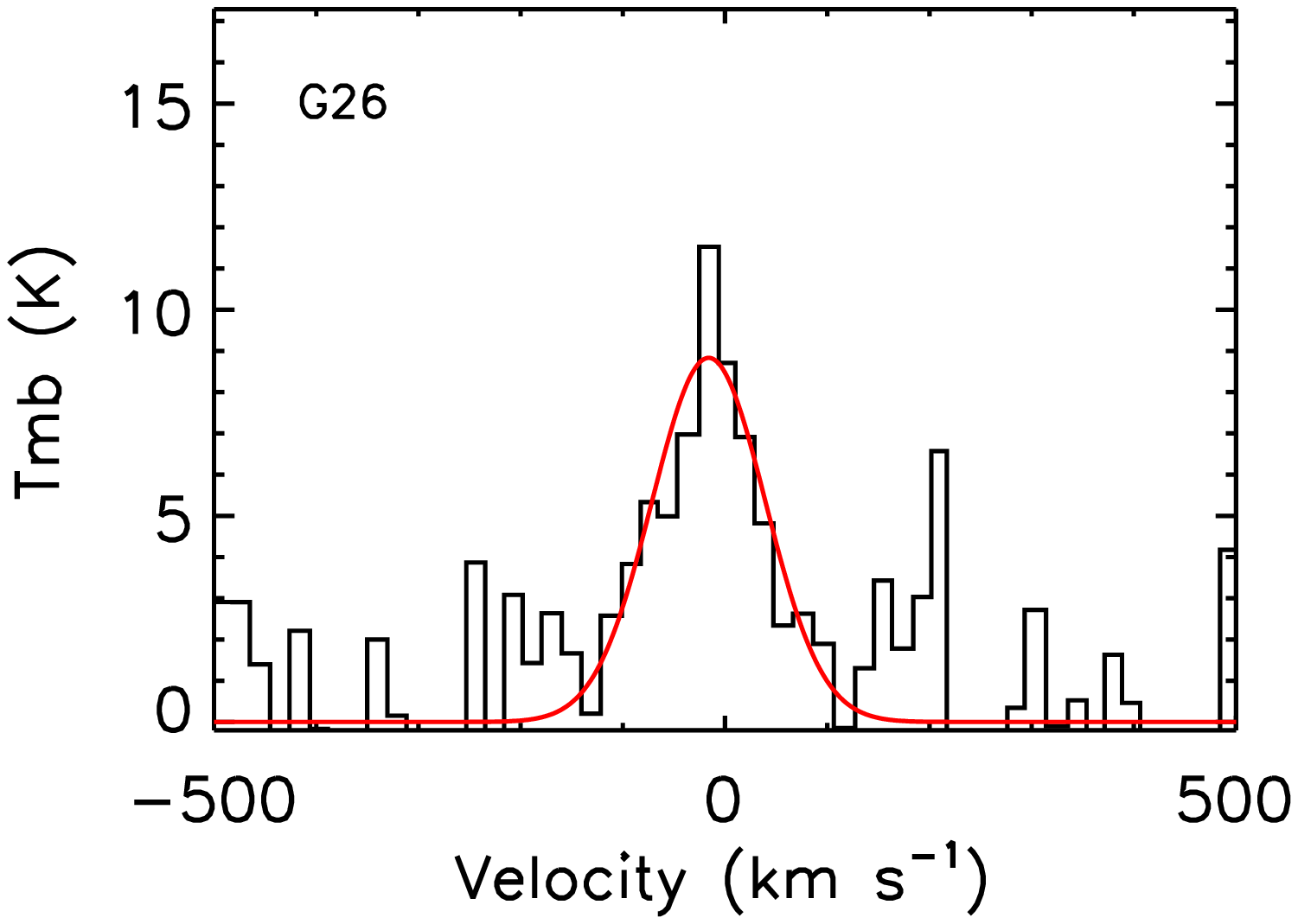}
   \includegraphics[width=5.7cm,clip=]{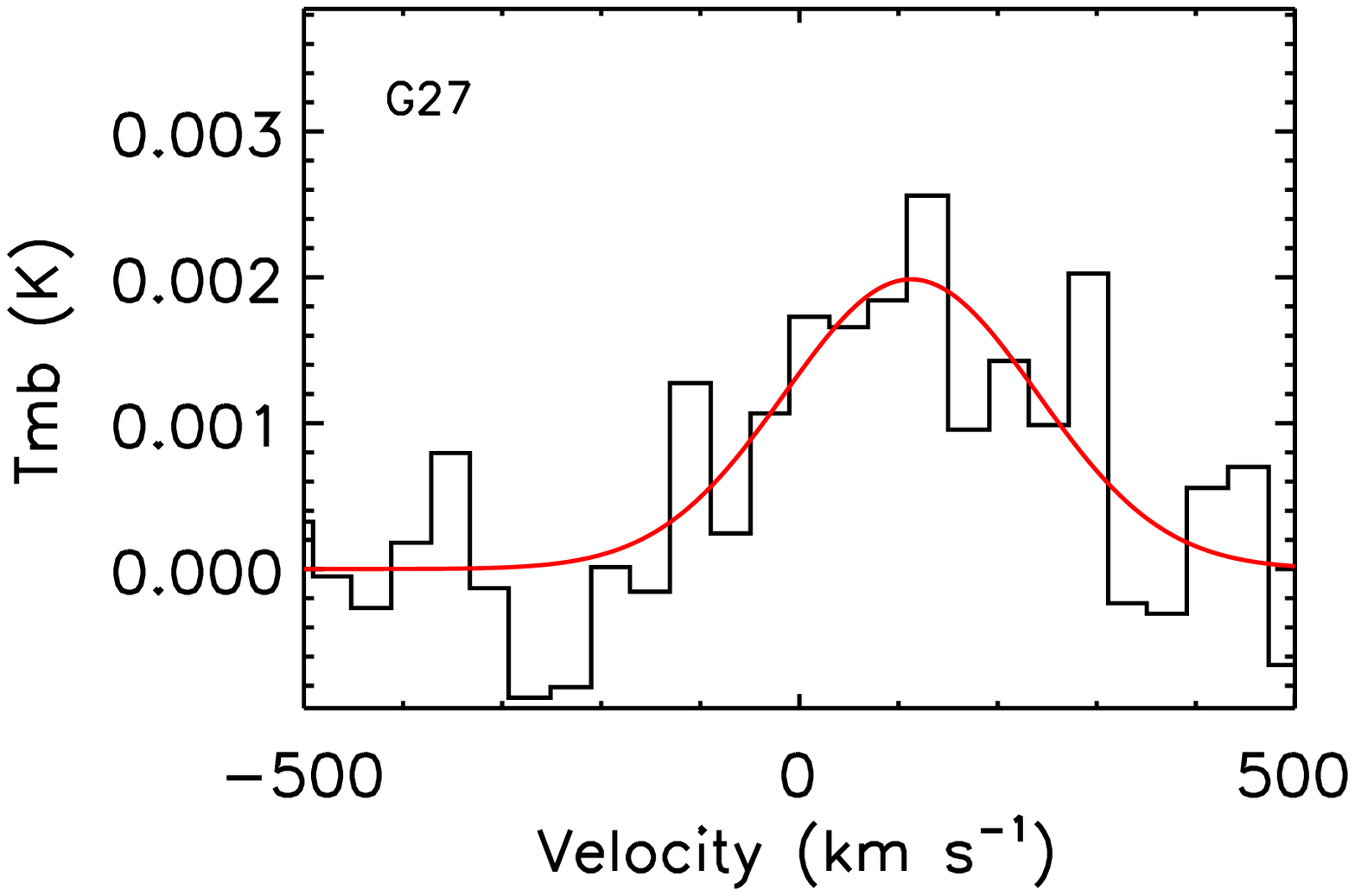}\\
     \vspace{-10pt}
  \hspace{-17pt}
  \includegraphics[width=5.7cm,clip=]{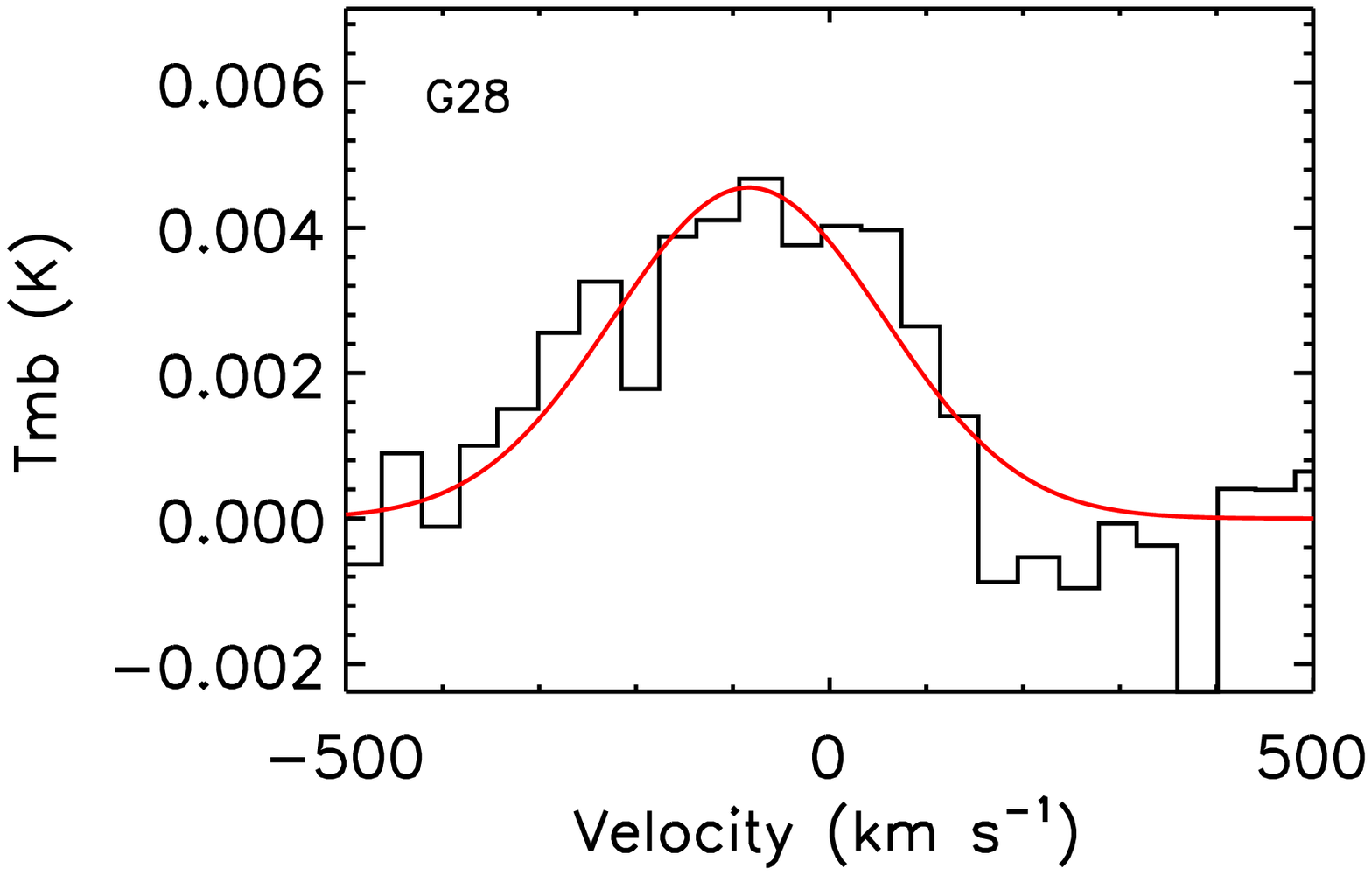}
   \includegraphics[width=5.7cm,clip=]{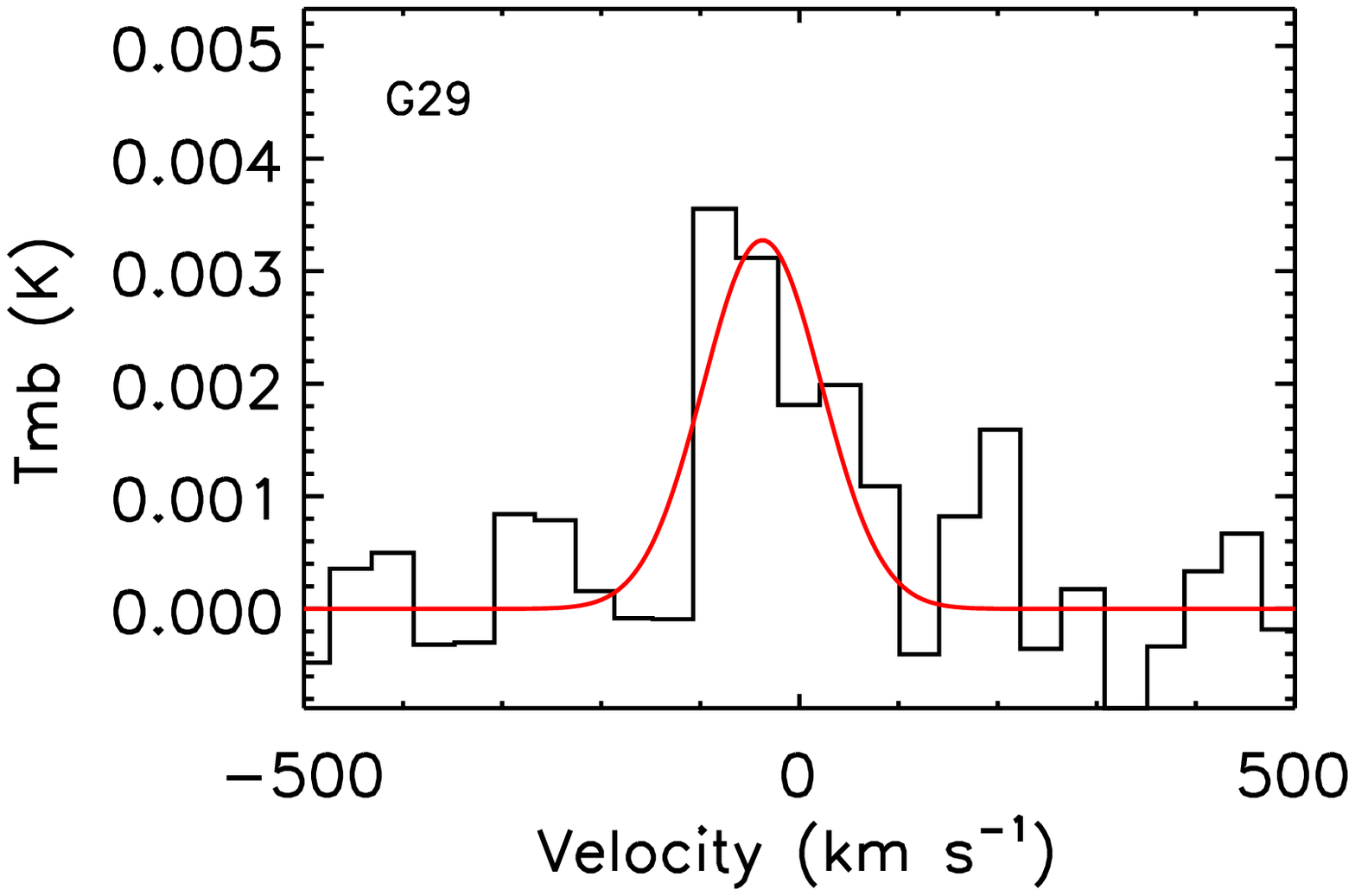}
    \includegraphics[width=5.7cm,clip=]{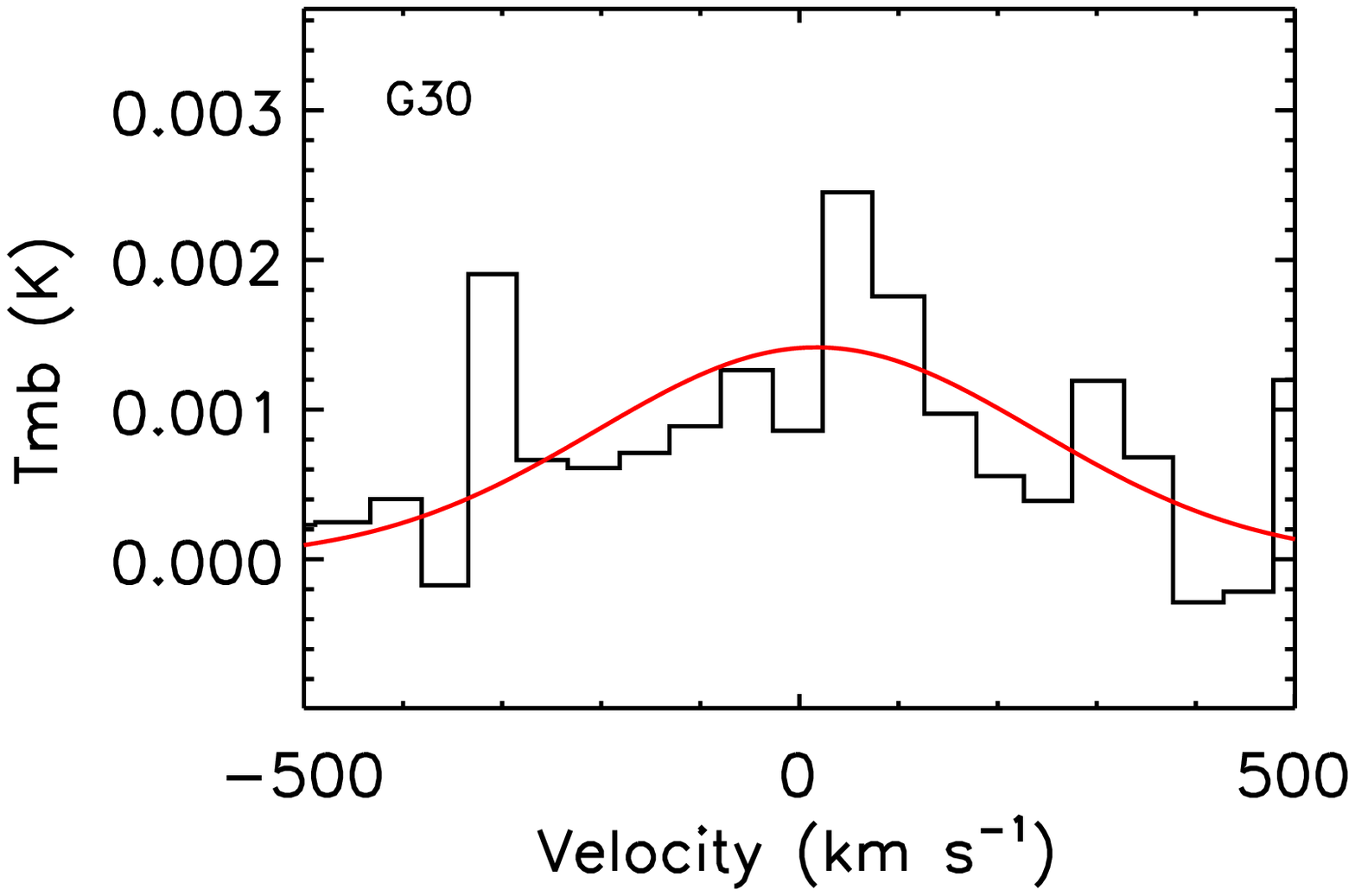}\\
  \caption{Continued.}
\end{figure*}
\addtocounter{figure}{-1}
\begin{figure*}
  \hspace{-15pt}
  \includegraphics[width=5.8cm,clip=]{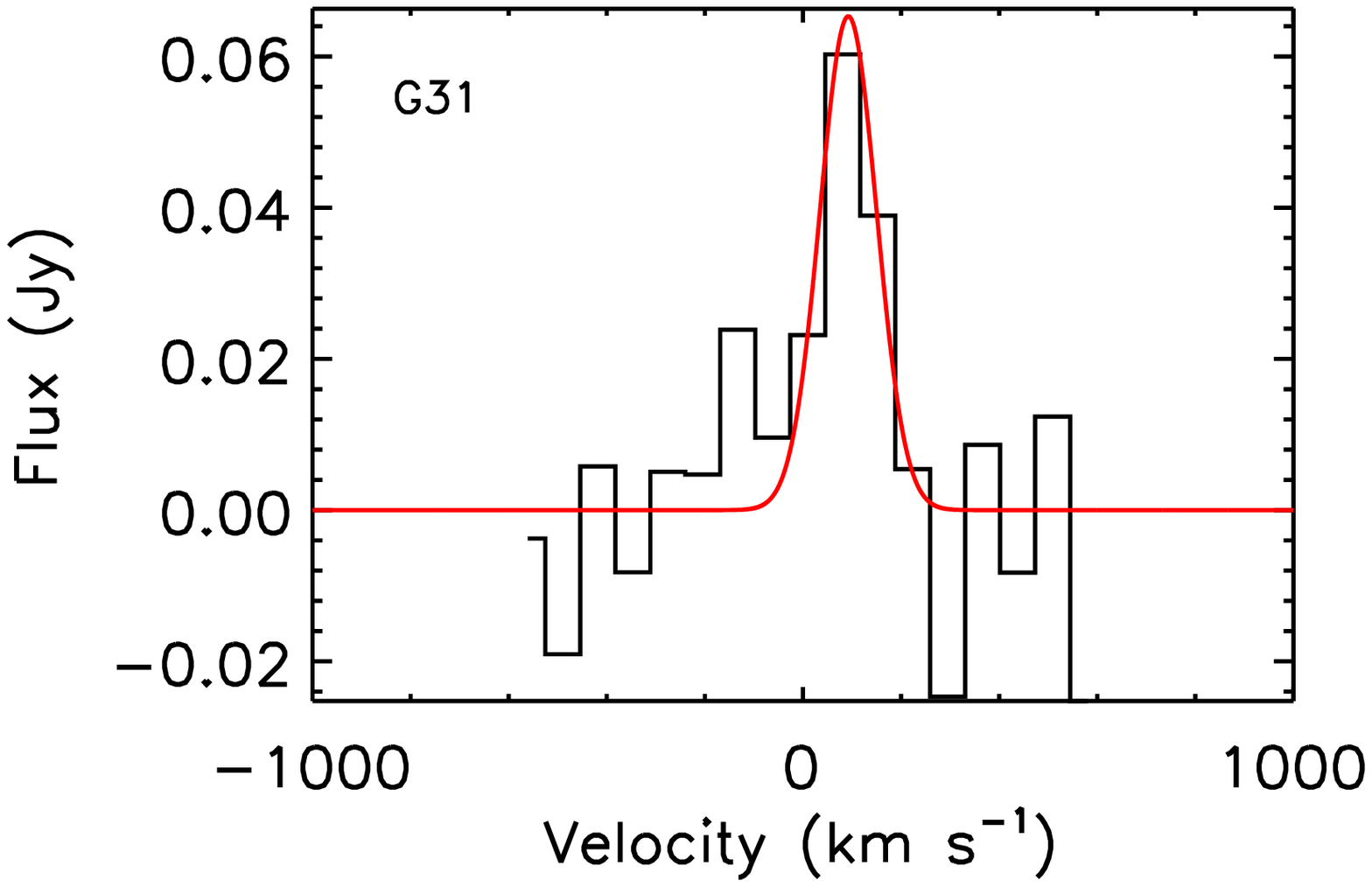}
  \includegraphics[width=5.8cm,clip=]{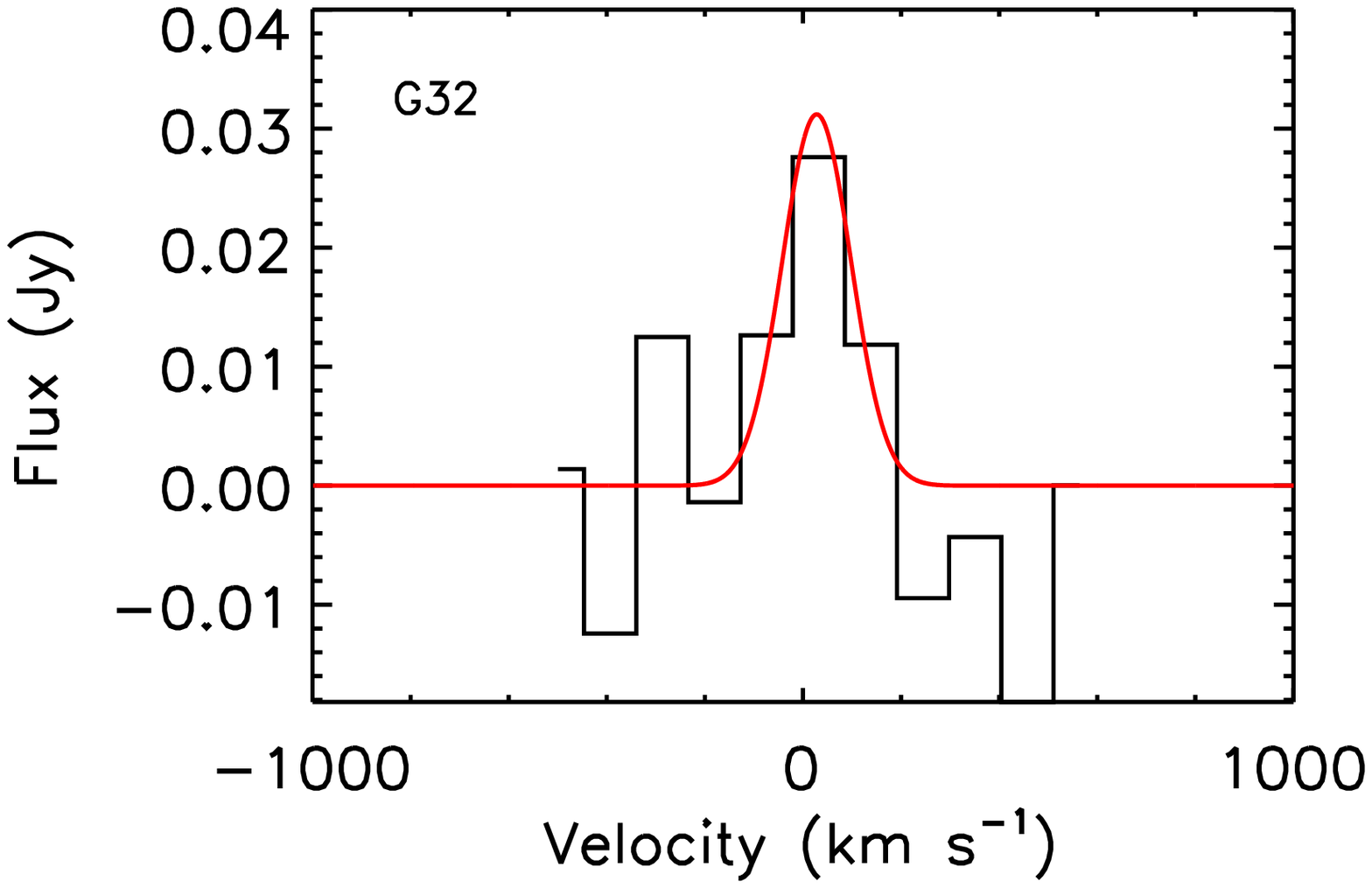}
  \includegraphics[width=5.8cm,clip=]{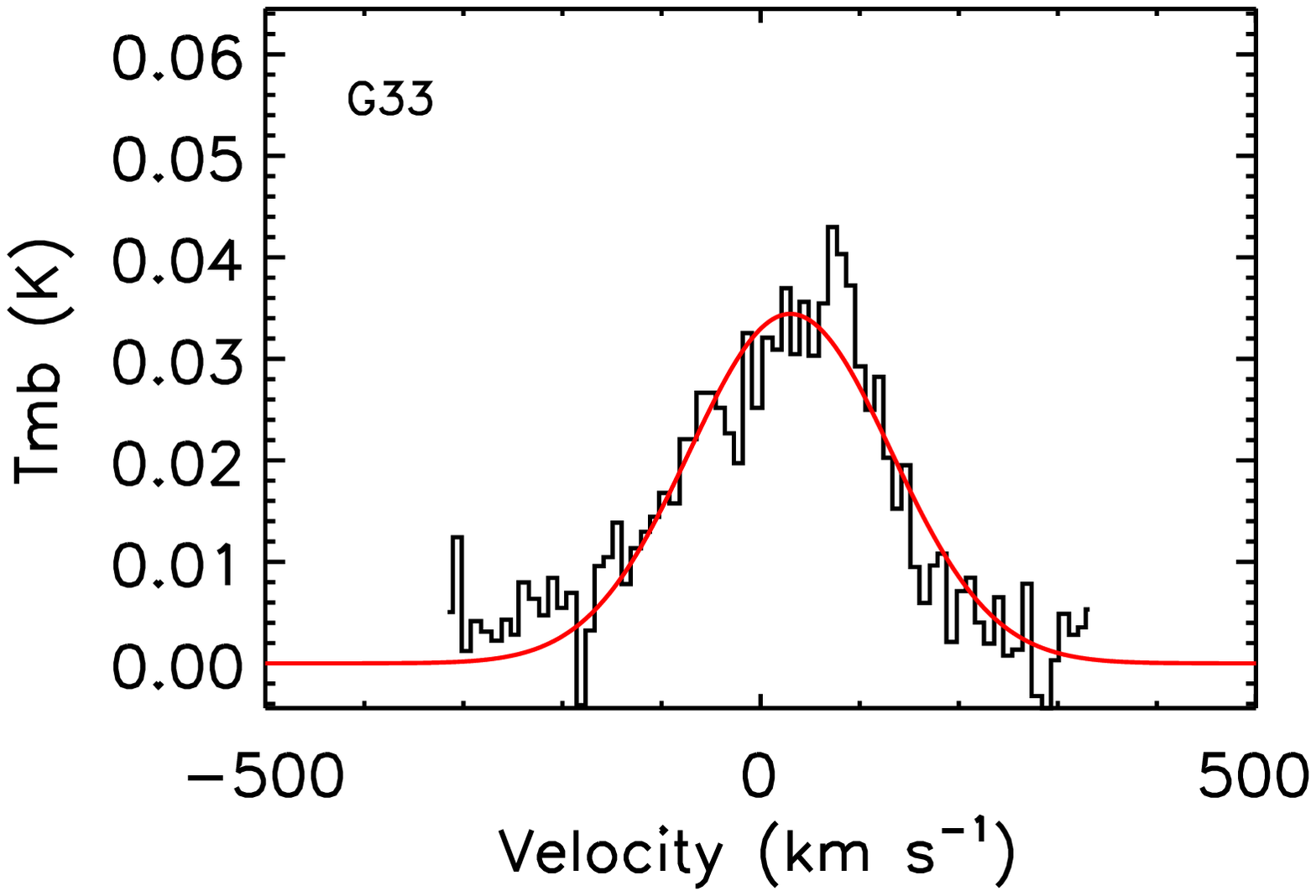}\\
  \vspace{-10pt}
  \hspace{-15pt}
   \includegraphics[width=5.8cm,clip=]{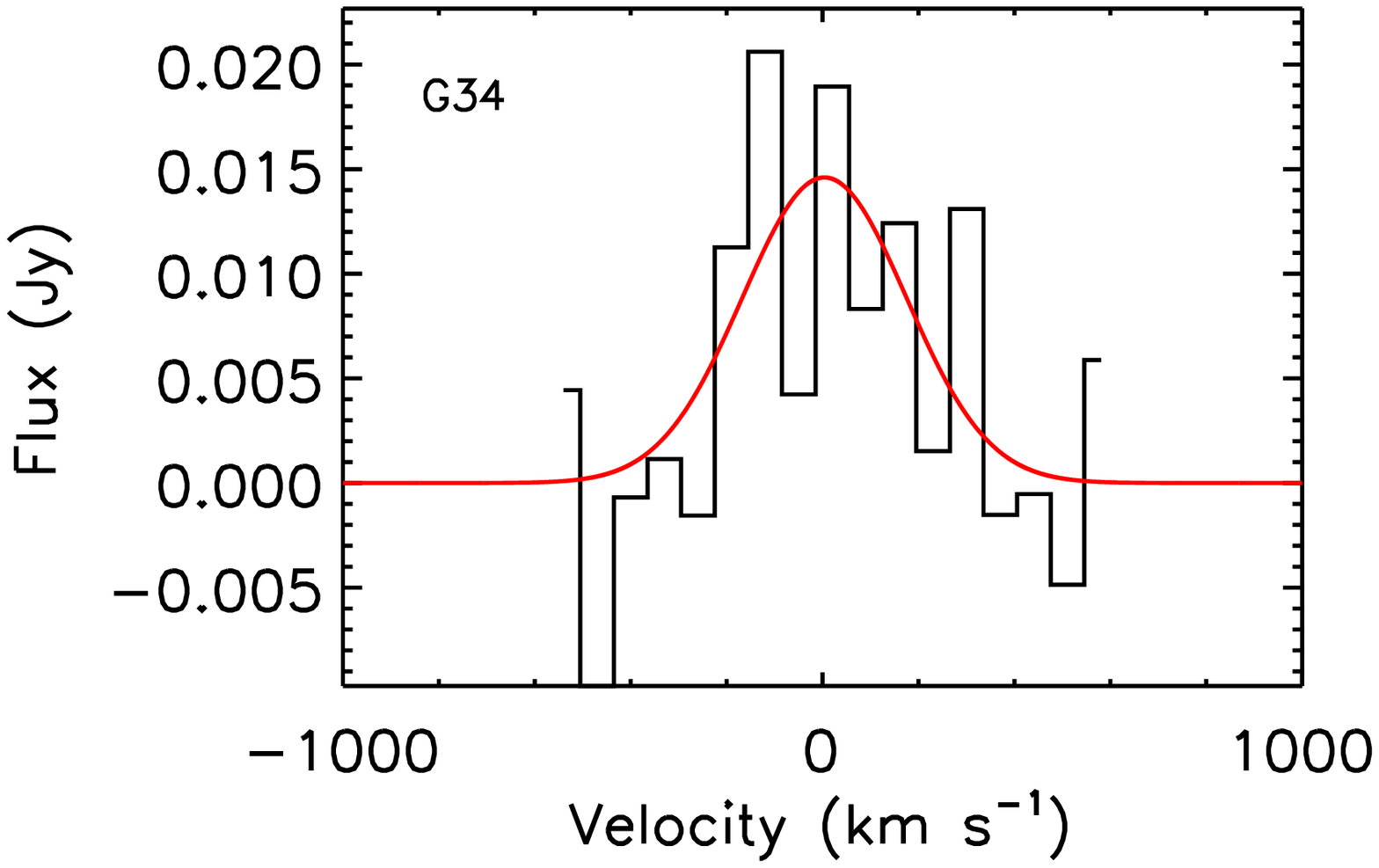}
  \includegraphics[width=5.8cm,clip=]{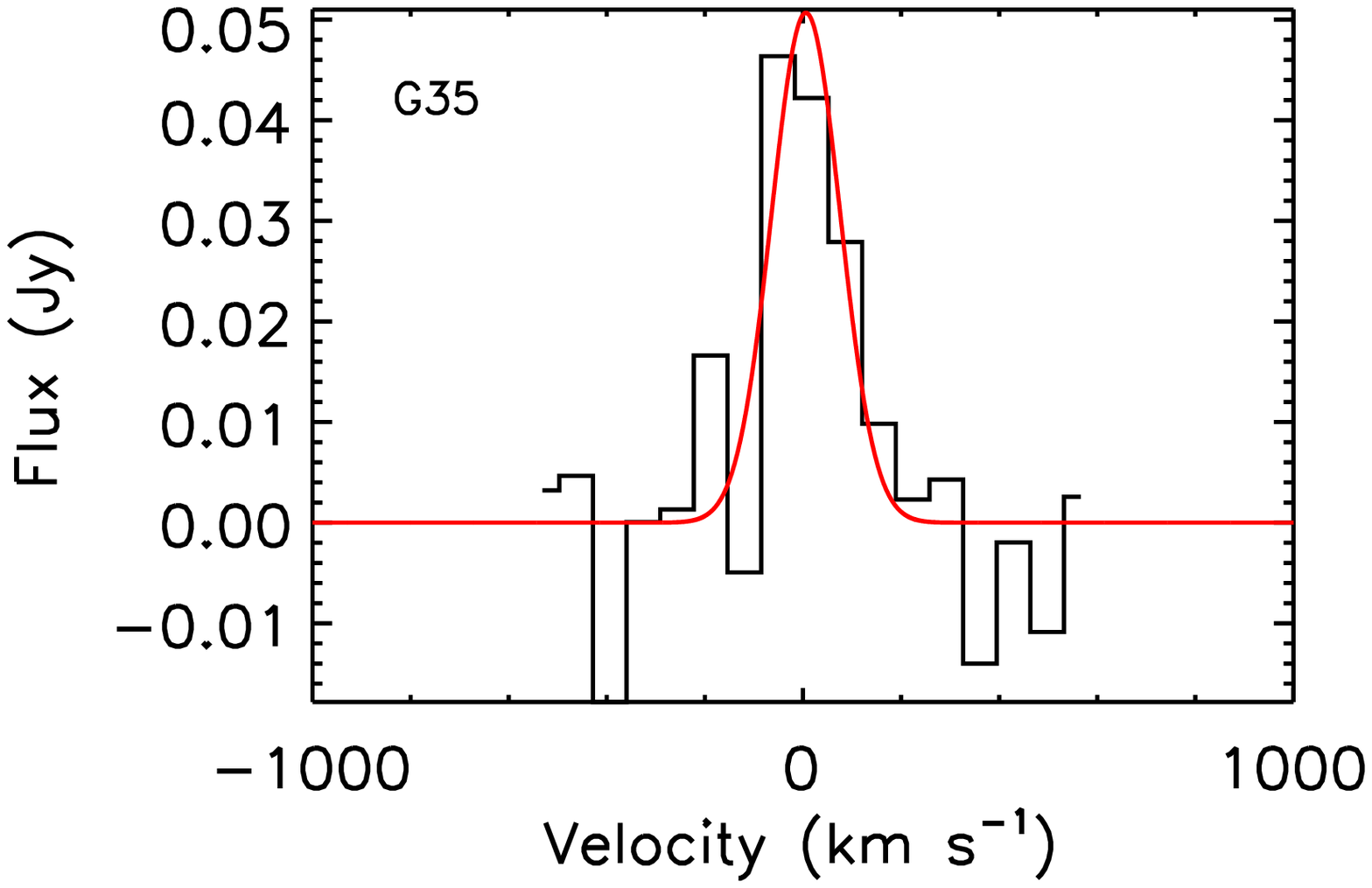}
  \includegraphics[width=5.8cm,clip=]{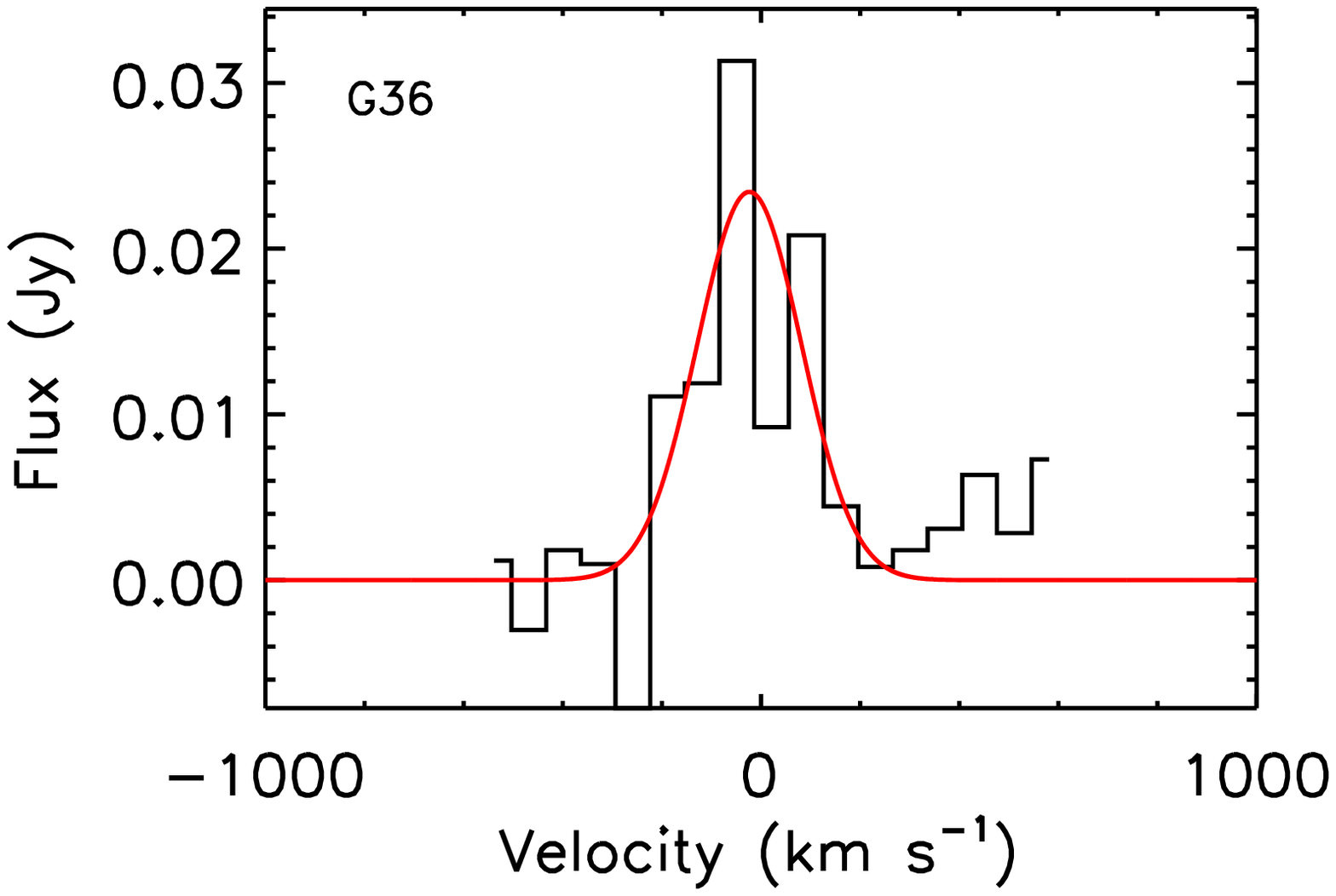}\\
   \vspace{-10pt}
  \hspace{-15pt}
   \includegraphics[width=5.8cm,clip=]{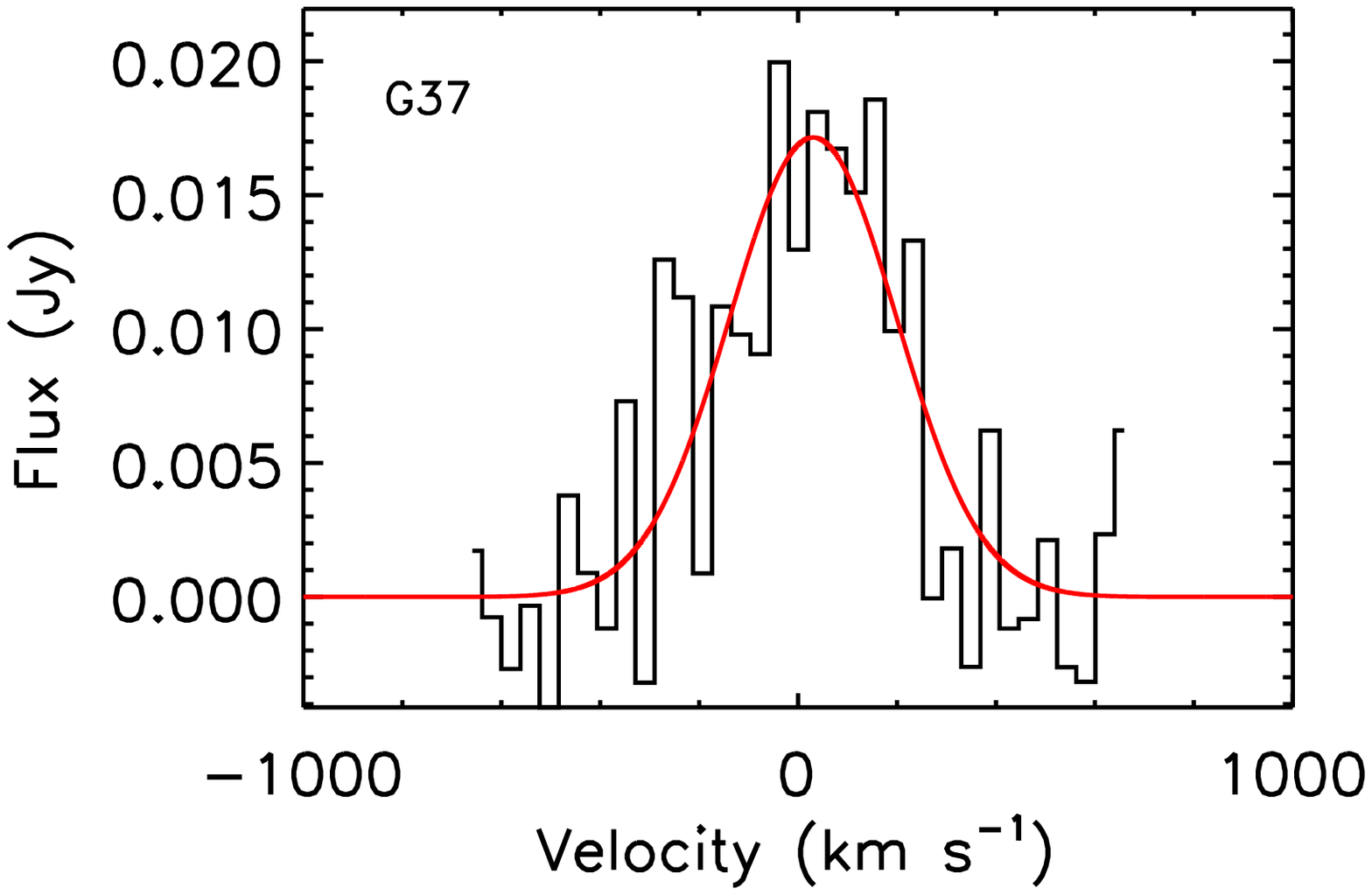}
  \includegraphics[width=5.8cm,clip=]{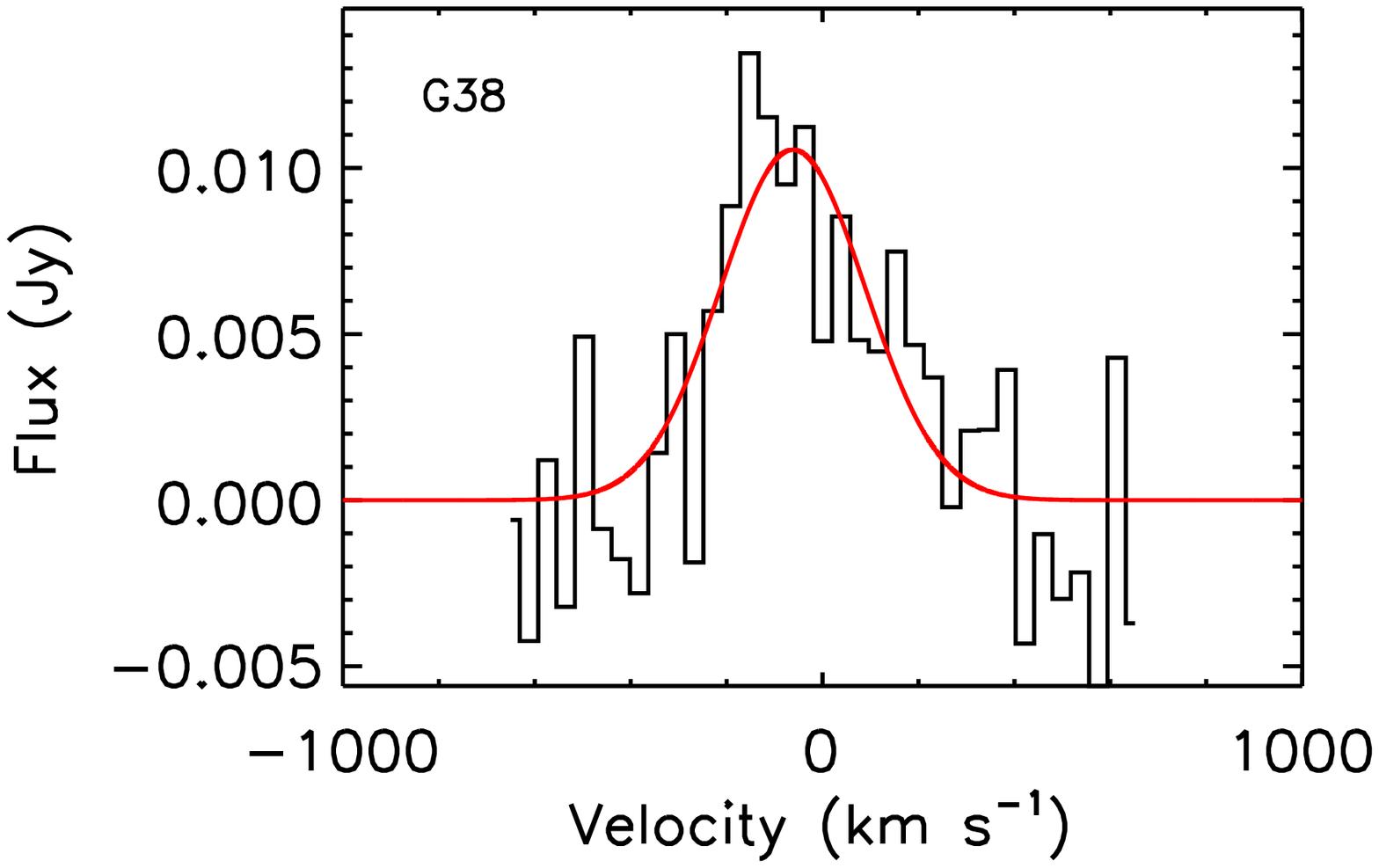}
  \includegraphics[width=5.8cm,clip=]{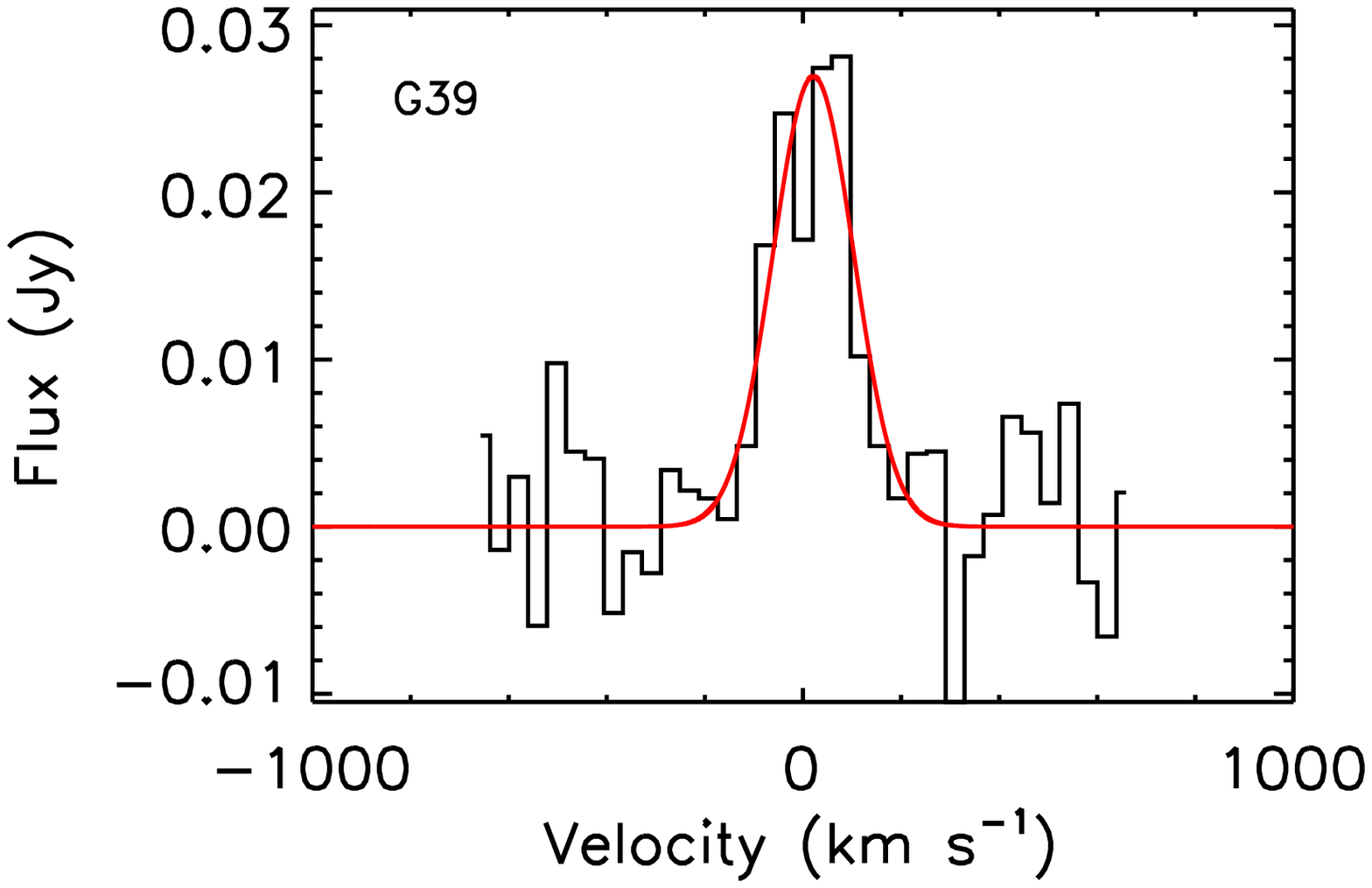}\\
   \vspace{-10pt}
  \hspace{-15pt}
   \includegraphics[width=5.8cm,clip=]{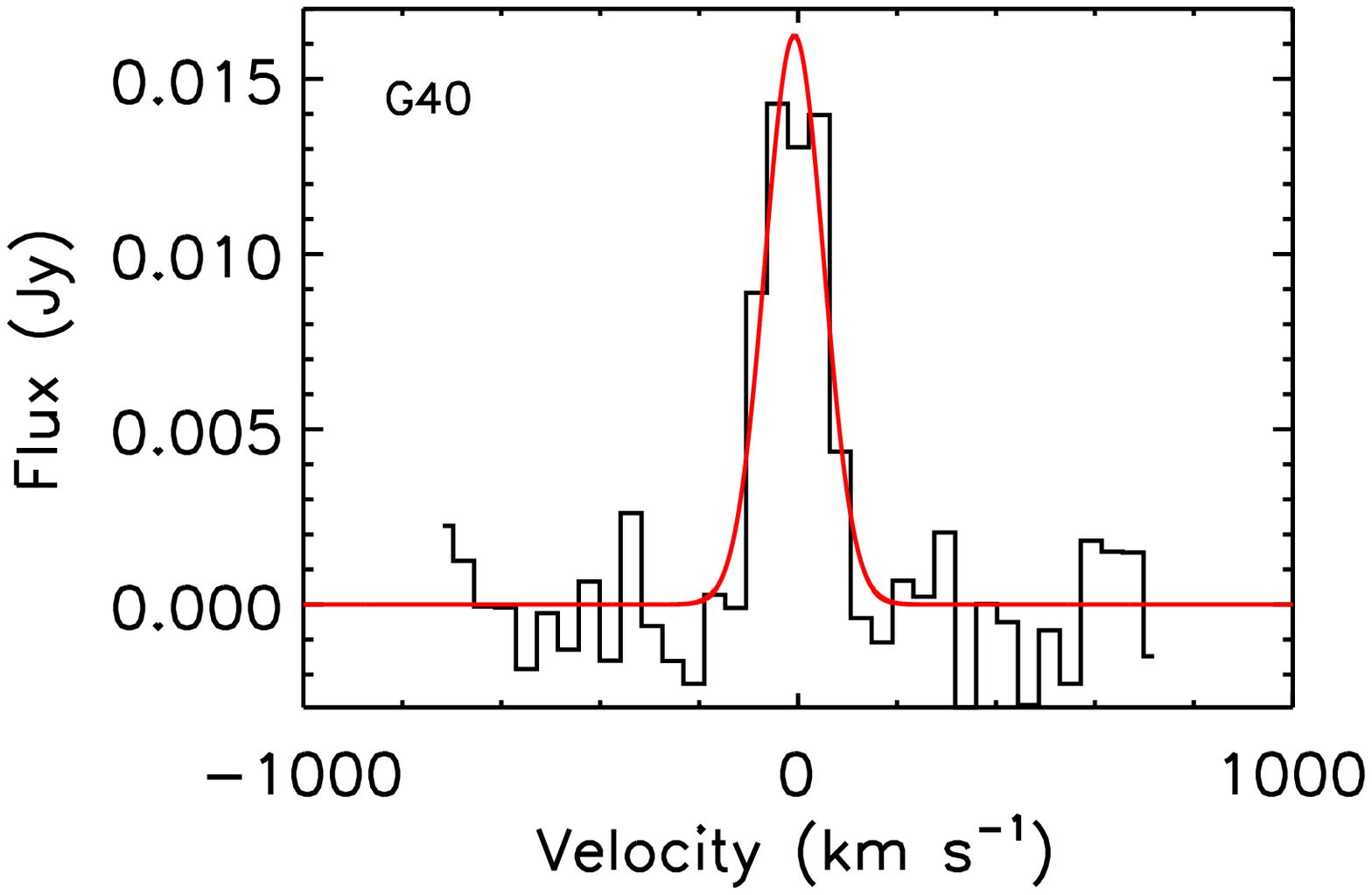}\\
  \caption{Continued.}
\end{figure*}
\label{lastpage}
\end{document}